\documentclass[preprint,prd,aps,showpacs,showkeys,nofootinbib]{revtex4}
\usepackage{amssymb}
\usepackage{amsmath}
\usepackage{graphicx}
\usepackage{float}
\usepackage{subfigure}
\usepackage{booktabs}
\usepackage{multirow}
\usepackage{appendix}
\usepackage{booktabs}
\usepackage{graphicx} 
\usepackage{amsmath,bm}
\usepackage{textcomp} 
\usepackage{fmtcount}
\usepackage{xcolor}

\renewcommand{\imath}{\mathrm{i}}

\begin{document}
\title{Lepton-flavor violation in the MRSSMSeesaw}

\author{Hao-Yi Liu$^{1,3,4}$\footnote{3338183428@qq.com},  Jin-Lei Yang$^{1,3,4}$\footnote{jlyang@hbu.edu.cn},Ke-Sheng Sun$^6$\footnote{sunkesheng@126.com;\;sunkesheng@bdu.edu.cn}, Tai-Fu Feng$^{1,2,3,4,5}$\footnote{fengtf@hbu.edu.cn}}

\affiliation{$^1$ Department of Physics, Hebei University, Baoding 071002, China}
\affiliation{$^2$Department of Physics, Guangxi University, Nanning, 530004, China}
\affiliation{$^3$ Hebei Key Laboratory of High-precision Computation and Application of Quantum Field Theory, Baoding, 071002, China}
\affiliation{$^4$ Hebei Research Center of the Basic Discipline for Computational Physics, Baoding, 071002, China}
\affiliation{$^5$ Department of Physics, Chongqing University, Chongqing 401331, China}
\affiliation{$^6$Department of Physics, Baoding University, Baoding, 071000,China}

\begin{abstract}
  The Minimal R-symmetric Supersymmetric Standard Model with Seesaw (MRSSMSeesaw) extends the MRSSM by incorporating right-handed neutrinos to generate neutrino masses via the Type-I seesaw mechanism. This work presents a detailed analysis of lepton flavor violation (LFV) processes, including $\ell_i \rightarrow \ell_j \gamma$, $\ell_i \rightarrow 3\ell_j$, and Higgs decays $h \rightarrow \ell_i \ell_j$, Based on the current experimental limitations, we carry out detailed parameter scanning and numerical calculations to analyse the effects of different sensitive parameters on LFV.The numerical results show that the non-diagonal elements involving the initial and final
  leptons are main sensitive parameters and LFV sources.

  \end{abstract}

  \date{\today}

  \maketitle
  
\section{Introduction}
  \indent\indent
Supersymmetry (SUSY) stands as one of the most theoretically compelling frameworks for physics beyond the Standard Model (BSM)\cite{Haag:1974qh}.However SUSY might not be realized in its minimal form, the Minimal Supersymmetry Standard Model(MSSM).Recent developments have highlighted the Minimal R-symmetric SUSY Standard Model (MRSSM) as a particularly promising alternative of Beyond-MSSM.It is based on a continuous unbroken $U(1)_R$ symmetry underDepartment of Physics, Baoding University, Baoding, 071000,China
which the superparticles are R-charged~\cite{Diessner:2014ksa}. It involves Dirac gauginos, N = 2 SUSY multiplets in the gauge and Higgs sectors, and supersoft SUSY breaking~\cite{Fox:2002bu,Kribs:2007ac,Itoyama:2011zi,Itoyama:2013sn,Itoyama:2013vxa}.Furthermore, the model can provide much more candidates for the dark matter comparing that with the MSSM~\cite{Buckley:2013sca,Chun:2009zx,Belanger:2009wf}. While at MRSSM,it doesn't have neutrinos mass,we then have extended the MRSSM  by incorporating right-handed neutrino fields, The right-handed neutrino field couples with the $S$ field. After the $S$ field acquires a vacuum expectation value, it gives mass to the right-handed neutrinos, and then the seesaw mechanism is employed to generate mass for the left-handed neutrinos~\cite{King:2013eh,Strumia:2006db,Mohapatra:2006gs,SHiP:2015vad,Schechter:1980gr}.

Low-energy lepton physics represents a promising frontier for fundamental discoveries in the coming years, as intriguing anomalies and deviations from SM predictions continue to accumulate in lepton-related observables.Among the various lepton flavour violating (LFV) processes studied in the literature, the most phenomenologically significant are the radiative decays $\ell^{-}\to \ell^{-}\gamma$, and $\ell^{-}\to \ell^{-}\ell^{-}\ell^{+}$, as their branching fractions can be measured with exceptional precision.In the table~\ref{tab1},we show the present experimental limits and future
sensitivities forDepartment of Physics, Baoding University, Baoding, 071000,China the LFV processes.Furthermore, we are alse interested here in the LFV of Higgs boson, $h\to e\mu$,$h\to e\tau$ and $h\to \mu \tau$.~\cite{Fok:2010vk,Sun:2019wii,Kotlarski:2019muo,Sun:2019qiw,Kotlarski:2019evq}.We have calculated the $\mu$,$\tau$ and Higgs LFV processes. And we discussed the impact of new parameters.

The paper is organized as follows. In Sec.II, the main ingredients of the MRSSMSeesaw are summarized briefly by introducing the superpotential and the general soft breaking terms. We present the analysis on the decay width of the rare LFV processes and the Higgs LFV in Sec.III. The numerical analyses are given in Sec.IV, and Sec.V gives a summary.

\begin{table*}
  \begin{tabular*}{\textwidth}{@{\extracolsep{\fill}}lll@{}}
  \hline
  LFV process & Present limit & Future sensitivity\\
  \hline
  $\mu\rightarrow e\gamma$ & $<4.2\times10^{-13}$~\cite{MEG:2016leq} & $\sim6\times10^{-14}$~\cite{Baldini:2013ke}\\
  $\mu\rightarrow 3e$ & $<1\times10^{-12}$~\cite{SINDRUM:1987nra} & $\sim10^{-16}$~\cite{Blondel:2013ia}\\
  $\tau\rightarrow e\gamma$ & $<3.3\times10^{-8}$~\cite{Georgi:1974sy} & $\sim10^{-8}-10^{-9}$~\cite{Langacker:1980js}\\
  $\tau\rightarrow 3e$ & $<2.7\times10^{-8}$~\cite{Haber:1984rc} & $\sim10^{-9}-10^{-10}$~\cite{Hayasaka:2013dsa}\\
  $\tau\rightarrow \mu\gamma$ & $<4.2\times10^{-8}\cite{Georgi:1974sy}$& $\sim10^{-8}-10^{-9}$~\cite{Hayasaka:2013dsa}\\
  $\tau\rightarrow 3\mu$ & $<2.1\times10^{-8}$~\cite{Haber:1984rc}& $\sim10^{-9}-10^{-10}$~\cite{Hayasaka:2013dsa}\\
  \hline
  \end{tabular*}
  \caption{Present limits and future sensitivities for the branching ratios for the LFV processes.}
  \label{tab1}
  \end{table*}

  \section{Details of the MRSSMSeesaw}

  \begin{table}[th]
    \begin{center}
      \begin{tabular}{ccccc} 
       \hline 
      chiral superfield & R-charge & spin 0 & spin \(\frac{1}{2}\) & \(U(1)_Y \otimes\, SU(2)_L \otimes\, SU(3)_C\)\\
      \midrule 
      \hline
      \(\hat{q}\) & 1 & \(\tilde{q}\) & \(q\) & \(\left(\frac{1}{6},{\mathbf 2},{\mathbf 3} \right) \)\\ 
      \(\hat{l}\) & 1 & \(\tilde{l}\) & \(l\) & \(\left(-\frac{1}{2},{\mathbf 2},{\mathbf 1}\right) \) \\ 
      \(\hat{H}_d\) & 0 & \(H_d\) & \(\tilde{H}_d\) & \(\left(-\frac{1}{2},{\mathbf 2},{\mathbf 1}\right) \) \\
      \(\hat{H}_u\) & 0 & \(H_u\) & \(\tilde{H}_u\) & \(\left(\frac{1}{2},{\mathbf 2},{\mathbf 1} \right) \) \\
      \(\hat{d}\) & 1 & \(\tilde{d}_R^*\) & \(d_R^*\) & \(\left(\frac{1}{3},{\mathbf 1},{\mathbf \overline{3}}\right) \) \\
      \(\hat{u}\) & 1 & \(\tilde{u}_R^*\) & \(u_R^*\) & \(\left(-\frac{2}{3},{\mathbf 1},{\mathbf \overline{3}}\right) \) \\
      \(\hat{e}\) & 1 & \(\tilde{e}_R^*\) & \(e_R^*\) & \(\left(1,{\mathbf 1},{\mathbf 1}\right) \) \\
      \midrule 
      \hline
      \(\hat{S}\) & 0 & \(S\) & \(\tilde{S}\) & \(\left(0,{\mathbf 1},{\mathbf 1}\right) \) \\
      \(\hat{T}\) & 0 & \(T\) & \(\tilde{T}\) & \(\left(0,{\mathbf 3},{\mathbf 1}\right) \) \\
      \(\hat{O}\) & 0 & \(O\) & \(\tilde{O}\) & \(\left(0,{\mathbf 1},{\mathbf 8}\right) \) \\
      \(\hat{R}_d\) & 2 & \(R_u\) & \(\tilde{R}_d\) & \(\left(\frac{1}{2},{\mathbf 2},{\mathbf 1}\right) \) \\
      \(\hat{R}_u\) & 2 & \(R_d\) & \(\tilde{R}_u\) & \(\left(-\frac{1}{2},{\mathbf 2},{\mathbf 1}\right) \)\\
      \(\hat{\nu}\) & 1 & \(\tilde{\nu}_R^*\) & \(\nu_R^*\) & \(\left(0,{\mathbf 1},{\mathbf 1}\right) \) 
      \\
      \\
      \hline
      vector superfield & R-charge & spin \(\frac{1}{2}\) & spin 1 & \(U(1)_Y \otimes\, SU(2)_L \otimes\, SU(3)_C\)\\
      \midrule 
      \hline
        $\hat B$ & 0 & $\tilde B^0$ & $B^0$ & \(\left(1,{\mathbf 1},{\mathbf 1}\right) \)\\
      $\hat W$ & 0 & $\tilde W^\pm$, $\tilde W^0$ & $W^\pm$, $W^0$ & \(\left(1,{\mathbf 3},{\mathbf 1}\right) \)\\
      $\hat g$ & 0 & $\tilde g$ & $g$ & \(\left(0,{\mathbf 1},{\mathbf 8}\right) \)
      \end{tabular} 
      \caption{Field content of the MRSSMSeesaw. R-charges in the second column correspond to the entire superfield.}
      \label{mrssm_field_content}
    \end{center}
    \end{table}
    The superpotential of the MRSSMSeesaw is structured as follows:
    
    \begin{eqnarray}
    \mathcal{W}_{MRSSMSeesaw} &=& \mu_d\hat{R}_d \hat{H}_d+\mu_u \hat{R}_u \hat{H}_u+\Lambda_d \hat{R}_d \hat{T}\hat{H}_d+\Lambda_u \hat{R}_u\hat{T}\hat{H}_u\nonumber\\
    &+&\lambda_d\hat{S}\hat{R}_d\hat{H}_d+\lambda_u\hat{S} \hat{R}_u\hat{H}_u-Y_d\hat{d}\hat{q}\hat{H}_d-Y_e\hat{e}\hat{l}\hat{H}_d+Y_u\hat{u}\hat{q}\hat{H}_u \nonumber\\ &+&Y_x\hat{\nu}\hat{\nu} \hat{S}+Y_\nu\hat{l}\hat{H}_u\hat{\nu}.
    \label{eq:W}
    \end{eqnarray}

    The first and second line represent the superpotential in MRSSM where $\hat{H}_u$ and $\hat{H}_d$ are the MSSM-like Higgs weak iso-doublets, $\hat{R}_u$ and $\hat{R}_d$ are the $R$-charged Higgs $SU(2)_L$ doublets and the corresponding Dirac higgsino mass parameters are denoted as $\mu_u$ and $\mu_d$. $Y_e$, $Y_u$ and $Y_d$ are the Yukawa couplings of charged lepton, up type quark and down type quark, respectively. $\lambda_u$, $\lambda_d$, $\Lambda_u$ and $\Lambda_d$ are parameters of Yukawa-like trilinear terms involving the singlet $\hat{S}$ and the triplet $\hat{T}$, which is given by
    \begin{equation}
    \hat{T} = \left(
    \begin{array}{cc}
    \hat{T}^0/\sqrt{2} &\hat{T}^+ \\
    \hat{T}^-  &-\hat{T}^0/\sqrt{2}\end{array}
    \right).\nonumber
     \end{equation}
    
In order to generate non-zero masses for Neutrinos via the Type-I Seesaw mechanism, we introduce three generations of Neutrino singlets and include the terms $Y_\nu\hat{l}\hat{H}_u\hat{\nu}$ and $Y_x\hat{\nu}\hat{\nu} \hat{S}$ in the superpotential Eq.~(\ref{eq:W}). The complete mass term for the right-Handed neutrino,allowed by the gauge symmetry, are $M_{\nu}\hat{\nu}\hat{\nu}+Y_x\hat{\nu}\hat{\nu}\hat{S}$. We argue that the right-handed Neutrino mass arises entirely from the VEV of the $\hat{S}$ field. This is equivalent to setting $M_{\nu}=0$.

For the phenomenological studies we present the soft breaking terms
\begin{eqnarray}
V_{SB,S} &=& m^2_{H_d}(|H^0_d|^2+|H^{-}_d|^2)+m^2_{H_u}(|H^0_u|^2+|H^{+}_u|^2)+(B_{\mu}(H^-_dH^+_u-H^0_dH^0_u)+h.c.)\nonumber\\
&+&m^2_{R_d}(|R^0_d|^2+|R^{+}_d|^2)+m^2_{R_u}(|R^0_u|^2+|R^{-}_u|^2)+m^2_T(|T^0|^2+|T^-|^2+|T^+|^2)\nonumber\\
&+&m^2_S|S|^2+ m^2_O|O^2|+\tilde{d}^*_{L,i} m_{q,{i j}}^{2} \tilde{d}_{L,j} +\tilde{d}^*_{R,i} m_{d,{i j}}^{2} \tilde{d}_{R,j}+\tilde{u}^*_{L,i}  m_{q,{i j}}^{2} \tilde{u}_{L,j}\nonumber\\
&+&\tilde{u}^*_{R,i}  m_{u,{i j}}^{2} \tilde{u}_{R,j}+\tilde{e}^*_{L,i} m_{l,{i j}}^{2} \tilde{e}_{L,j}+\tilde{e}^*_{R,{i}} m_{r,{i j}}^{2} \tilde{e}_{R,{j}} +\tilde{\nu}^*_{L,i} m_{l,{i j}}^{2} \tilde{\nu}_{L,j}+\tilde{\nu}^*_{R,i} m_{\nu,{i j}}^{2} \tilde{\nu}_{R,j}.
\end{eqnarray}

All trilinear scalar couplings involving Higgs bosons to squarks and sleptons are forbidden due to the R-symmetry. The soft-breaking terms, which describe the Dirac mass terms for the gauginos and the interaction terms between the adjoint scalars and the auxiliary D-fields of the corresponding gauge multiplet, take the form~\cite{Fox:2002bu}
\begin{equation}
V_{SB,DG}=M^B_D(\tilde{B}\tilde{S}-\sqrt{2}\mathcal{D}_B S)
+M^W_D(\tilde{W}^a\tilde{T}^a-\sqrt{2}\mathcal{D}_W^a T^a)
+M^O_D(\tilde{g}\tilde{O}-\sqrt{2}\mathcal{D}_g^a O^a)+h.c.,
\label{}
\end{equation}
where $\tilde{B}$, $\tilde{W}$ and $\tilde{g}$ are usually MSSM Weyl fermions, $M^B_D$, $M^W_D$ and $M^O_D$ are the mass of bino, wino and gluino, respctively.

For the scalar sector, after the electroweak symmetry breaking, 
(neutral) scalar components of $\hat H_d, \hat H_u, \hat{S}$ and  $\hat T$ acquire vacuum expectation values (VEVs), which we parametrize as:
\begin{align} 
H_d^0=& \, \textstyle{\frac{1}{\sqrt{2}}} (v_d + \phi_{d}+i  \sigma_{d}) \;,& 
H_u^0=& \, \textstyle{\frac{1}{\sqrt{2}}} (v_u + \phi_{u}+i  \sigma_{u}) \;, \nonumber \\ 
T^0  =& \, \textstyle{\frac{1}{\sqrt{2}}} (v_T + \phi_T +i  \sigma_T)   \;,&
S^0   = & \, \textstyle{\frac{1}{\sqrt{2}}} (v_S + \phi_S +i  \sigma_S)   \;.
\label{eq:vevs}
\end{align}
Since R-Higgs bosons carry R-charge 2, they do not develop VEVs.As for the color octet.The masses of the CP-even ($O_S$) and CP-odd ($O_A$) components of the sgluon field $O \equiv (O_S + \imath O_A)/\sqrt{2}$ are split as
\begin{align}
  \label{eq:sgluons_mass_splitting1}
  m_{O_S}^2 =& m_O^2 + 4 (M_O^D)^2 ,\\
  \label{eq:sgluons_mass_splitting2}
    m_{O_A}^2 =& m_O^2 .
\end{align}
And Sneutrino slip into CP-even and CP-odd parts:
\begin{align} 
    \tilde\nu_L=& \, \textstyle{\frac{1}{\sqrt{2}}} (\phi_{L}+i  \sigma_{L}) \;,& 
    \tilde\nu_R=& \, \textstyle{\frac{1}{\sqrt{2}}} (\phi_{R}+i  \sigma_{R}) \; 
    \label{eq:Sneutrino}
    \end{align}

After EWSB,the $\mu$-terms results from the VEVs of scalar singlet and triplet, and we define these abbreviations for clarity:
\begin{align}
\mu_i^{\text{eff,}\pm}&
=\mu_i+\frac{\lambda_iv_S}{\sqrt2}
\pm\frac{\Lambda_iv_T}{2}
,
&\mu_i^{\text{eff,}0}&
=\mu_i+\frac{\lambda_iv_S}{\sqrt2}
,&i=u,d.
\end{align}

The minimization conditions for the scalar potential, or tadpole equations, then read
\begin{equation}\label{eq:tadpoles}
0= t_{d} = t_{u} = t_{T} = t_{S},
\end{equation}

Where the tree-level tadpoles $t_i\equiv\frac{\partial
V^{EW}}{\partial \phi_i}$ are

\begin{align}
  \label{eq:tadpole_equation1}
  0 = & v \cos\beta  \left [4 m_Z^2 \cos 2 \beta  - 8 g_1 M^D_B v_S + 8 g_2
     M^D_W v_T+4 \lambda_D^2 v_S^2+ 4 \sqrt{2} \lambda_D v_S
     (\Lambda_D v_T + 2 \mu_D)\right .\\
     \nonumber
     & \left. +2 (\Lambda_D v_T + 2 \mu_D)^2+8
    m_{H_d}^2 \right ] ,\\
    \label{eq:tadpole_equation2}
    0 = & v \sin \beta \left(- 4 m_Z^2 \cos 2 \beta + 8 g_1 M^D_B v_S - 8 g_2
     M^D_W v_T+4 \lambda_U^2 v_S^2-  4 \sqrt{2} \lambda_U
     \Lambda_U v_S v_T \right. \\
     \nonumber
     & \left.+ 8 \sqrt{2} \lambda_U \mu_u v_S + 2
     \Lambda_U^2 v_T^2-8 \Lambda_U \mu_u v_T + 8 m_{H_u}^2 + 8 \mu_u^2 \right),\\
    \label{eq:tadpole_equation3}
     0 = & v_T \left[ 4 (M_W^D)^2 + m_T^2 \right] + \left[ 2 g_2 M^D_W + \Lambda_d ( \sqrt{2} \lambda_d v_S + \Lambda_d v_T + 2 \mu_d ) \right] v^2 \cos^2 \beta  \\
     & - \frac{1}{4} \left[2 g_2 M^D_W + \Lambda_u (\sqrt{2} \lambda_u v_S - \Lambda_u v_T + 2 \mu_u ) \right] v^2 \sin^2 \beta \nonumber,
     \\
     \label{eq:tadpole_equation4}
     0 = & v_S \left[4 (M^D_B)^2 + m_S^2\right] + \frac{1}{4} \left[- 2 g_1 M^D_B + \lambda_d (2 \lambda_d v_S + \sqrt{2} \Lambda_d v_T + 2 \sqrt{2} \mu_d) \right] v^2 \cos^2 \beta\\
     & + \frac{1}{4} \left[ 2 g_1 M^D_B + \lambda_u (2 \lambda_u v_S + 2\sqrt{2} \mu_u - \sqrt{2}\Lambda_u v_T)) \right] v^2 \sin^2 \beta \nonumber.
     \end{align}
     
 We introduce the variables $v=\sqrt{v_u^2+v_d^2}$ and $\tan{\beta}=v_u/v_d$, analogous to the MSSM.
Those equations are then solved analytically to obtain $m_{H_d}^2$, $m_{H_u}^2$, $m_S^2$ and $m_T^2$.

Using the tadpole eqs~(\ref{eq:tadpoles}), the pseudo-scalar Higgs boson mass matrix in the basis $(\sigma_d,\sigma_u,\sigma_S,\sigma_T)$ 
has a simple form (in Landau gauge)
\begin{equation}
\mathcal{M}_A=
\begin{pmatrix}
 B_\mu \frac{v_u}{v_d} & B_\mu & 0 & 0 \\
 B_\mu &  B_\mu \frac{v_d}{v_u} & 0 & 0 \\
 0 & 0 & m_S^2+\frac{\lambda_d^2 v_d^2+\lambda_u^2 v_u^2 }{2} & \frac{\lambda_d\Lambda_d v_d^2-\lambda_u\Lambda_u v_u^2}{2 \sqrt{2}} \\
 0 & 0 & \frac{\lambda_d\Lambda_d v_d^2-\lambda_u\Lambda_u v_u^2}{2 \sqrt{2}}& m_T^2+ \frac{\Lambda_d^2 v_d^2+\Lambda_u^2 v_u^2 }{4}\\
\end{pmatrix}
\;.
\end{equation}

Mass for Z bosen is the  same as usual,but the mass of W boson has  contribution from the triplet VEV:
\begin{equation}
m_Z^2 = \frac{g^2_1+g^2_2}{4} v^2\;,\qquad m_W^2= \frac{g^2_2}{4} v^2+g^2_2 v_T^2\;,\qquad 
\hat\rho_{\text{tree}} = 1 + \frac{4 v_T^2}{v^2}\;.
\label{eq:bosonmasses}
\end{equation}

The scalar Higgs boson mass matrix in the weak basis $(\phi_d,\phi_u,\phi_S,\phi_T)$  is given by
\begin{equation}\label{eq:scalarmassmatrix}
\mathcal{M}_{H^0}=
\begin{pmatrix}
\mathcal{M}_{\text{MSSM}}
&\mathcal{M}_{21}^T\\
\mathcal{M}_{21} &
\mathcal{M}_{22}
\\
\end{pmatrix},
\end{equation}
with the  sub-matrices ($c_\beta=\cos\beta$, $s_\beta=\sin\beta$, etc)
\begin{align}
\mathcal{M}_{\text{MSSM}}&= 
\begin{pmatrix}
 m_Z^2 c_\beta^2+m_A^2 s_\beta^2 \; & -(m_Z^2 + m_A^2)s_\beta c_\beta  \\
  -(m_Z^2 + m_A^2)s_\beta c_\beta \; &  m_Z^2 s_\beta^2+m_A^2 c_\beta^2\\
\end{pmatrix}
\;, \notag\\
\mathcal{M}_{22}&= 
\begin{pmatrix} 4 (M_B^D)^2+m_S^2+\frac{\lambda_d^2 v_d^2+\lambda_u^2 v_u^2}{2} \;
& \frac{\lambda_d \Lambda_d v_d^2-\lambda_u \Lambda_u v_u^2}{2 \sqrt{2}} \\
 \frac{\lambda_d \Lambda_d v_d^2-\lambda_u \Lambda_u v_u^2}{2 \sqrt{2}} \;
 & 4 (M_W^D)^2+m_T^2+\frac{\Lambda_d^2 v_d^2+\Lambda_u^2 v_u^2}{4}\\
\end{pmatrix}
\,, \notag\\
\mathcal{M}_{21}&= 
\begin{pmatrix}
 v_d ( \sqrt{2}\lambda_d \mu_u^{\text{eff,}-} -g_1 M_B^D )\; & \;
v_u (\sqrt{2} \lambda_u\mu_u^{\text{eff,}-} +g_1 M_B^D) \\
v_d ( \Lambda_d \mu_u^{\text{eff,}+} + g_2 M_W^D) \;& - 
 v_u (\Lambda_u  \mu_u^{\text{eff,}+} + g_2 M_W^D) \\
\end{pmatrix}
\;.\notag
\end{align}


The charged Higgs boson mass matrix in the weak basis $(H^{-*}_d,H^{+}_u,T^{-*},T^+)$  is given as
\begin{equation}
\mathcal{M}_{H^\pm}=
\begin{pmatrix}
\mathcal{M}_{\text{MSSM},\pm}
&\mathcal{M}_{21,\pm}^T\\
\mathcal{M}_{21,\pm} &
\mathcal{M}_{22,\pm}
\\
\end{pmatrix},
\end{equation}
with the  sub-matrices
\begin{align}
\mathcal{M}_{\text{MSSM},\pm}&= 
\begin{pmatrix}
 m_{H^\pm}^2 c_\beta^2-2 v_T (g_2 M^D_W+\Lambda_d\mu_d^{\text{eff,}0})\; &
  m_{H^\pm}^2s_\beta c_\beta  \\
  m_{H^\pm}^2s_\beta c_\beta \; &  m_{H^\pm}^2c_\beta^2+2 v_T (g_2 M^D_W+ \Lambda_u\mu_u^{\text{eff,}0})\\
\end{pmatrix}
\;, \notag\\
\mathcal{M}_{22,\pm}&=
\begin{pmatrix} 
  2 (M^D_W)^2+ m_T^2 \;
 & 2 (M^D_W)^2 \\
 2 (M^D_W)^2 \;
 & 2 (M^D_W)^2+ m_T^2\\
\end{pmatrix}
 \\
&+\begin{pmatrix} 
\frac{ g_2^2 v_T^2+\Lambda_d^2 v^2c_j^2}{2}
 -\frac{g_2^2v^2\cos 2 j }{4} \;
 &-\frac{g_2^2 v_T^2}{2} \\
 -\frac{g_2^2 v_T^2}{2}\; 
 &
\frac{ g_2^2 v_T^2+\Lambda_u^2 v^2s_j^2}{2}
 +\frac{g_2^2v^2 \cos 2 j}{4} \\
\end{pmatrix}
\;, \notag\\
\mathcal{M}_{21,\pm}&=
\begin{pmatrix}
\frac{ v_d}{2\sqrt{2}}
(2\Lambda_d \mu_d^{\text{eff,}-}+2{g_2 M^D_W }+{v_T g_2^2}) \; &
\frac{ v_u}{2\sqrt{2}}
(2\Lambda_u \mu_u^{\text{eff,}-}+2{ g_2 M^D_W}+{v_T g_2^2}) \\
\frac{ v_d}{2\sqrt{2}}
(2\Lambda_d \mu_d^{\text{eff,}+}+2{ g_2 M^D_W}-{v_T g_2^2}) \;  &
\frac{ v_u}{2\sqrt{2}}
(2\Lambda_u\mu_u^{\text{eff,}+}+2{ g_2 M^D_W}-v_T g_2^2 ) \\
\end{pmatrix},
\;\notag
\end{align}

Where the charged Higgs mass parameter $m^2_{H^{\pm}}=m^2_A+\frac{g_2^2v^2}{4}$ reads as in the MSSM.

The mass matrix of neutral R-Higgs bosons is
\begin{equation}
\mathcal{M}_R = \left (
\begin{matrix}
  m^2_{R_d R_d} & \frac{1}{4} \left ( \Lambda_u \Lambda_d - 2 \lambda_u \lambda_d \right ) v_u v_d \\
  \frac{1}{4} \left ( \Lambda_u \Lambda_d - 2 \lambda_u \lambda_d \right ) v_u v_d & m^2_{R_u R_u}
\end{matrix}
\right ),
\end{equation}
with
\begin{align*}
 &m^2_{R_d R_d}=
 m_{R_d}^2+ (\mu_d^{\text{eff,}+})^2+g_1 M^D_B v_S- g_2M^D_W v_T
+\textstyle{\frac{1}{8}} [(g_1^2+g_2^2) (v_u^2- v_d^2)+4 \lambda_d^2 v_d^2+2 \Lambda_d^2 v_d^2],\\
&m^2_{R_u R_u}=m_{R_u}^2+(\mu_u^{\text{eff,}-})^2- g_1 M^D_B v_S+ g_2 M^D_W v_T
+\textstyle{\frac{1}{8}} [(g_1^2 +g_2^2 )(v_d^2-v_u^2)+4 \lambda_u^2 v_u^2+2 \Lambda_u^2  v_u^2].\\
\end{align*}
Wihch is diagonalized by an orthogonal matrix with mixing angle $\theta_R$.

The charged R-Higgs do not mix because of the R-symmetry and the mass eigenvalues are given as
\begin{align}
& m^2_{R^+_1}=
 m_{R_d}^2+  (\mu_u^{\text{eff,}+})^2+g_1 M^D_B v_S+ g_2M^D_W v_T
+\textstyle{\frac{1}{8}} [(g_1^2-g_2^2) (v_u^2- v_d^2)+4 \Lambda_d^2 v_d^2] \notag\\
&m^2_{R^+_2}= m_{R_u}^2 + (\mu_d^{\text{eff,}-})^2 - g_1 M^D_B v_S- g_2 M^D_W v_T
+\textstyle{\frac{1}{8}} [(g_1^2 -g_2^2 )(v_d^2-v_u^2)+4 \Lambda_u^2  v_u^2]\;.
\end{align}

In the weak basis of eight neutral  electroweak two-component fermions: $ {\xi}_i=({\tilde{B}}, \tilde{W}^0, \tilde{R}_d^0, \tilde{R}_u^0)$,  $\zeta_i=(\tilde{S}, \tilde{T}^0, \tilde{H}_d^0, \tilde{H}_u^0) $ with R-charges $+1$, $-1$ respectively, the neutralino mass matrix takes the form as

\begin{equation} 
\label{eq:neut-massmatrix}
m_{{\chi}} = \left( 
\begin{array}{cccc}
M^{D}_B &0 &-\frac{1}{2} g_1 v_d  &\frac{1}{2} g_1 v_u \\ 
0 &M^{D}_W &\frac{1}{2} g_2 v_d  &-\frac{1}{2} g_2 v_u \\ 
- \frac{1}{\sqrt{2}} \lambda_d v_d  &-\frac{1}{2} \Lambda_d v_d  & - \mu_d^{\text{eff,}+} &0\\ 
\frac{1}{\sqrt{2}} \lambda_u v_u  &-\frac{1}{2} \Lambda_u v_u  &0 & \mu_u^{\text{eff,}-}
\end{array} 
\right) .
 \end{equation} 
The transformation to a diagonal mass matrix and mass eigenstates $\kappa_i$ and $\psi_i$ is performed by two unitary mixing matrices \(N^1\) and \(N^2\) as
\begin{align} \nonumber
N^{1,*} m_{{\chi}} N^{2,\dagger} &= m^{diag}_{{\chi}} \,,
&
{\xi}_i&=\sum_{j=1}^4 N^{1,*}_{ji}{\kappa}_j\,,
&
\zeta_i=\sum_{j=1}^4 N^{2,*}_{ij}{\psi}_j\,,
\end{align} 
and physical four-component Dirac Neutralinos are constructed as
\begin{equation}
{\chi}_i=\left(\begin{array}{c}
\kappa_i\\
{\psi}^{*}_i\end{array}\right)\qquad i=1,2,3,4.
\end{equation}

The mass matrix of Charginos in the weak basis of eight charged two-component
fermions breaks  into two $(2\times2)$ submatrices. 
The first, in the
basis \( (\tilde{T}^-, \tilde{H}_d^-), (\tilde{W}^+, \tilde{R}_d^+) \)
of spinors with R-charge equal to electric charge,
 takes the form of
\begin{equation} 
m_{{\chi}^+} = \left( 
\begin{array}{cc}
g_2 v_T  + M^{D}_W \; &\frac{1}{\sqrt{2}} \Lambda_d v_d \\ 
\frac{1}{\sqrt{2}} g_2 v_d \; &+ \mu_d^{\text{eff,}-}
\end{array} 
\right) 
\label{eq:cha1-massmatrix}
 \end{equation} 
The diagonalization and transformation to mass eigenstates $\lambda^\pm_i$ is
performed by two unitary matrices \(U^1\) and \(V^1\)  as
\begin{align} 
U^{1,*} m_{{\chi}^+} V^{1,\dagger} &= m^{diag}_{{\chi}^+} \,,
&\tilde{T}^- &= \sum_{j=1}^2U^{1,*}_{j 1}\lambda^-_{{j}}\,,& 
\tilde{H}_d^- &= \sum_{j=1}^2U^{1,*}_{j 2}\lambda^-_{{j}}\,,\\ 
&&
\tilde{W}^+ &= \sum_{j=1}^2V^{1,*}_{1 j}\lambda^+_{{j}}\,,&
R_d^+ &= \sum_{j=1}^2V^{1,*}_{2 j}\lambda^+_{{j}}
\end{align} 
and the corresponding physical four-component charginos are built as 
\begin{equation}
{\chi}^+_i=\left(\begin{array}{c}
\lambda^+_i\\
\lambda^{-*}_i\end{array}\right)\qquad i=1,2.
\end{equation}
The second submatrix, in the basis $ (\tilde{W}^-, R_u^-)$, $(\tilde{T}^+, \tilde{H}_u^+) $
 of spinors with R-charge equal to minus electric charge, reads
\begin{equation} 
m_{{\rho}^-} = \left( 
\begin{array}{cc}
- g_2 v_T  + M^{D}_W \;&\frac{1}{\sqrt{2}} g_2 v_u \\ 
- \frac{1}{\sqrt{2}} \Lambda_u v_u \; &  - \mu_u^{\text{eff,}+} \end{array} 
\right) 
\label{eq:cha2-massmatrix}
 \end{equation} 
The diagonalization and transformation to mass eigenstates $\eta^\pm_i$  is
performed by \(U^2\) and \(V^2\) as 
\begin{align} 
U^{2,*} m_{{\rho}^-} V^{2,\dagger} & = m^{diag}_{{\rho}^-} \,,&
\tilde{W}^- & = \sum_{j=1}^2U^{2,*}_{j 1}\eta^-_{{j}}\,, &
R_u^- & = \sum_{j=1}^2U^{2,*}_{j 2}\eta^-_{{j}}\\ 
&&\tilde{T}^+ & = \sum_{j=1}^2V^{2,*}_{1 j}\eta^+_{{j}}\,, &
\tilde{H}_u^+ & = \sum_{j=1}^2V^{2,*}_{2 j}\eta^+_{{j}}
\end{align} 
and the corresponding physical four-component charginos are built as 
\begin{equation}
{\rho}^-_i=\left(\begin{array}{c}
\eta^-_i\\
\eta^{+*}_i\end{array}\right)\qquad i=1,2.  \end{equation}
Moverover the mass matrix of CP-odd Sneutrino, in basis($\sigma^i_{L},\sigma^j_{R}$),($\sigma^i_{L},\sigma^j_{R}$) are 
\begin{equation} 
m^2_{\tilde\nu^I} = \left( 
\begin{array}{cc}
m_{\text{$\sigma^i_L$}\text{$\sigma^i_L$}} &m_{\text{$\sigma^i_R$}\text{$\sigma^i_L$}}\\ 
m_{\text{$\sigma^i_L$}\text{$\sigma^i_R$}} &m_{\text{$\sigma^i_R$}\text{$\sigma^i_R$}}\end{array} 
\right), 
 \end{equation} 
\begin{align} 
m_{\text{$\sigma^i_L$}\text{$\sigma^i_L$}} &= \frac{1}{2}v_u^2Y_{\nu}Y_{\nu}^{T}+m_l^2+{\bf 1}\Big((-g_1v_sM^{B}_D+g_2v_TM^{W}_D )+\frac{1}{8}(g_1^2+g_2^2)(-v_u^2+v_d^2) \Big),\\ 
m_{\text{$\sigma^j_R$}\text{$\sigma^i_L$}} &=  -v_s v_u {{Y_x  Y_{\nu}}}, \\ 
m_{\text{$\sigma^i_L$}\text{$\sigma^j_R$}} &= -v_s v_u {{Y_x^{T}  Y_{\nu}^{T}}} ,\\ 
m_{\text{$\sigma^j_R$}\text{$\sigma^j_R$}} &= \frac{1}{2}v_{u}^{2} {{Y_{\nu}^{T}  Y_\nu}}  +  {m_{\nu}^2}  + 2 v_{s}^{2} {{Y_x}^2}. 
\end{align} 
This matrix is diagonalized by \(Z^I\): 
\begin{equation} 
Z^I m^2_{\nu^I} Z^{I,\dagger} = m^{dia}_{2,\nu^I},
\end{equation} 
with 
\begin{align} 
\text{$\sigma^i_L$}_i = \sum_{j}Z^{I,*}_{j i}\nu^I_{{j}}\,, \hspace{1cm} 
\text{$\sigma^j_R$}_i = \sum_{j}Z^{I,*}_{j i}\nu^I_{{j}} \qquad i,j=1,2,3.
\end{align}

And the mass matrix for CP-even Sneutrino in basis\( \left(\text{$\phi^i_L$}, \text{$\phi^j_R$}\right), \left(\text{$\phi^i_L$}, \text{$\phi^j_R$}\right) \) are 
 
\begin{equation} 
m^2_{\tilde\nu^R} = \left( 
\begin{array}{cc}
m_{\text{$\phi^i_L$}\text{$\phi^i_L$}} &m_{\text{$\phi^j_R$}\text{$\phi^i_L$}} \\ 
m_{\text{$\phi^i_L$}\text{$\phi^j_R$}}  &m_{\text{$\phi^j_R$}\text{$\phi^j_R$}} \end{array} 
\right), 
 \end{equation} 
\begin{align} 
m_{\text{$\phi^i_L$}\text{$\phi^i_L$}} &= \frac{1}{2}v_u^2Y_{\nu}Y_{\nu}^{T}+m_l^2+{\bf 1}\Big((-g_1v_sM^{B}_D+g_2v_TM^{W}_D )+\frac{1}{8}(g_1^2+g_2^2)(-v_u^2+v_d^2) \Big),\\ 
m_{\text{$\phi^j_R$}\text{$\phi^i_L$}} &= -v_s v_u {{Y_x  Y_{\nu}}}, \\ 
m_{\text{$\phi^i_L$}\text{$\phi^j_R$}} &=-v_s v_u {{Y_x^{T}  Y_{\nu}^{T}}}, \\ 
m_{\text{$\phi^j_R$}\text{$\phi^j_R$}} &= \frac{1}{2}v_{u}^{2} {{Y_{\nu}^{T}  Y_\nu}}  +  {m_{\nu}^2}  + 2 v_{s}^{2} {{Y_x}^2}.
\end{align} 
This matrix is diagonalized by \(Z^R\): 
\begin{equation} 
Z^R m^2_{\nu^R} Z^{R,\dagger} = m^{dia}_{2,\nu^R}, 
\end{equation} 
with 
\begin{align} 
\text{$\phi_L$}_i = \sum_{j}Z^{R,*}_{j i}\nu^R_{{j}}\,, \hspace{1cm} 
\text{$\phi_R$}_i= \sum_{j}Z^{R,*}_{j i}\nu^R_{{j}}\qquad i,j=1,2,3.
\end{align} 

Since we have introduced right-handed Neutrinos, The Neutrino mass matrix is extended to a $6\times6$ matrix  consisting of the blocks given in eqs~(\ref{eq:45}). The mass matrix in basis\( \left(\nu_L, \nu_R^*\right), \left(\nu_L, \nu_R^*\right) \)  can be written as 
 
\begin{equation} 
m_{\nu} = \left( 
\begin{array}{cc}
0 &\frac{1}{\sqrt{2}} v_u Y_\nu \\ 
\frac{1}{\sqrt{2}} v_u Y_{\nu}^{T}  &\sqrt{2} v_s Y_x \end{array} 
\right),\label{eq:45} 
\end{equation} 
This matrix is diagonalized by \(U^V_{\nu}\): 
\begin{equation} 
U^{V,*}_{\nu} m_{\nu} U_{\nu}^{V,\dagger} = m^{dia}_{\nu},
\end{equation} 
with 
\begin{align} 
\nu_{L,{i}} = \sum_{j}U^{V,*}_{{\nu},{j i}}\lambda_{\nu,{j}}\,, \hspace{1cm} 
\nu_{R,{i}} = \sum_{j}U_{\nu,{j i}}^{V}\lambda^*_{\nu,{j}}.
\end{align} 

The three generations of right-handed Neutrinos first acquire  Majorana masses through the breaking of the S-field. Then, via a Yukawa left-right mixing coupling, the observed small masses for the left-handed Neutrinos are produced by the type-I seesaw mechanism.


\section{Lepton Flavor Kit }
\indent\indent

In this section, we investigate the amplitude and branching ratio of $\ell^{-}_{i} \to \ell^{-}_{j}\gamma$ , $\ell^{-}_{i}\to \ell^{-}_{j}\ell^{-}_{j}\ell^{+}_{j}$ and $ h \to \ell_i^{+} \ell_j^{-}$.
\subsection{Rare decay $\ell^{-}_{i} \to \ell^{-}_{j}\gamma$}

\begin{figure}
  \centering
  \subfigure[]{\label{fig:sub1}\includegraphics[width=0.4\textwidth]{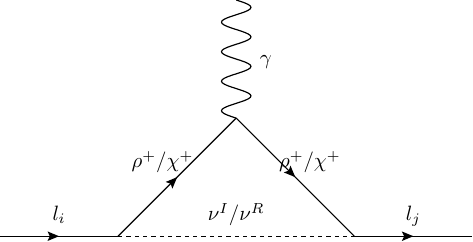}}
  \subfigure[]{\label{fig:sub1}\includegraphics[width=0.4\textwidth]{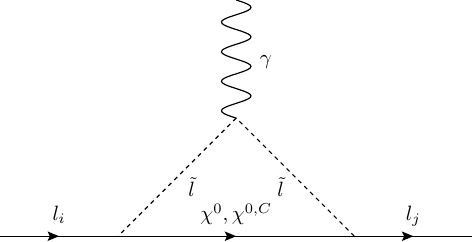}}
  \caption{(a) represents the contributions from Charginos and  Sneutrino loops and (b) represents the contributions from charged Sleptons and Neutrinos loops.}
  \label{fig:llgama}
\end{figure}

At first, the off-shell amplitude for $l_i^-\rightarrow l_j^-\gamma$  is generally written as Eq.~(\ref{eq:langllgama}), where $P_L=(1-\gamma_5)/2$, $P_R=(1+\gamma_5)/2$.
The corresponding Feynman diagrams are depicted in Fig.~\ref{fig:llgama} . 

\begin{equation} 
  L_{\ell \ell\gamma}=e\overline{\ell_{j}}\left[ \gamma^\mu (K_1^L P_L+K_1^R P^R)+im_{\ell_{i}}\sigma^{\mu\nu}q_{\nu} (K_2^L P_L+K_2^RP_R)  \right]\ell_{i}A_{\mu}+h.c.
  \label{eq:langllgama}
  \end{equation}   
  Using the amplitude Eq~(\ref{eq:langllgama}), the decay width for $l_i^-\rightarrow l_j^-\gamma$  can be obtained easily as \cite{Hisano:1995cp}:
  
  \begin{equation}
  \Gamma \left( \ell^{-}_{i} \to \ell^{-}_{j}\gamma \right) = \frac{\alpha m_{\ell_i}^5}{4} \left( |K_2^L|^2 + |K_2^R|^2 \right) 
  \label{eq:Deacyllgamma}
  \end{equation}
  where $\alpha$ is the fine structure constant and the dipole Wilson coefficients $K_2^{L,R}$. And the branching ratio for $l_i^-\rightarrow l_j^-\gamma$ is
  \begin{eqnarray}
  &&Br(l_i^-\rightarrow l_j^-\gamma)=\frac{\Gamma(l_i^-\rightarrow l_i^-\gamma)}{\Gamma_{l_i^-}},
  \end{eqnarray}
  where $\Gamma_{l_i^-}$ is the total decay width of the lepton $l_i^-$. In the numerical calculation, we use $\Gamma_\mu\approx2.996\times10^{-19}{\rm GeV}$ for the muon and $\Gamma_\tau\approx2.265\times10^{-12}{\rm GeV}$ for the tauon~\cite{ParticleDataGroup:2024cfk}

\subsection{Rare decay $\ell^{-}_{i}\to \ell^{-}_{j}\ell^{-}_{j}\ell^{+}_{j}$} 
\begin{figure}
  \centering
  \subfigure[]{\label{fig:sub1}\includegraphics[width=0.35\textwidth]{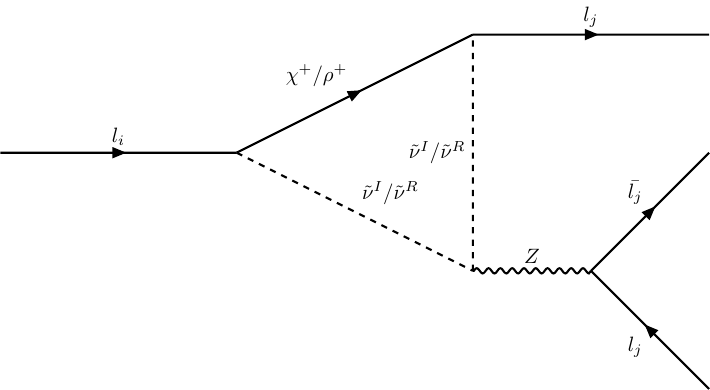}}
  \subfigure[]{\label{fig:sub2}\includegraphics[width=0.35\textwidth]{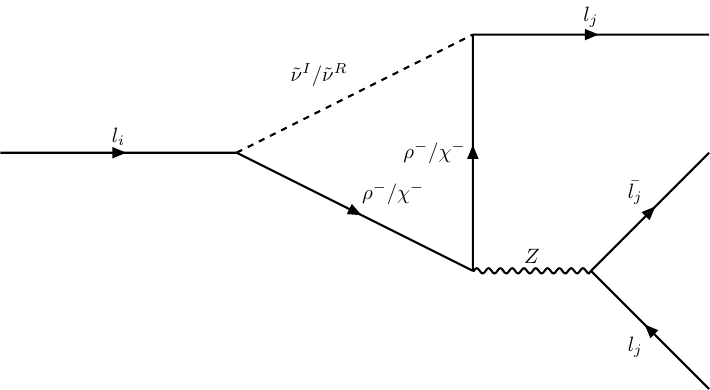}}
  \subfigure[]{\label{fig:sub3}\includegraphics[width=0.35\textwidth]{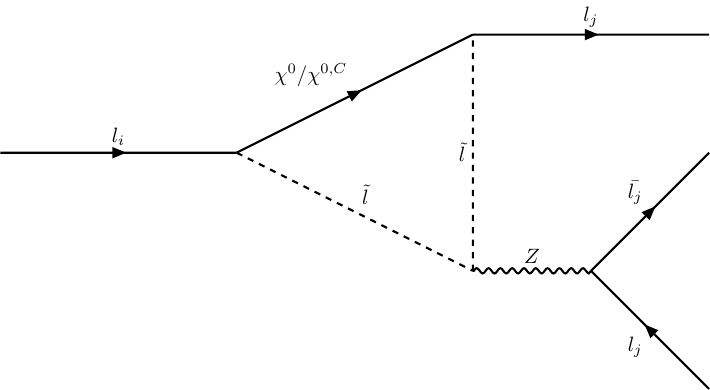}}
  \subfigure[]{\label{fig:sub4}\includegraphics[width=0.35\textwidth]{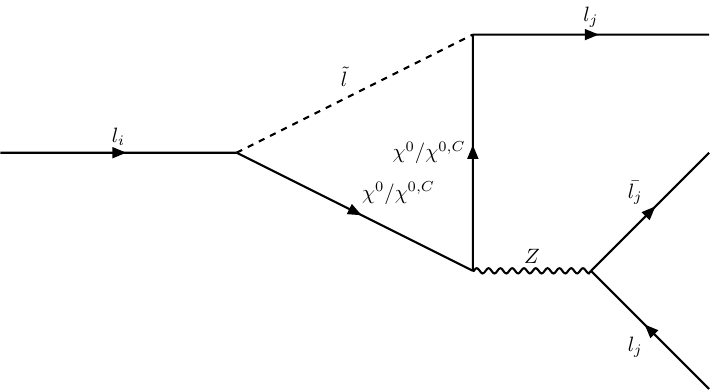}}

  \subfigure[]{\label{fig:sub4}\includegraphics[width=0.23\textwidth]{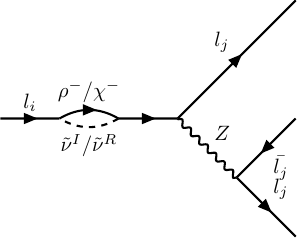}}
  \subfigure[]{\label{fig:sub4}\includegraphics[width=0.23\textwidth]{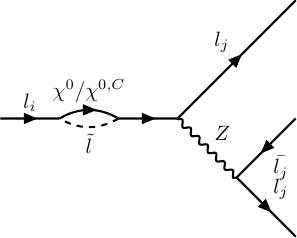}}
  \subfigure[]{\label{fig:sub4}\includegraphics[width=0.18\textwidth]{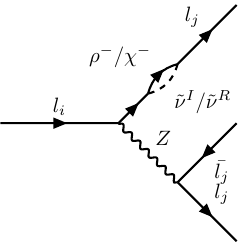}}
  \subfigure[]{\label{fig:sub4}\includegraphics[width=0.18\textwidth]{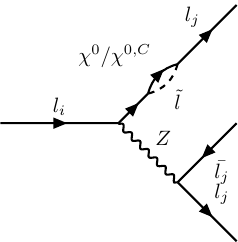}}

  \caption{The Z-penguin diagrams for $\ell^{-}_{i}\to \ell^{-}_{j}\ell^{-}_{j}\ell^{+}_{j}$.}
  \label{fig:l3lYZ}
\end{figure}

\begin{figure}
  \centering
  \subfigure[]{\label{fig:sub1}\includegraphics[width=0.35\textwidth]{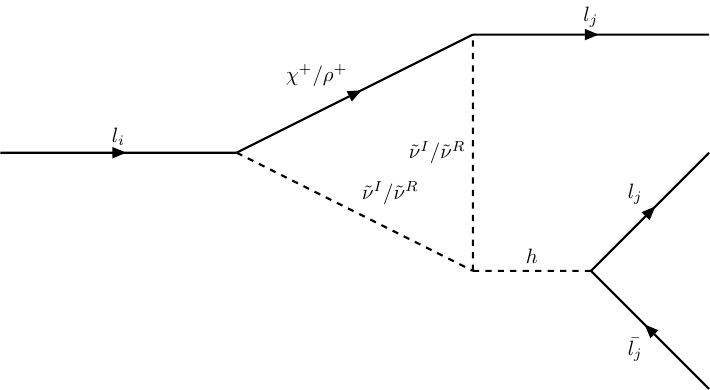}}
  \subfigure[]{\label{fig:sub2}\includegraphics[width=0.35\textwidth]{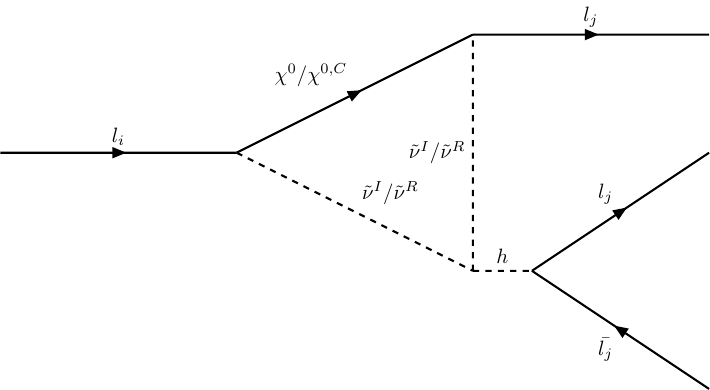}}
  \subfigure[]{\label{fig:sub3}\includegraphics[width=0.35\textwidth]{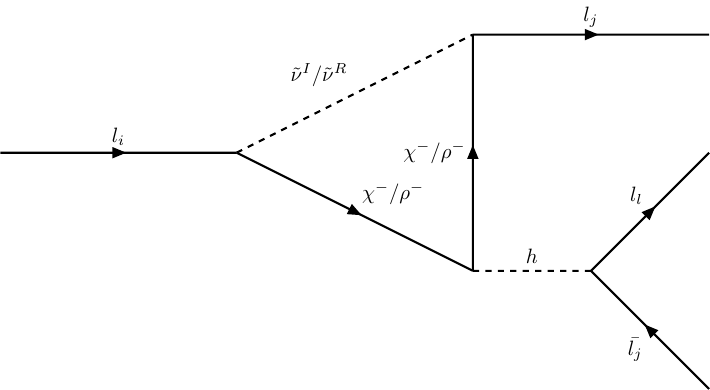}}
  \subfigure[]{\label{fig:sub4}\includegraphics[width=0.35\textwidth]{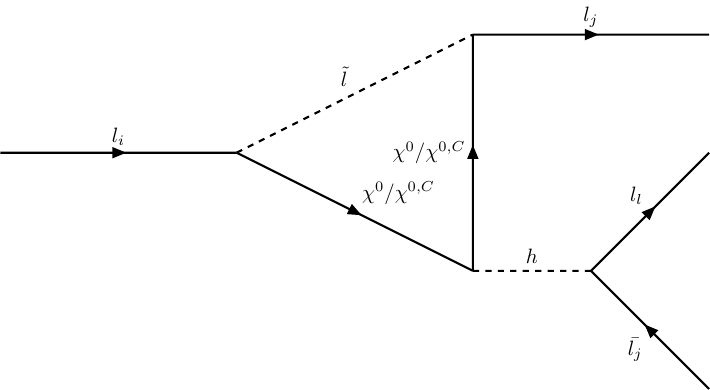}}

  \subfigure[]{\label{fig:sub4}\includegraphics[width=0.23\textwidth]{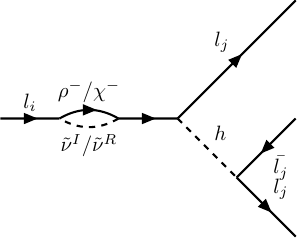}}
  \subfigure[]{\label{fig:sub4}\includegraphics[width=0.23\textwidth]{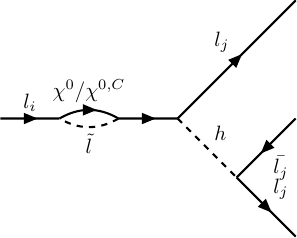}}
  \subfigure[]{\label{fig:sub4}\includegraphics[width=0.18\textwidth]{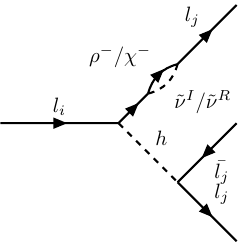}}
  \subfigure[]{\label{fig:sub4}\includegraphics[width=0.18\textwidth]{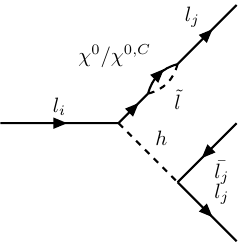}}
  \caption{The H-penguin diagrams for $\ell^{-}_{i}\to \ell^{-}_{j}\ell^{-}_{j}\ell^{+}_{j}$.}
  \label{fig:l3lH}
\end{figure}

\begin{figure}
  \centering
  \subfigure[]{\label{fig:sub1}\includegraphics[width=0.45\textwidth]{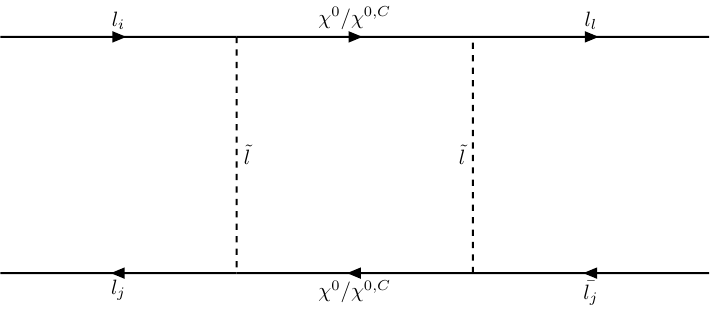}}
  \subfigure[]{\includegraphics[width=0.45\textwidth]{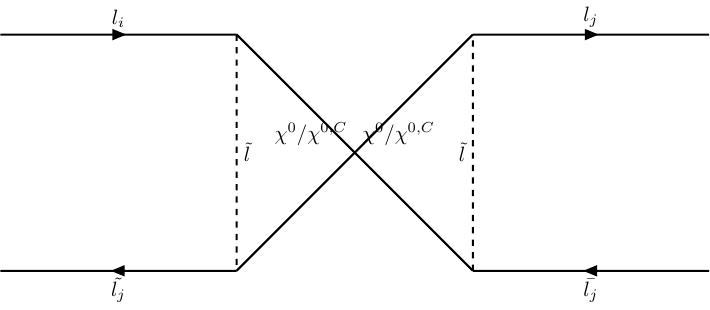}}

  \subfigure[]{\label{fig:sub2}\includegraphics[width=0.45\textwidth]{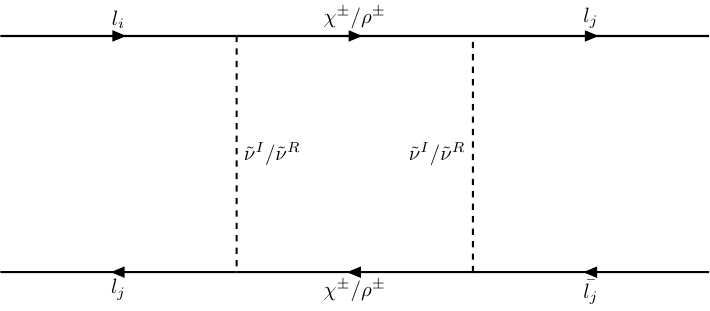}}
  \caption{ Box-type diagrams for the process $\ell^{-}_{i}\to \ell^{-}_{j}\ell^{-}_{j}\ell^{+}_{j}$.(a) ,(b)represents the contributions from neutral fermions Neutralinos and charge scalars loops Sleptons and (c) represents the contributions from
  charged fermions Charginos and neutral scalars Sneutrinos loops.}
  \label{fig:l3lbox}
\end{figure}

For the process  $\ell^{-}_{i}\to \ell^{-}_{j}\ell^{-}_{j}\ell^{+}_{j}$,the effective amplitude includes the contributions from Penguin-type and Box-type diagrams~\cite{Arganda:2005ji,Sun:2019qiw}. Fig.~\ref{fig:l3lYZ} and Fig.~\ref{fig:l3lH}
represents the contributions from Z boson and Higgs bosen. Moverover the Box-type diagrams of  $\ell^{-}_{i}\to \ell^{-}_{j}\ell^{-}_{j}\ell^{+}_{j}$ decay shown as Fig.~\ref{fig:l3lbox}

The four fermions operator can be written as~\cite{Ilakovac:2012sh,Arganda:2005ji} 
  \begin{equation} 
  L_{4\ell}=\sum_{I=S,V,T |X ,Y=L,R}A_{XX}^I\overline{\ell_{j}}\Gamma_IP_X\ell_{i}\overline{\ell_{\delta}}\Gamma_IP_Y\ell_{\gamma}+h.c,
  \label{eq:4ferion}
  \end{equation} 
  where $I = \{S, V, T\}, X, Y = \{L, R\}, \Gamma_S = 1, \Gamma_V = \gamma_\mu \text{ and } \Gamma_T = \sigma_{\mu\nu}.$
 Using the Eq.~(\ref{eq:4ferion}), the decay width are~\cite{Hisano:1995cp,Ilakovac:2012sh,Arganda:2005ji} 
  \begin{eqnarray}
  \Gamma \left( \ell^{-}_{i}\to \ell^{-}_{j}\ell^{-}_{j}\ell^{+}_{j} \right) &=& \frac{m_{\ell_i}^5}{512 \pi^3} \left[ e^4 \, \left( \left| K_2^L \right|^2 + \left| K_2^R \right|^2 \right) \left( \frac{16}{3} \log{\frac{m_{\ell_i}}{m_{\ell_j}}} - \frac{22}{3} \right) \right. \label{L3Lwidth} \\
  &+& \frac{1}{24} \left( \left| A_{LL}^S \right|^2 + \left| A_{RR}^S \right|^2 \right) + \frac{1}{12} \left( \left| A_{LR}^S \right|^2 + \left| A_{RL}^S \right|^2 \right) \nonumber \\
  &+& \frac{2}{3} \left( \left| \hat A_{LL}^V \right|^2 + \left| \hat A_{RR}^V \right|^2 \right) + \frac{1}{3} \left( \left| \hat A_{LR}^V \right|^2 + \left| \hat A_{RL}^V \right|^2 \right) + 6 \left( \left| A_{LL}^T \right|^2 + \left| A_{RT}^T \right|^2 \right) \nonumber \\
  &+& \frac{e^2}{3} \left( K_2^L A_{RL}^{S \ast} + K_2^R A_{LR}^{S \ast} + c.c. \right) - \frac{2 e^2}{3} \left( K_2^L \hat A_{RL}^{V \ast} + K_2^R \hat A_{LR}^{V \ast} + c.c. \right) \nonumber \\
  &-& \frac{4 e^2}{3} \left( K_2^L \hat A_{RR}^{V \ast} + K_2^R \hat A_{LL}^{V \ast} + c.c. \right) \nonumber \\
  &-& \left. \frac{1}{2} \left( A_{LL}^S A_{LL}^{T \ast} + A_{RR}^S A_{RR}^{T \ast} + c.c. \right) - \frac{1}{6} \left( A_{LR}^S \hat A_{LR}^{V \ast} + A_{RL}^S \hat A_{RL}^{V \ast} + c.c. \right) \right]  \nonumber \, .
  \end{eqnarray}
Here we have defined  
\begin{eqnarray}
\hat A^V_{XY}=A^V_{XY}+e^2K_1^X
\end{eqnarray}  
The mass of the leptons in the final state has been neglected in this formula, with the exception of the dipole terms $K_2^{L,R}$,where an infrared divergence would otherwise occur due to the massless photon propagator~\cite{Porod:2014xia}. And the branching ratio for $\ell^{-}_{i}\to \ell^{-}_{j}\ell^{-}_{j}\ell^{+}_{j} $ is
\begin{eqnarray}
&&Br(\ell^{-}_{i}\to \ell^{-}_{j}\ell^{-}_{j}\ell^{+}_{j} )=\frac{\Gamma(\ell^{-}_{i}\to \ell^{-}_{j}\ell^{-}_{j}\ell^{+}_{j} )}{\Gamma_{l_i^-}},
\end{eqnarray}
  
\subsection{Higgs decay $ h \to \ell_i^{+} \ell_j^{-}$}

Then turn to the Higgs LFV deacy . The Lagrangian of $h-\ell-\ell$ interaction is given by

  \begin{equation} 
    L_{\ell \ell h}=\overline{\ell_{j}} (S_1^L P_L+S_1^R P^R)\ell_{i}h.
    \end{equation}

\begin{figure}
    \centering
    \subfigure[]{\label{fig:sub1}\includegraphics[width=0.2\textwidth]{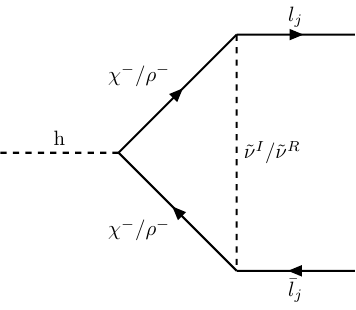}}
    \subfigure[]{\label{fig:sub2}\includegraphics[width=0.2\textwidth]{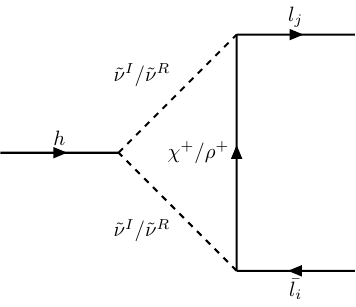}}
    \subfigure[]{\label{fig:sub3}\includegraphics[width=0.2\textwidth]{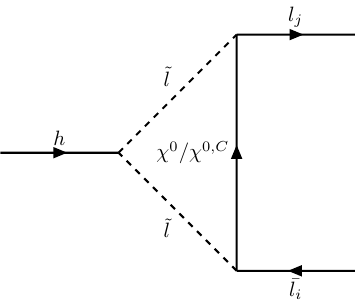}}
    \subfigure[]{\label{fig:sub3}\includegraphics[width=0.2\textwidth]{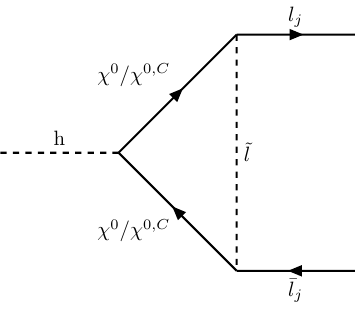}}
   
    \subfigure[]{\label{fig:sub1}\includegraphics[width=0.2\textwidth]{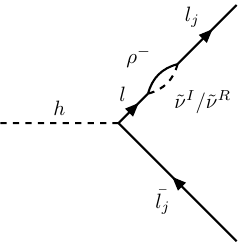}}
    \subfigure[]{\label{fig:sub2}\includegraphics[width=0.2\textwidth]{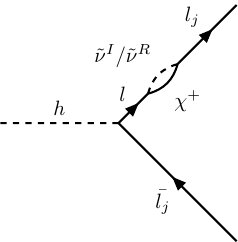}}
    \subfigure[]{\label{fig:sub3}\includegraphics[width=0.2\textwidth]{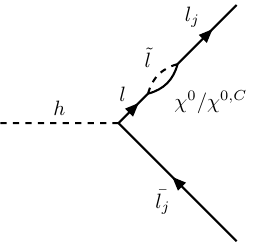}}

    \subfigure[]{\label{fig:sub3}\includegraphics[width=0.2\textwidth]{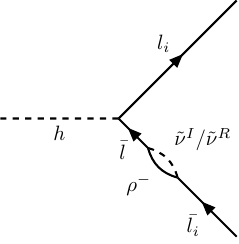}}
    \subfigure[]{\label{fig:sub1}\includegraphics[width=0.2\textwidth]{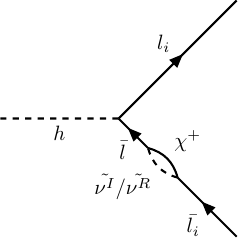}}
    \subfigure[]{\label{fig:sub2}\includegraphics[width=0.2\textwidth]{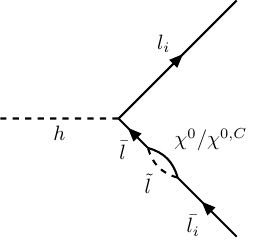}}

    \caption{One-loop diagrams for the Higgs LFV in the MRSSMSeesaw model}
    \label{fig:h2l}
  \end{figure}

Then the Higgs LFV decay widths can be written as~\cite{Arganda:2004bz}:
 \begin{eqnarray}
    \Gamma \left( h \to \ell_i^{+} \ell_j^{-} \right) &\equiv& \Gamma \left( h \to \ell_i \bar \ell_j \right) + \Gamma \left( h \to \bar \ell_i \ell_j \right) = \\
    && \frac{1}{16 \pi m_h} \left[ \left(1-\left(\frac{m_{\ell_i} + m_{\ell_j}}{m_h}\right)^2\right)\left(1-\left(\frac{m_{\ell_i} - m_{\ell_j}}{m_h}\right)^2\right)\right]^{1/2} \nonumber \\
    && \times \left[ \left( m_h^2 - m_{\ell_i}^2 - m_{\ell_j}^2 \right) \left( |S_L|^2 + |S_R|^2 \right)_{i j} - 4 m_{\ell_i} m_{\ell_j} \text{Re}(S_L S_R^\ast)_{i j} \right] \nonumber \\
    && + (i \leftrightarrow j) \nonumber.
    \label{eq:deacyh}
\end{eqnarray}
The branching ratio for $h_i\rightarrow \ell_i^+\ell_j^-$ is
  \begin{eqnarray}
  &&Br(h_i\rightarrow \ell_i^+\ell_j^-)=\frac{\Gamma(h_i\rightarrow \ell_i^+\ell_j^-)}{\Gamma_{h}},
  \end{eqnarray}
  where $\Gamma_{H}$ is the total decay width of  Higgs $h$. In the numerical calculation, we use $\Gamma_{h}\approx3.7 ~ {\rm MeV}$~\cite{ParticleDataGroup:2024cfk}.  
And  Higgs LFV diagrams is shown as plot in Fig.~\ref{fig:h2l}.



\section{Constraints and numerical analysis}
To delve into the parameter space of our model, we have encode the MRSSMSeesaw  into the SARAH~\cite{Porod:2014xia,Bagnaschi:2022zvd,Benakli:2022gjn,Goodsell:2020rfu,Goodsell:2018tti,Goodsell:2017pdq,Staub:2013tta}. It results in the generation of a Fortran code that interfaces with the SPheno~\cite{Porod:2003um,Porod:2011nf,Staub:2011dp} library, enabling the computation of the particle spectrum, decay processes, and flavour processes including rare LFV ones
  
We use BSMArt~\cite{Goodsell:2023iac,Faraggi:2023jzm} version 1.5,a python program for the exploration of parameter spaces of BSM Model including SPheno~\cite{Porod:2003um,Porod:2011nf,Staub:2011dp} ,HiggsTools~\cite{Bahl:2022igd} and so on. We have designated the unknown parameters as variables for scanning within BSMArt~\cite{Goodsell:2023iac,Faraggi:2023jzm}. The SM input parameters are consistently configured as follows~\cite{ParticleDataGroup:2024cfk}: 
\begin{align*}
  m_t &= 172.89~\text{GeV}\,,\; 
  \hat m^{\overline{\text{MS}}}_b(\hat m_b) = 4.18~\text{GeV}\,,\; 
  M_Z = 91.1887~\text{GeV}\,,\; \\
  m_\tau &= 1.776690~\text{GeV}\,,\; 
  \alpha^{-1} = 137.035999139\,,\; 
  \alpha^{\overline{\text{MS}}}_s(m_Z) = 0.1187\,,\; \\
  G_F &= 1.166370 \times 10^{-5}~\text{GeV}^{-2}\,,\\
  m_e &= 0.511 \times 10^{-3}~\text{GeV}\,,\; 
  m_\mu = 0.105~\text{GeV}\,.
\end{align*}
  
   We fix $M^D_O=1.5~\text{TeV}$ ,
  $m_{O}^2=m^2_{R_u}= m^2_{R_d}= (1~\text{TeV})^2 $,$v_S=-1~\text{TeV}$ , $Y_x=-1$ , $Y_\nu=10^{-6}$ and for soft breaking Sleptons mass parameters, we set $m_{\tilde{l}}=m_{\tilde{e}}=m_{\tilde{\nu}}=1~\text{TeV}$. We then scan  within the ranges:
  
  \begin{gather}
     \lambda_d ; \lambda_u;  \Lambda_d;\Lambda_u \in [-3,3],
     \tan\beta \in [1,50],\\ M_B^D;M_W^D \in [100,3000],
    v_T \in [-4,4], \mu_u,\mu_d \in [100,20000]
   \end{gather}

   To obtain the result we implement a Markov-Chain Monte-Carlo (MCMC) scan in {\sc BSMArt}~\cite{Goodsell:2023iac,Faraggi:2023jzm} using a likelihood function based on: the Higgs mass(a gaussian likelihood peaked at $125.2$ GeV with variance $3$ GeV); the $W$ boson mass (a gaussian likelihood peaked at $80.3559$ GeV with variance $30$ MeV);  Higgs data from {\sc HiggsTools}~\cite{Bahl:2022igd} : the $p$--value for the SM-like Higgs  
  
  The following constrains are considered in the scanning~\cite{ParticleDataGroup:2024cfk}
  \begin{enumerate}
    \item  $h_1$ is chosen to be SM Higgs boson with $m_1=125.2\pm3.0$ Gev.
    \item Measurements of the W boson mass with $m_W=80.3692\pm0.003$ GeV.
    \item The $STU$ parameters with $S =-0.04\pm0.10~, T = 0.01\pm0.12~,U =0.01\pm0.09$.
    \item All the points  must be allowed by HiggsBounds.
  
  \end{enumerate}
  
Considering these limitations, we adopt the following parameters in the numerical calculation:

\begin{table}[h]
    \centering
    
    \begin{tabular}{cccccccc|cccccc} 
    \hline
    $\tan\beta$ & $\lambda_D$ & $\lambda_U$ & $\Lambda_D$ & $\Lambda_U$ & $M_D^B$ & $M_D^W$ & $v_T$ & $m_h$ & $m_{\chi^0}$ & $S$ & $T$ & $U$ & HBresult \\ \hline
    39.35 & 0.15 & -0.35 & 0.66 & -2.14 & 141.44 & 339.66 & -0.52 & 125.10 & 146.41 &0.05 & 0.12 & -0.002 & 1 \\ \hline
    \end{tabular}
    \caption{Benchmark points. Dimensionful parameters are given in GeV, Higgs sector of the benchmark points was checked against existing experimental data using HiggsTools.}
    \label{tab:14col_example}
\end{table}

Concerning the Neutrino mass term we newly introduced , there are two possible neutrino mass orderings: the first is the normal ordering (NO), and the second is the inverted ordering (IO). The effective mass matrix of the left-handed neutrinos is given in the usual seesaw approximation as~\cite{ParticleDataGroup:2024cfk,King:2013eh,T2K:2013ppw,T2K:2019bcf,IceCube-PINGU:2014okk,Fernandez-Martinez:2016lgt,Hyper-Kamiokande:2016srs}
\begin{eqnarray}
m_\nu=-\frac{1}{2}v_u^2Y_\nu^T\cdot (\sqrt{2}Y_x v_s)^{-1}\cdot Y_\nu,
  \label{eq:effneutrino}
\end{eqnarray}
 Since the neutrino mass matrix is complex symmetric, Eq.~(\ref{eq:effneutrino}) is diagonalized by~\cite{ParticleDataGroup:2024cfk}
\begin{eqnarray}
  \hat{m}_\nu=U^T\cdot m_\nu\cdot U ,
    \label{eq:dianeu}
  \end{eqnarray}
where $U$ is the Pontecorvo-Maki-Nakagawa-Sakata(PMNS) matrix~\cite{Pontecorvo:1967fh}.~Inverting Eq.~(\ref{eq:effneutrino}), $Y_\nu$ can be written as~\cite{Casas:2001sr}
\begin{eqnarray}
  Y_\nu=\sqrt{2}\frac{i}{v_u}\sqrt{\sqrt{2}Y_x v_s}\cdot R\cdot \sqrt{\hat{m}_\nu}\cdot U^\dagger,
    \label{eq:Yv}
  \end{eqnarray}
  where $\hat{m}_{\nu}$ is the diagonal matrix with $m_i$ eigenvalues and $R$ in general is a complex orthogonal matrix. Note that in the special case we set $R =1$. There are 3 light neutrino masses in $\hat{m}_{\nu}$ and the 3 heavy ``right-handed'' neutrino masses in $\hat{M}_R=\sqrt{2}Y_x v_s$.
  For $U$ we will use the standard form~\cite{ParticleDataGroup:2024cfk}
  \begin{eqnarray}
  U = \begin{pmatrix}
    c_{12}c_{13} & s_{12}c_{13} & s_{13}e^{-i\delta} \\
    -s_{12}c_{23} - c_{12}s_{23}s_{13}e^{i\delta} & c_{12}c_{23} - s_{12}s_{23}s_{13}e^{i\delta} & s_{23}c_{13} \\
    s_{12}s_{23} - c_{12}c_{23}s_{13}e^{i\delta} & -c_{12}s_{23} - s_{12}c_{23}s_{13}e^{i\delta} & c_{23}c_{13}
    \end{pmatrix} \times
  \begin{pmatrix}  
  e^{i\alpha_1/2} & 0 & 0 \\
  0 & e^{i\alpha_2/2} & 0 \\
  0 & 0 & 1
  \end{pmatrix},
  \end{eqnarray}
  with $c_{ij} = \cos\theta_{ij}$ and $s_{ij} = \sin\theta_{ij}$. 

  \begin{table}
    \centering
    \begin{tabular}{c c c}
      \hline
      \textbf{Observable} & \textbf{Normal Hierarchy} & \textbf{Inverted Hierarchy} \\
      \hline
      $\theta_{12}~(^\circ)$ & $33.41^{+0.75}_{-0.72}$ & $33.41^{+0.75}_{-0.72}$ \\
      $\theta_{23}~(^\circ)$ & $49.1^{+1.0}_{-1.3}$ & $49.5^{+0.9}_{-1.2}$ \\
      $\theta_{13}~(^\circ)$ & $8.54^{+0.11}_{-0.12}$ & $8.57^{+0.12}_{-0.11}$ \\
      $\Delta m_{12}^2~ (10^{-5}\, \mathrm{eV}^2)$ & $7.41^{+0.21}_{-0.20}$ & $7.41^{+0.21}_{-0.20}$ \\
      $\Delta m_{32}^2~ (10^{-3}\, \mathrm{eV}^2)$ & $2.437^{+0.028}_{-0.027}$ & $-2.498^{+0.028}_{-0.027}$ \\
      $\delta_{CP}~(^\circ)$ & $197^{+42}_{-25}$ & $286^{+27}_{-32}$ \\
      \hline
    \end{tabular}
    \caption{Neutrino oscillation parameters\cite{ParticleDataGroup:2024cfk} }
    \label{tab:neutrino}
\end{table}
For NO , we set
\begin{align*}
  \begin{cases}
  m_{\nu,1} = 0.01 \, \text{eV}, \\
  m_{\nu,2} = \sqrt{m_1^2 + \Delta m_{21}^2} \approx 0.0123 \, \text{eV}, \\
  m_{\nu,3} = \sqrt{m_1^2 + \Delta m_{31}^2} \approx 0.0506 \, \text{eV}.
  \end{cases}
  \end{align*} 
 
  For IO, we set  
  \begin{align*}
  \begin{cases}
  m_{\nu,3} = 0.01 \, \text{eV}, \\
  m_{\nu,1} = \sqrt{m_3^2 + |\Delta m_{32}^2|} \approx 0.0509 \, \text{eV}, \\
  m_{\nu,2} = \sqrt{m_1^2 + \Delta m_{21}^2} \approx 0.0516 \, \text{eV}.
  \end{cases}
  \end{align*}

\begin{figure}
  \centering
  \subfigure[]{\label{fig:sub1}\includegraphics[width=0.32\textwidth]{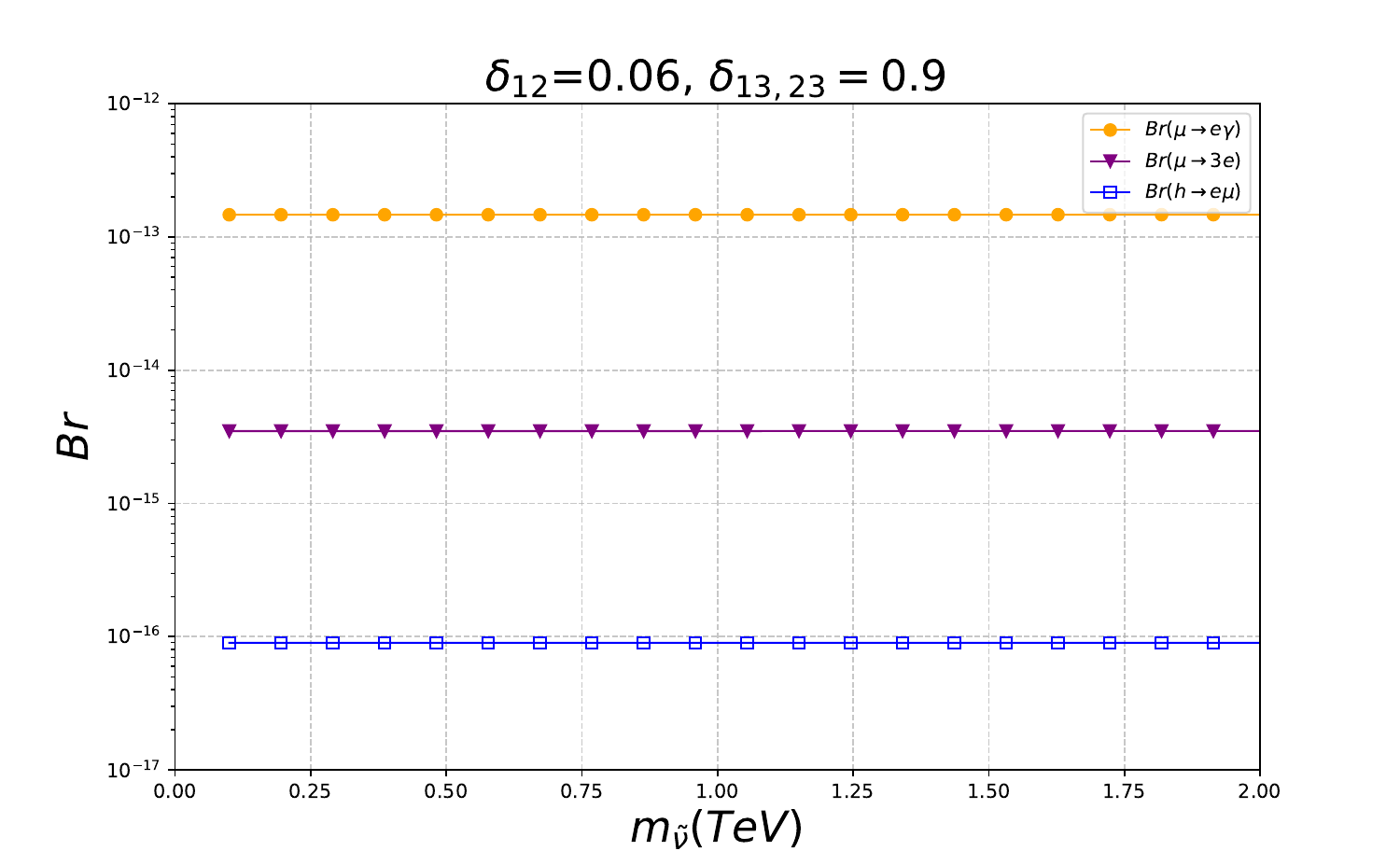}}
  \subfigure[]{\label{fig:sub1}\includegraphics[width=0.32\textwidth]{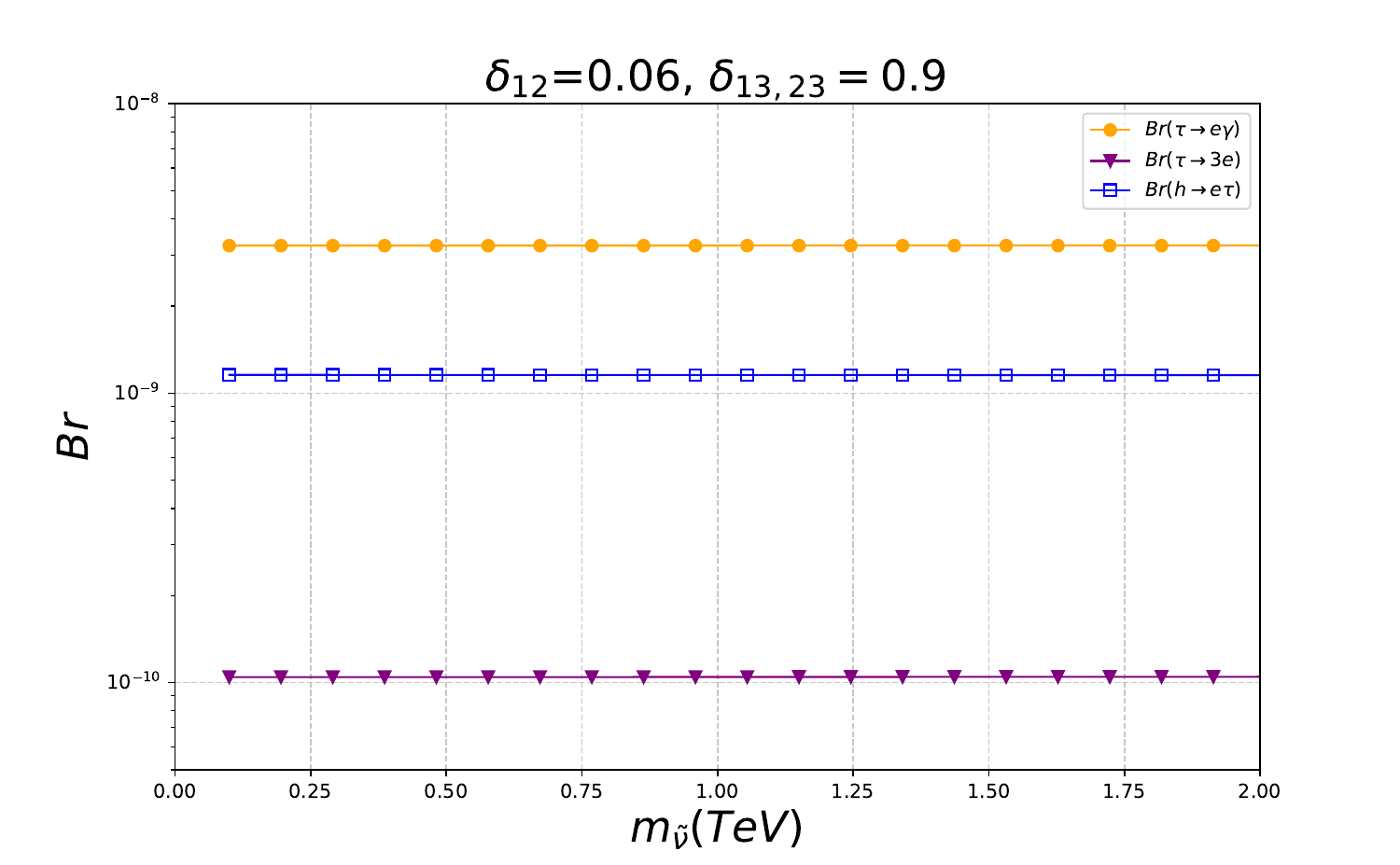}}
  \subfigure[]{\label{fig:sub1}\includegraphics[width=0.32\textwidth]{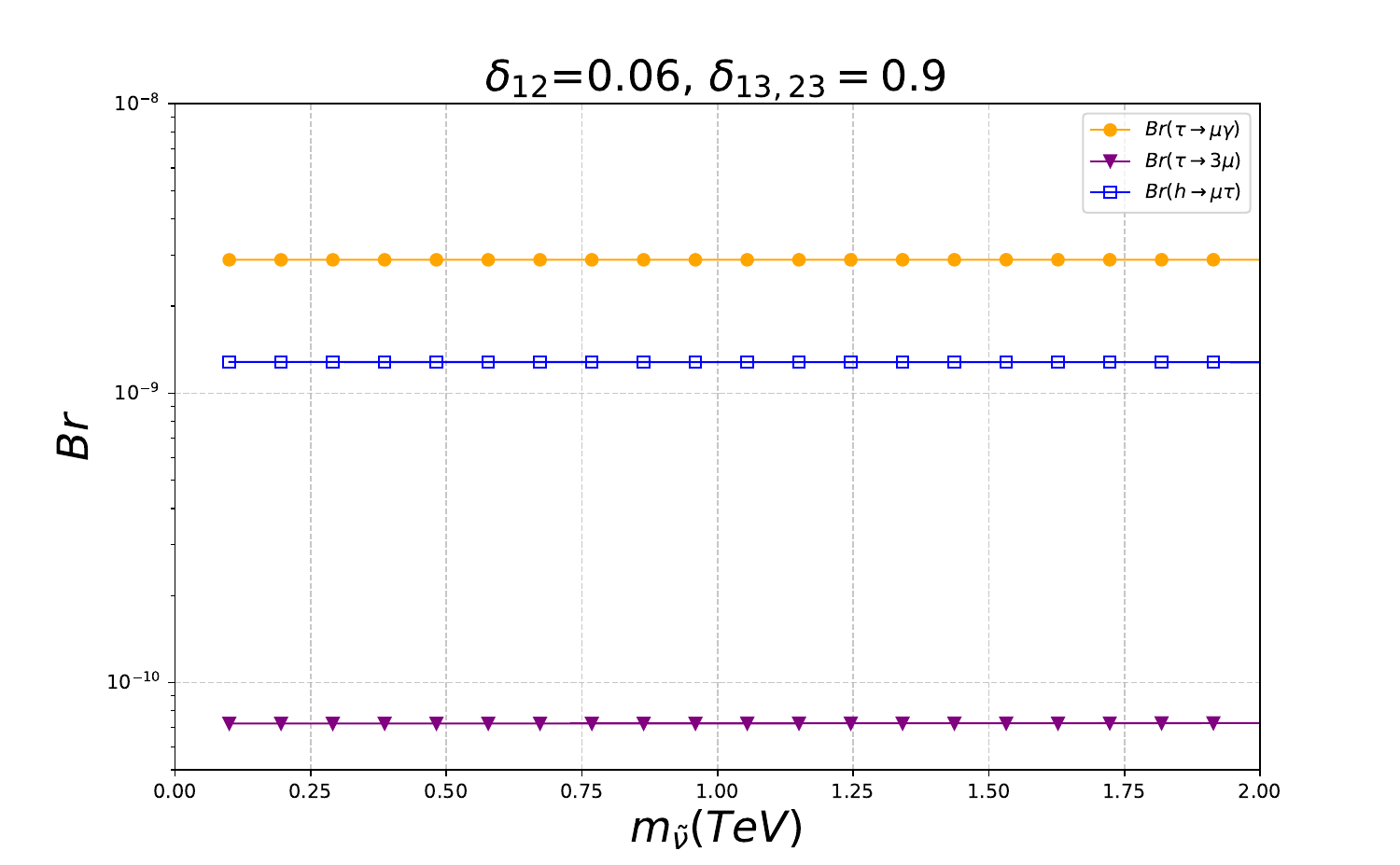}}

  \subfigure[]{\label{fig:sub1}\includegraphics[width=0.32\textwidth]{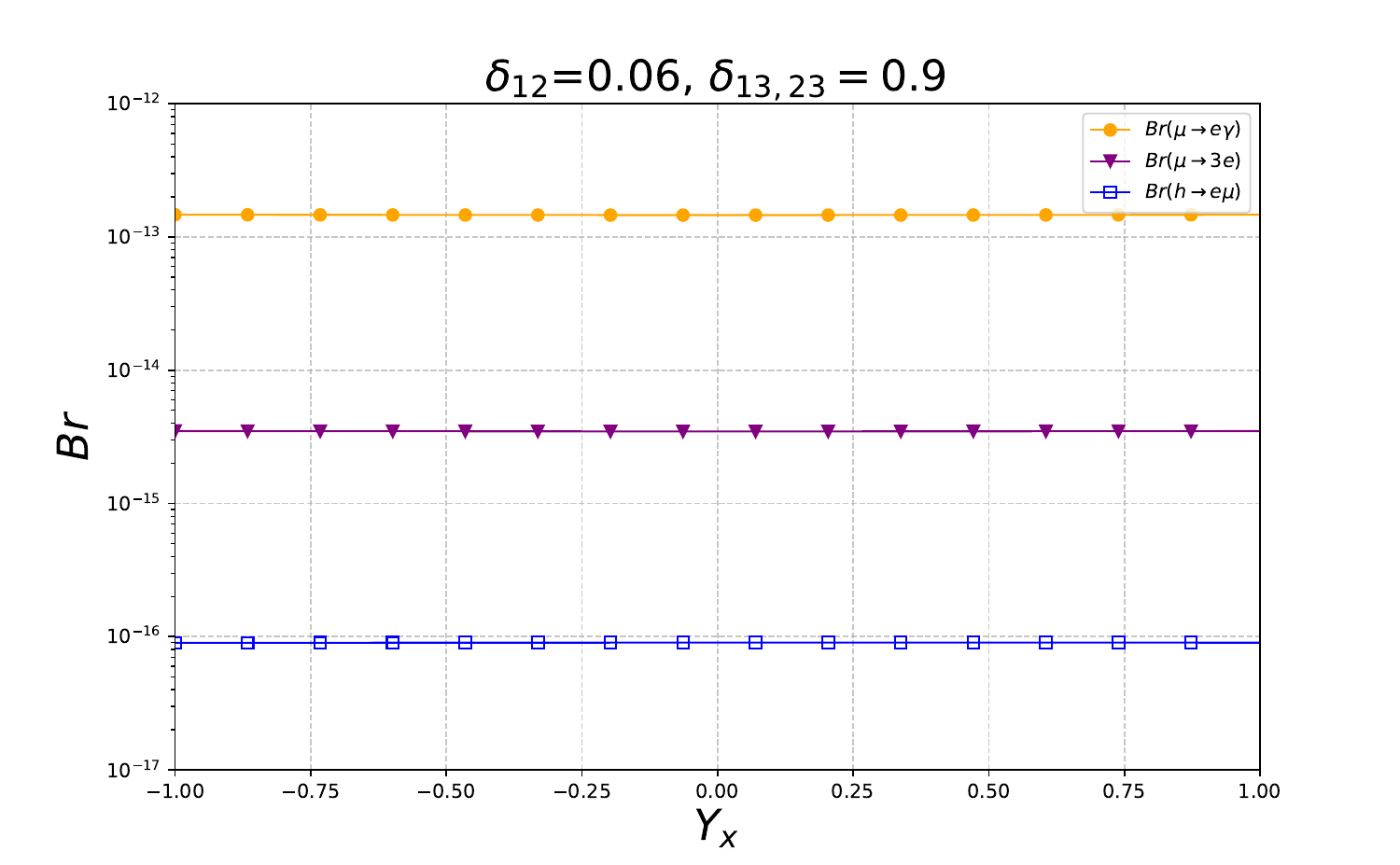}}
  \subfigure[]{\label{fig:sub1}\includegraphics[width=0.32\textwidth]{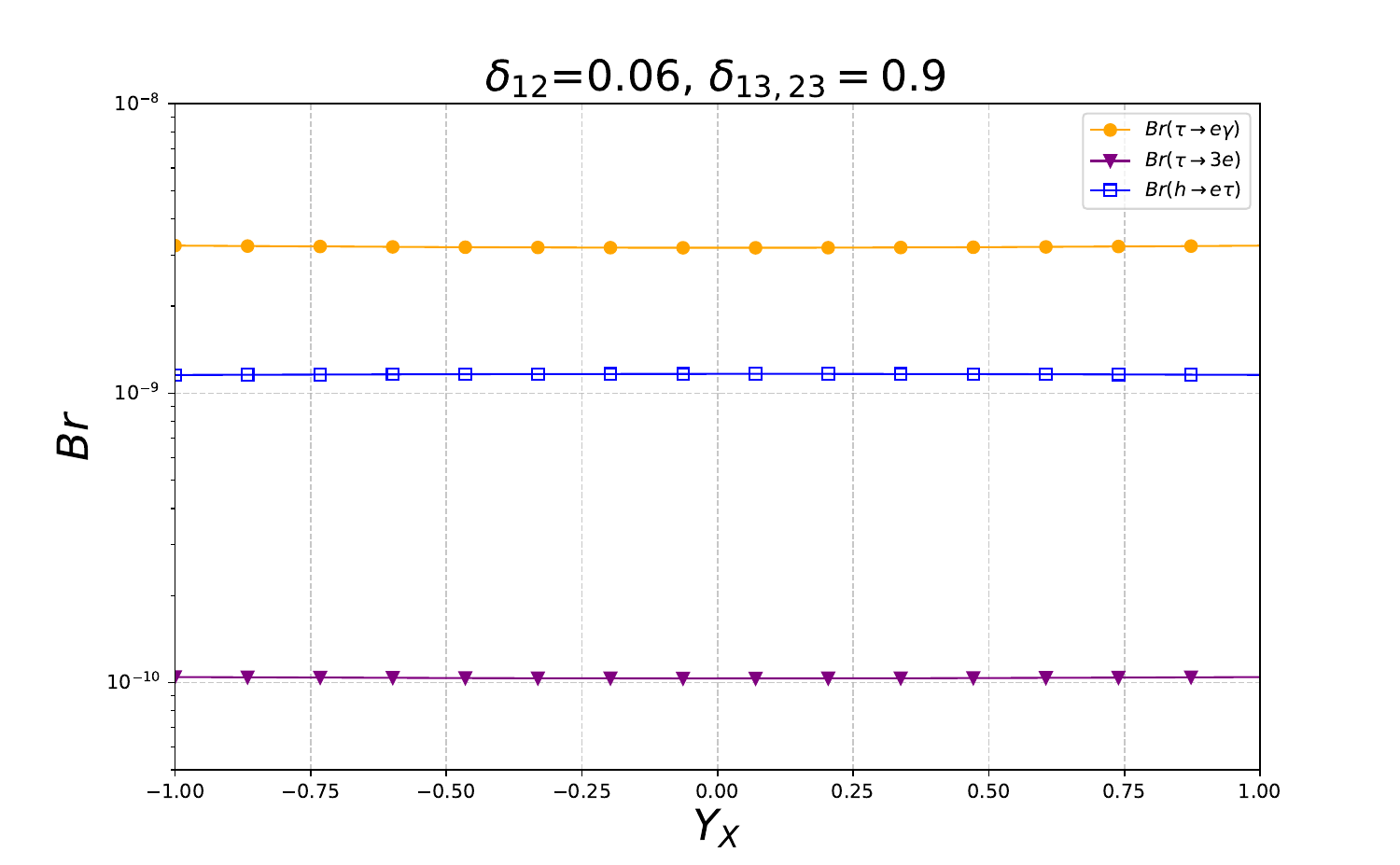}}
  \subfigure[]{\label{fig:sub1}\includegraphics[width=0.32\textwidth]{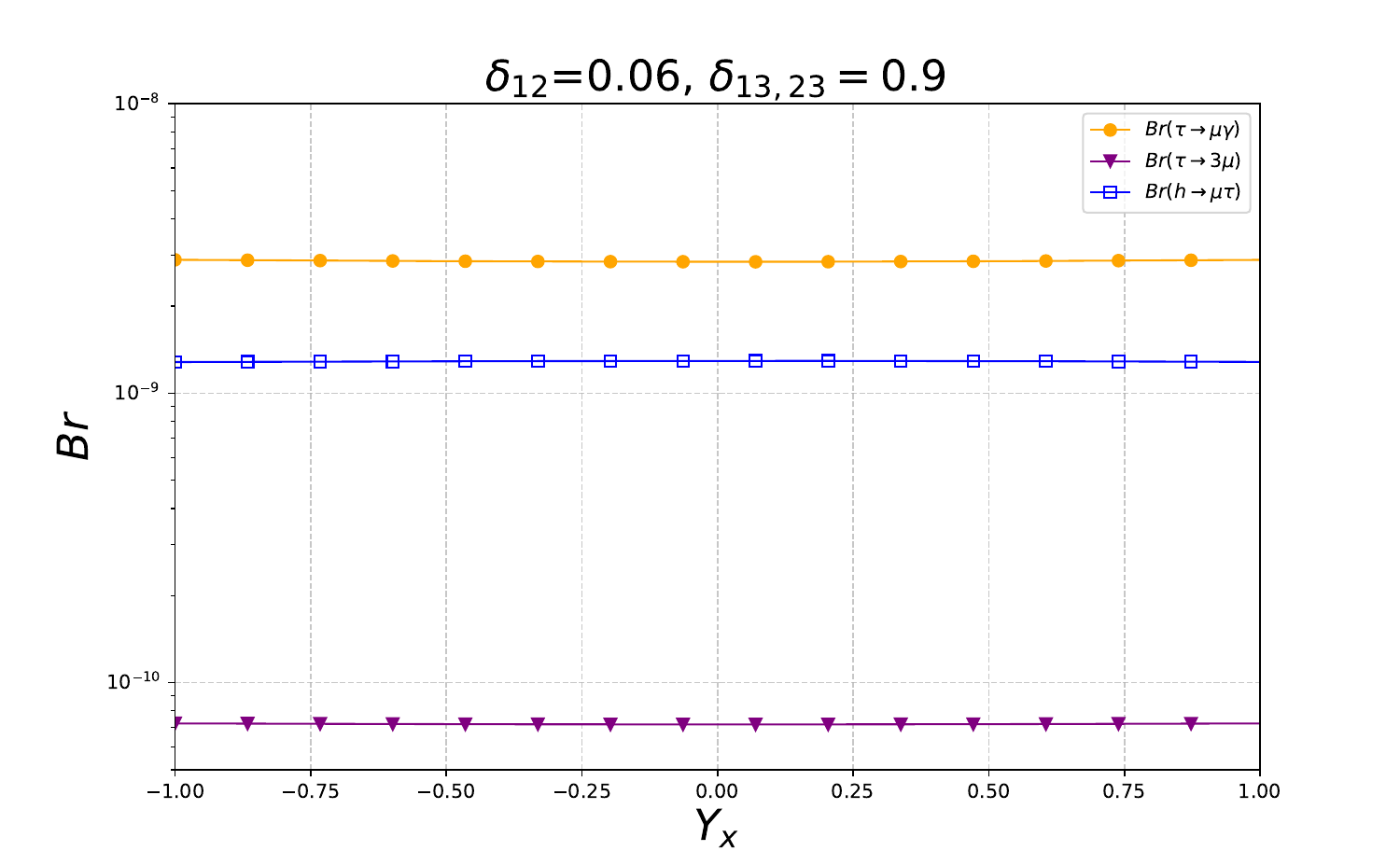}}

  \caption{The influence of new  parameters $Y_x$ and $m_{\tilde{\nu}}$ to LFV, $\delta_{ij}$ is defined as in  Eq.~(\ref{eq:delta}) }
  \label{fig:new  add parameters}
\end{figure}

We consider the right-handed Neutrino  masses to be at the TeV scale. Furthermore, we find that under this assumption, the Yukawa coupling \( Y_\nu \) is constrained to be on the order of \( 10^{-7} \). The newly introduced RHNs contribute to LFV processes via \( Y_\nu \), rendering the impact of the right-handed Neutrino coupling \( Y_X \) on LFV negligible. The corresponding  diagrams are shown in Fig.~\ref{fig:new  add parameters}. Since \( Y_X \) has a negligible effect, we simply set \( v_S = -1\,\text{TeV} \) and \( Y_X = -1 \) for simplicity.
Then use  Eq~(\ref{eq:Yv}), we get two different $Y_\nu$:
$$
Y_\nu^{\text{NO}}=\begin{pmatrix}
  4.06\times 10^{-7}& 
  3.84\times 10^{-8}+0.75 \times10^{-8}i& 
  -2.92\times 10^{-7}+3.25 \times10^{-8}i\\
  3.84\times 10^{-8}-3.75 \times10^{-8}i & 
  1.15\times 10^{-6}& 
  4.49\times 10^{-7}-1.08\times10^{-8}i \\
  -2.92\times 10^{-7}-3.25\times10^{-8}i & 
  4.49\times 10^{-7}+1.08\times10^{-8}i & 
  9.52\times 10^{-7}
  \end{pmatrix},
$$

$$
Y_\nu^{\text{IO}}=\begin{pmatrix}
1.53\times10^{-6}& 
9.46\times10^{-8}-2.80\times10^{-8}i& 
7.56\times10^{-8}-2.42\times10^{-8}i\\
9.46\times10^{-8}+2.80\times10^{-8}i& 
1.07\times10^{-6} & 
-4.19\times10^{-7}-2.11\times10^{-10}i\\
7.56\times10^{-8}+2.42\times10^{-8}i & 
-4.19\times10^{-7}+2.11\times10^{-10}i& 
1.19\times10^{-6}
\end{pmatrix}.
$$

For the soft breaking Slepton mass matrices $ M_{\tilde{l}} $  and $ M_{\tilde{e}} $, we introduce the slepton flavor mixings, which take into account the off-diagonal terms

\begin{eqnarray}
M_{\tilde{l}} = \begin{pmatrix}
1 & \delta_{12} & \delta_{13} \\
 \delta_{21} & 1 & \delta_{23} \\
\delta_{31} & \delta_{32} & 1
\end{pmatrix}m_{\tilde{l}},~M_{\tilde{e}} = \begin{pmatrix}
1 & \delta_{12} & \delta_{13} \\
\delta_{21} & 1 & \delta_{23} \\
\delta_{31} & \delta_{32} & 1
\end{pmatrix}m_{\tilde{e}}.
\label{eq:delta}
\end{eqnarray}

\begin{figure}
  \centering
  \subfigure[]{\label{fig:sub1}\includegraphics[width=0.3\textwidth]{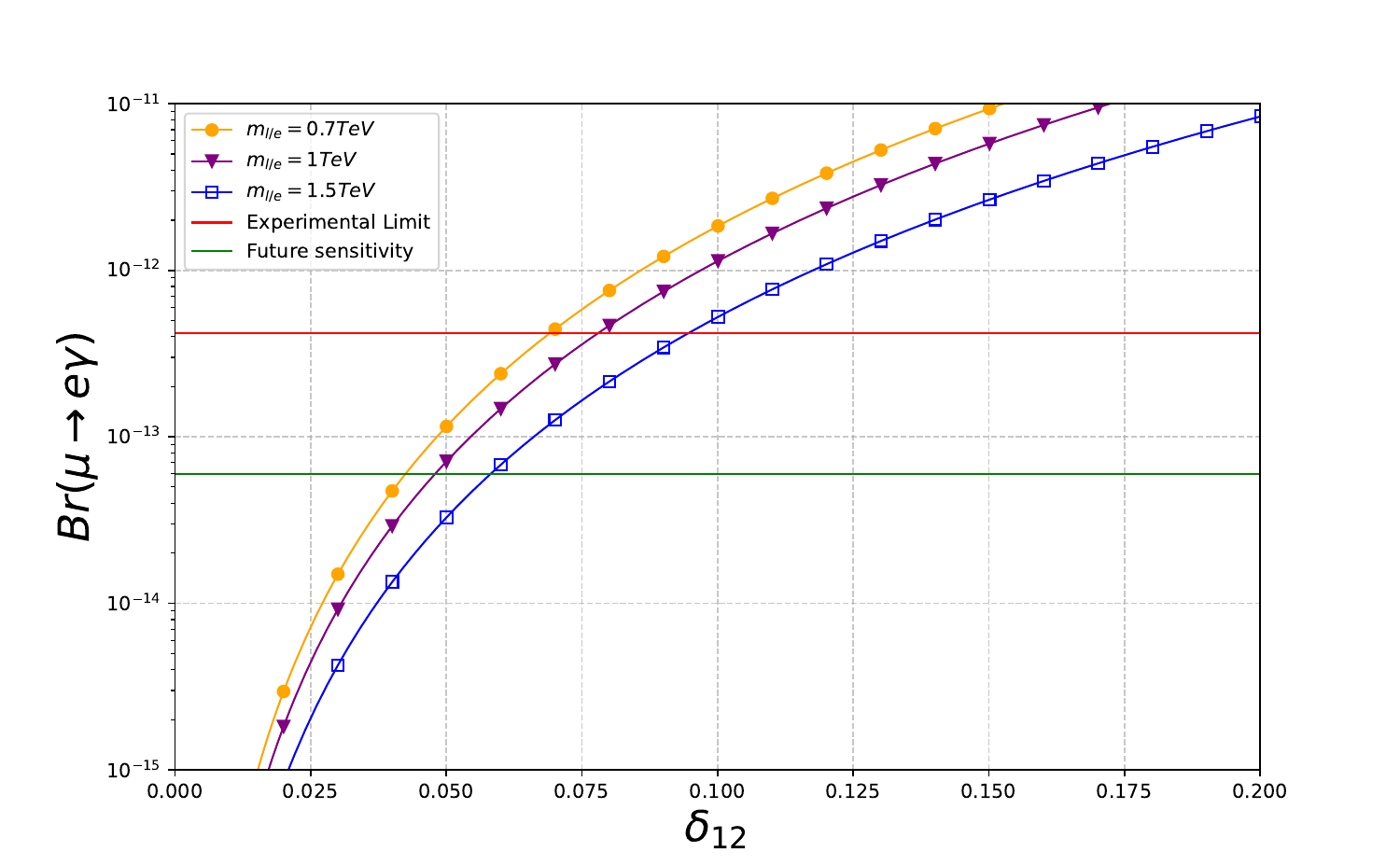}}
  \subfigure[]{\label{fig:sub2}\includegraphics[width=0.3\textwidth]{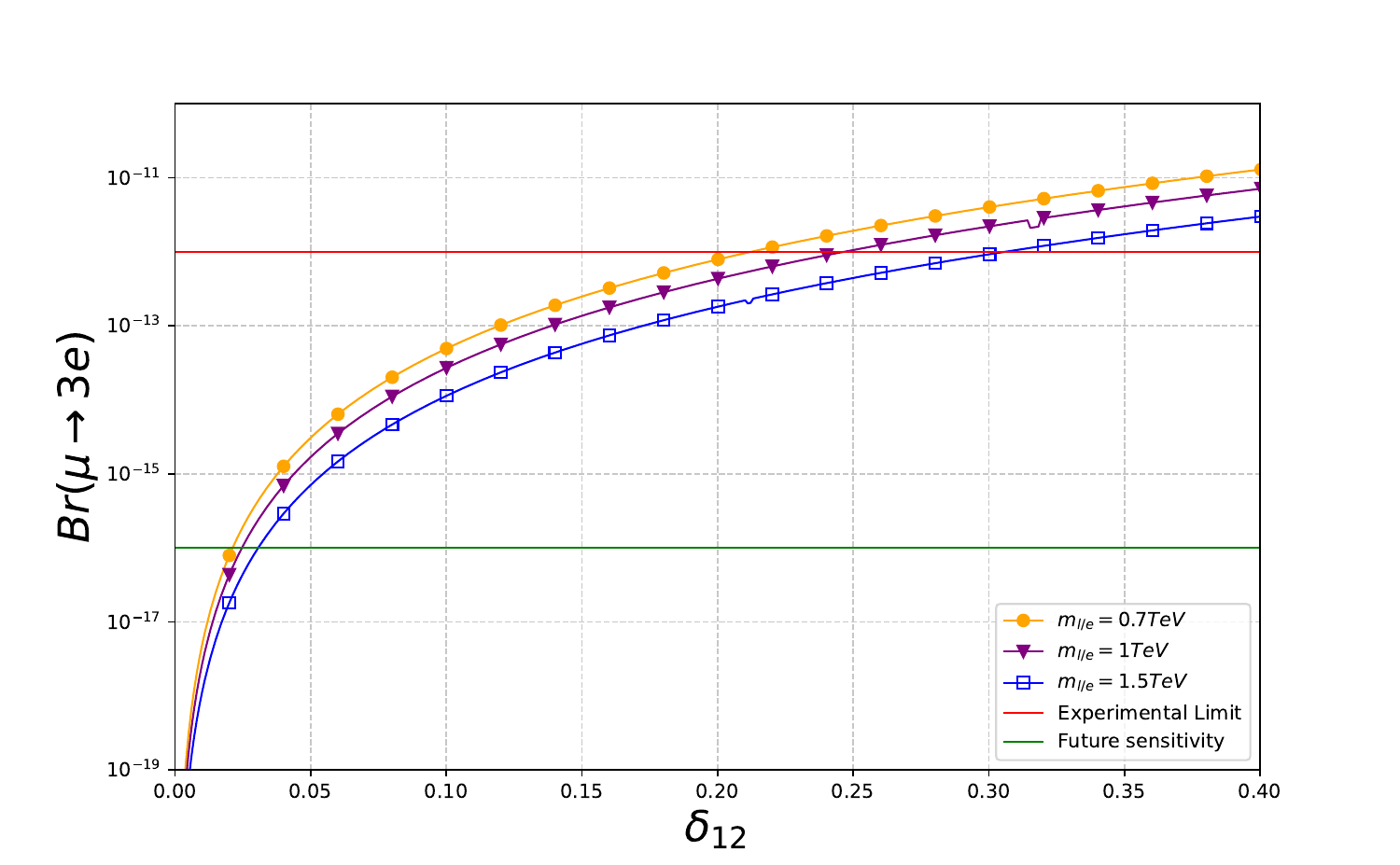}}
  \subfigure[]{\label{fig:sub3}\includegraphics[width=0.3\textwidth]{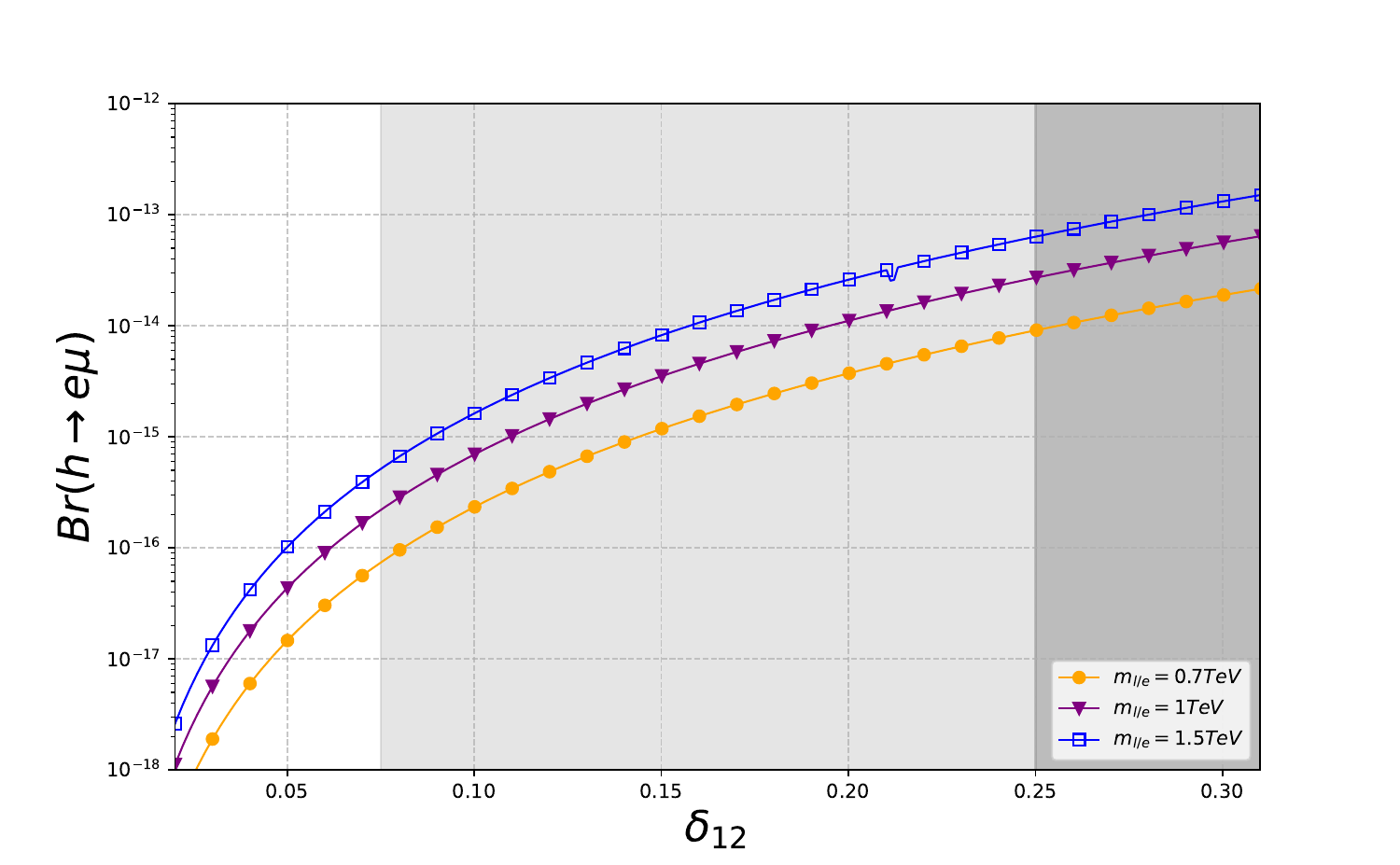}}
  \caption{The influence of $\delta_{12}$ and $m_{\tilde{l}/\tilde{e}}$ to Br($\mu\to e \gamma$), Br($\mu \to 3e$) and Br($h\to e\mu$), where the red and green lines denote the present limits and future sensitivities respectively, (c).Light gray represents the limit for $\mu\to e\gamma$, while dark gray represents the limit for $\mu\to 3e$.}
  \label{fig:delta12}
\end{figure}

\begin{figure}
  \centering
  \subfigure[]{\label{fig:sub1}\includegraphics[width=0.3\textwidth]{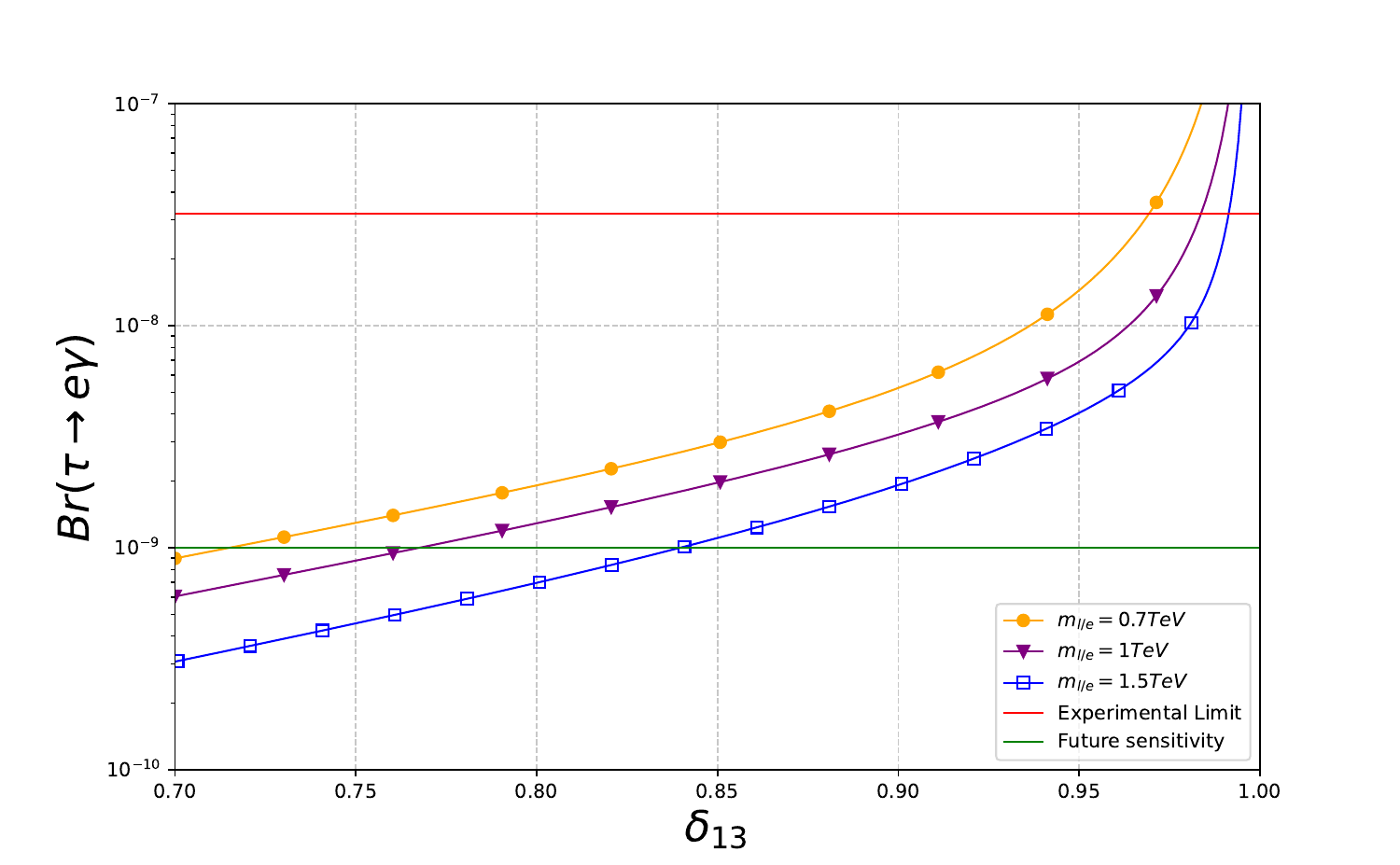}}
  \subfigure[]{\label{fig:sub2}\includegraphics[width=0.3\textwidth]{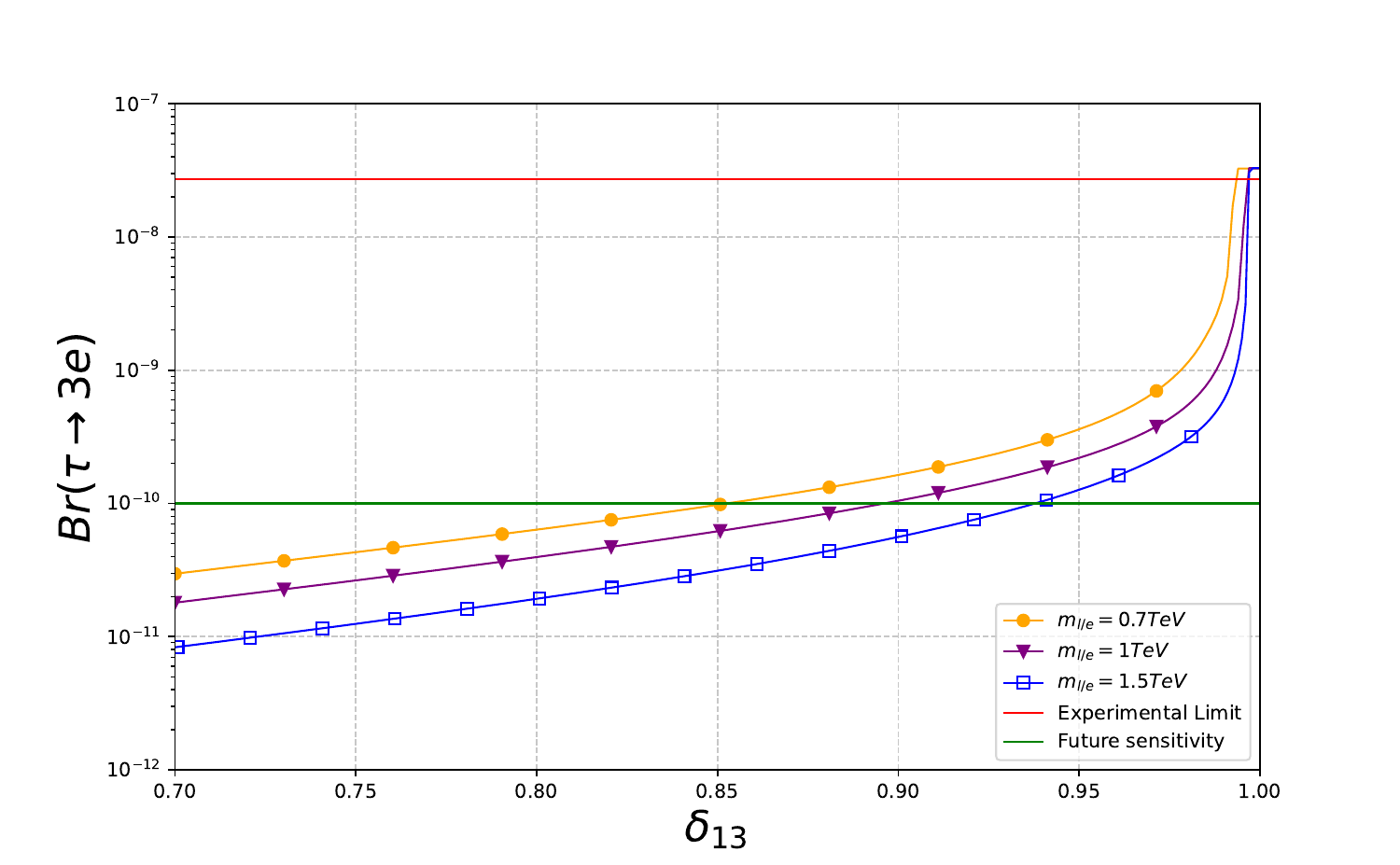}}
  \subfigure[]{\label{fig:sub3}\includegraphics[width=0.3\textwidth]{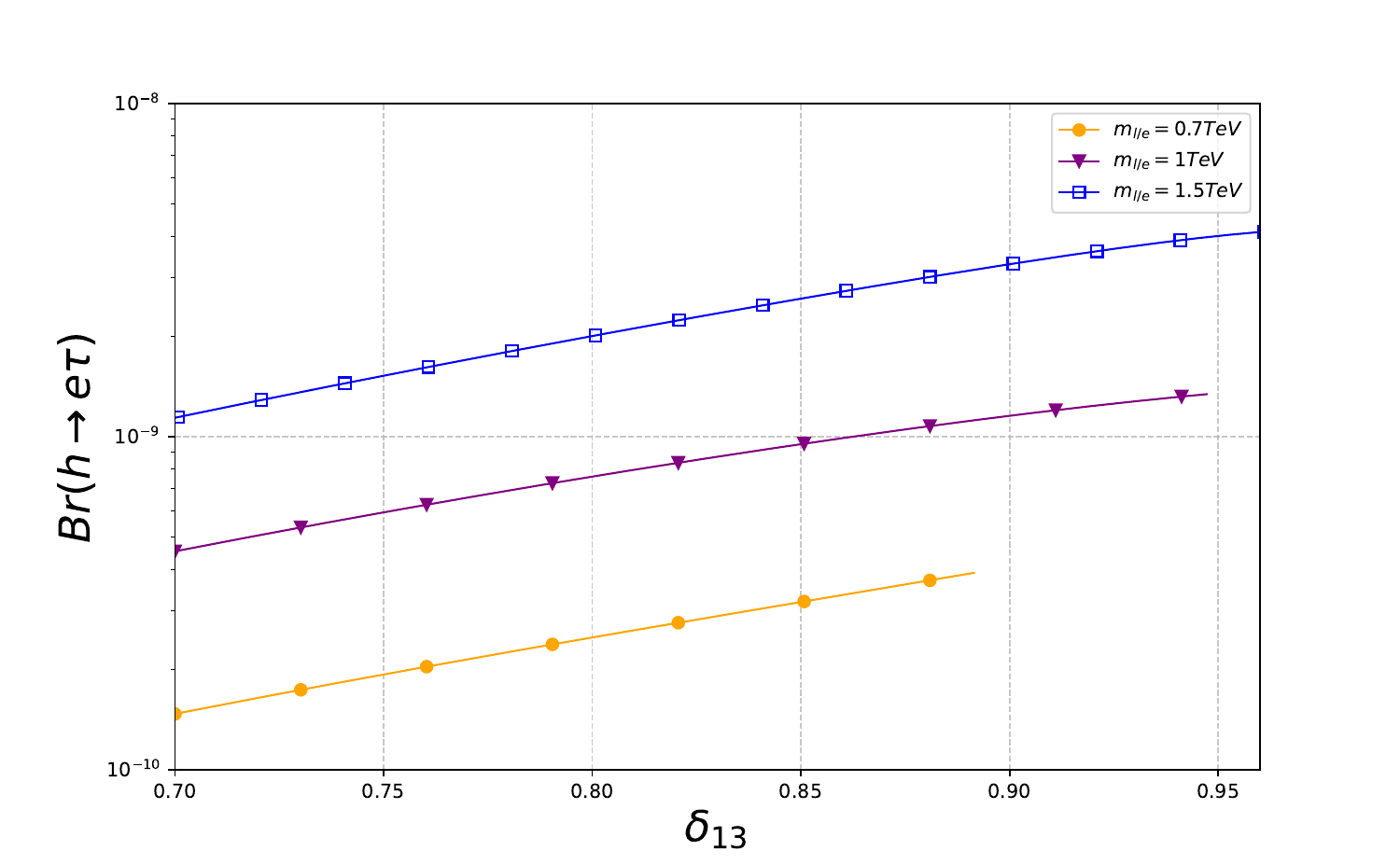}}
  \caption{The influence of $\delta_{13}$ and $m_{\tilde{l},\tilde{e}}$ to Br($\tau\to e \gamma$) , Br($\tau\to 3e$) and Br($h\to e\tau$) where the red and green lines
  denote the present limits and future sensitivities respectively }
  \label{fig:delta13}
\end{figure}

\begin{figure}
  \centering
  \subfigure[]{\label{fig:sub1}\includegraphics[width=0.3\textwidth]{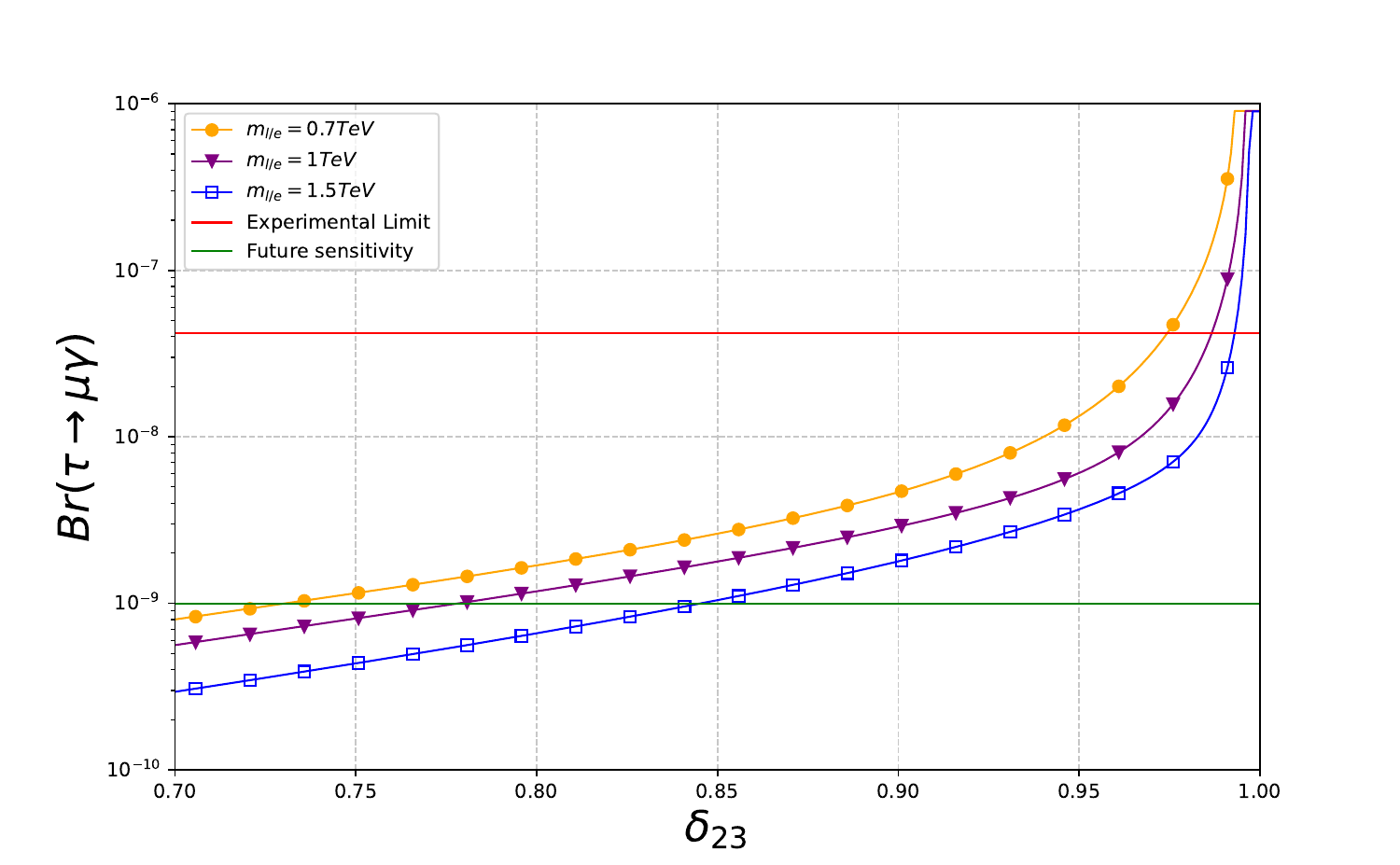}}
  \subfigure[]{\label{fig:sub2}\includegraphics[width=0.3\textwidth]{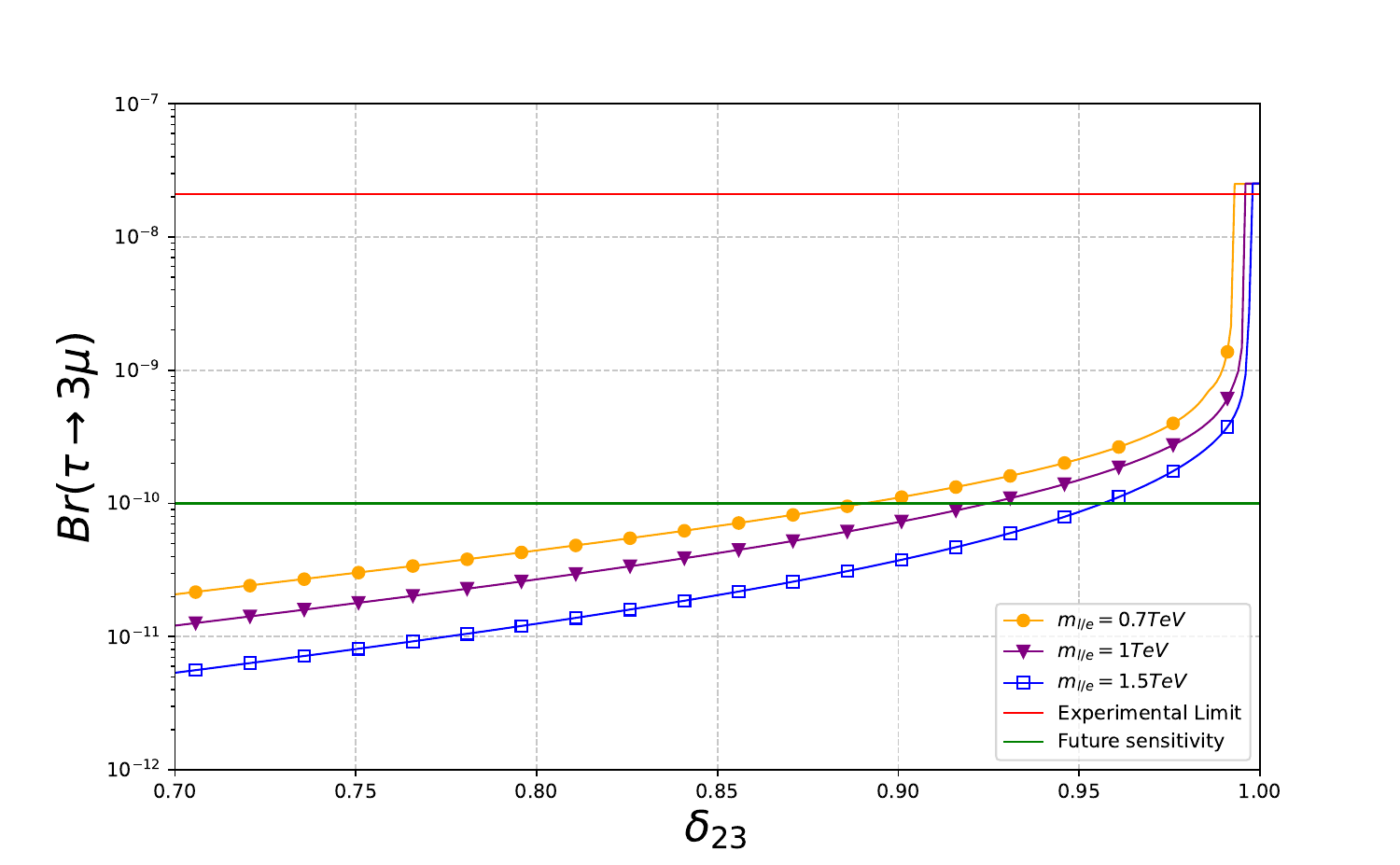}}
  \subfigure[]{\label{fig:sub3}\includegraphics[width=0.3\textwidth]{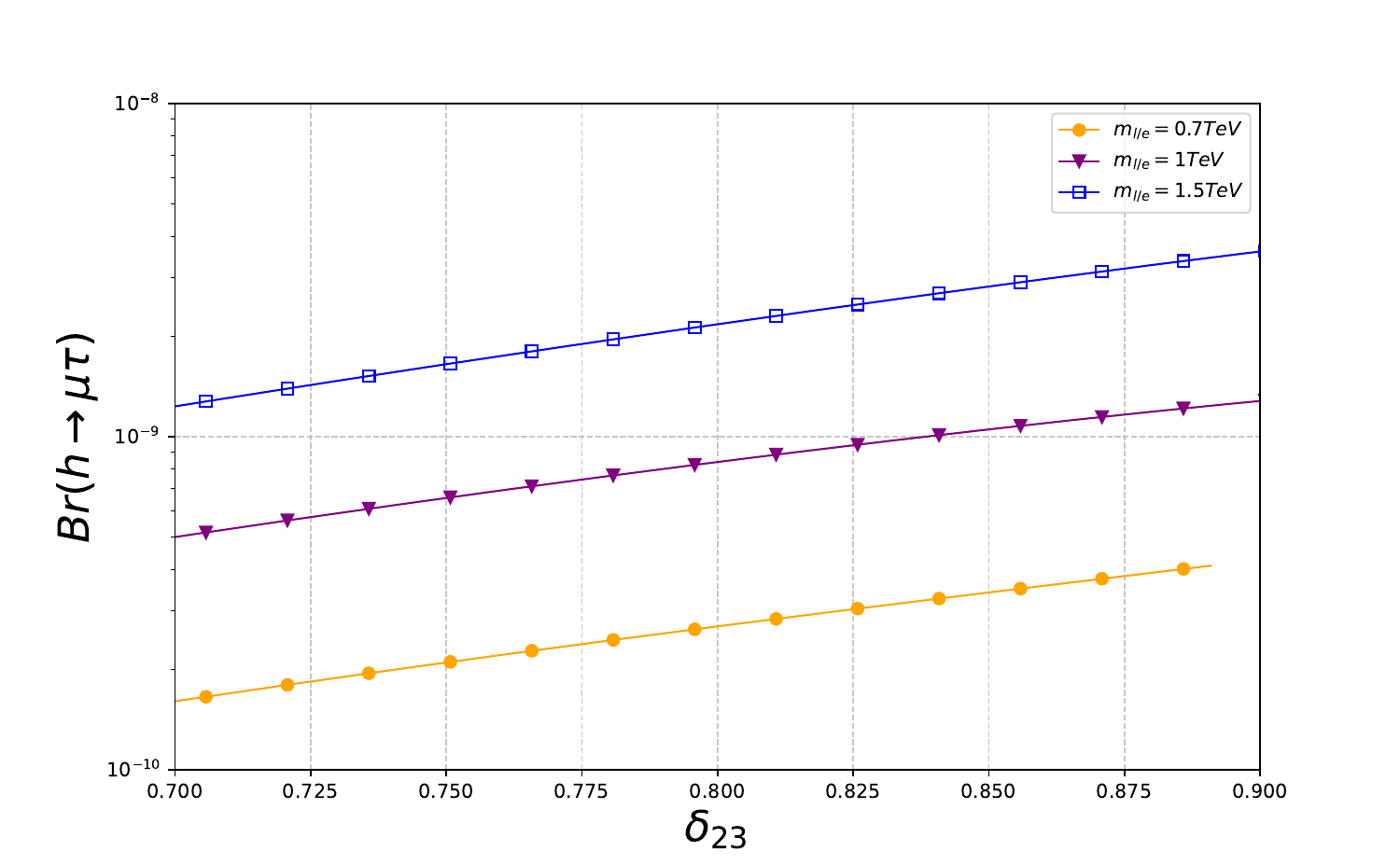}}
  \caption{The influence of $\delta_{23}$ and $m_{\tilde{l}/\tilde{e}}$ to Br($\tau\to \mu \gamma$), Br($\tau \to 3\mu$) and  Br($h\to \mu\tau$) where the red and green lines denote the present limits and future sensitivities respectively.}
  \label{fig:delta23}
\end{figure}

Then we plot Br($\mu\to e \gamma$), Br($\mu \to 3e$) and Br($h\to e\mu$) versus $\delta_{12}=\delta_{21}$ for $\delta_{13} =\delta_{31}= \delta_{23}=\delta_{32} = 0$ in Fig.~\ref{fig:delta12}. In Fig.~\ref{fig:delta13} we picture Br($\tau \to e\gamma$) and Br($\tau \to 3 e$) versus $\delta_{13}=\delta_{31}$ for $\delta_{12}=\delta_{21}=\delta_{23}=\delta_{32}=0$ and Br($\tau \to \mu\gamma$) and Br($\tau\to 3 \mu$) versus $\delta_{23}=\delta_{32}$ for  $\delta_{12}=\delta_{21}=\delta_{13}=\delta_{31}=0$ are drawn in Fig.~\ref{fig:delta23}.

It is obvious that the LFV rates increase with the increasing of Slepton mixing parameters. Fig.~\ref{fig:delta12} shows that  the present experimental limit bound of BR($\mu\to e\gamma$) constrains $\delta_{12}<0.075$, which also coincides with the present experimental limit bound of Br($\mu\to 3e$) , Due to the constraints of $\delta_{12}$ boundary condition, we obtained Higgs LFV Br($h\to e\mu$) at $10^{-15}$. In addition, from Fig.~\ref{fig:delta13} we can see that Br($\tau\to e \gamma$) can reach the corresponding present experimental limit bounds but Br($\tau \to 3e $) can't. However, the high future experimental sensitivities still keep a hope to detect Br($\tau\to e \gamma$) and Br($\tau \to 3e $) and we get the Br($h\to e\tau$) at $10^{-9}$. Moverover the Br($\tau\to \mu \gamma$) can reach the corresponding present experimental limit bounds, but Br($\tau\to 3\mu$)  can't reach. And we investigated Higgs LFV Br($h\to \mu \tau$) at $10^{-9}$.

\begin{figure}
  \centering
  \subfigure[]{\label{fig:sub1}\includegraphics[width=0.12\textwidth]{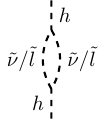}}
  \subfigure[]{\label{fig:sub2}\includegraphics[width=0.11\textwidth]{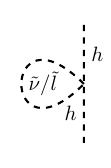}}
  \subfigure[]{\label{fig:sub3}\includegraphics[width=0.3\textwidth]{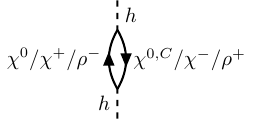}}
  \caption{The influence of Charginos, Neutralinos,  Sneutrinos and Sleptons  in Higgs self-energies. }
  \label{fig:self}
\end{figure}

Notice that, we must consider the loops contribution by Sleptons, Neutralinos and Charginos for Higgs mass, that the diagrams are shown as Fig.~\ref{fig:self}  

\begin{figure}
  \centering
  \subfigure[]{\label{fig:sub1}\includegraphics[width=0.32\textwidth]{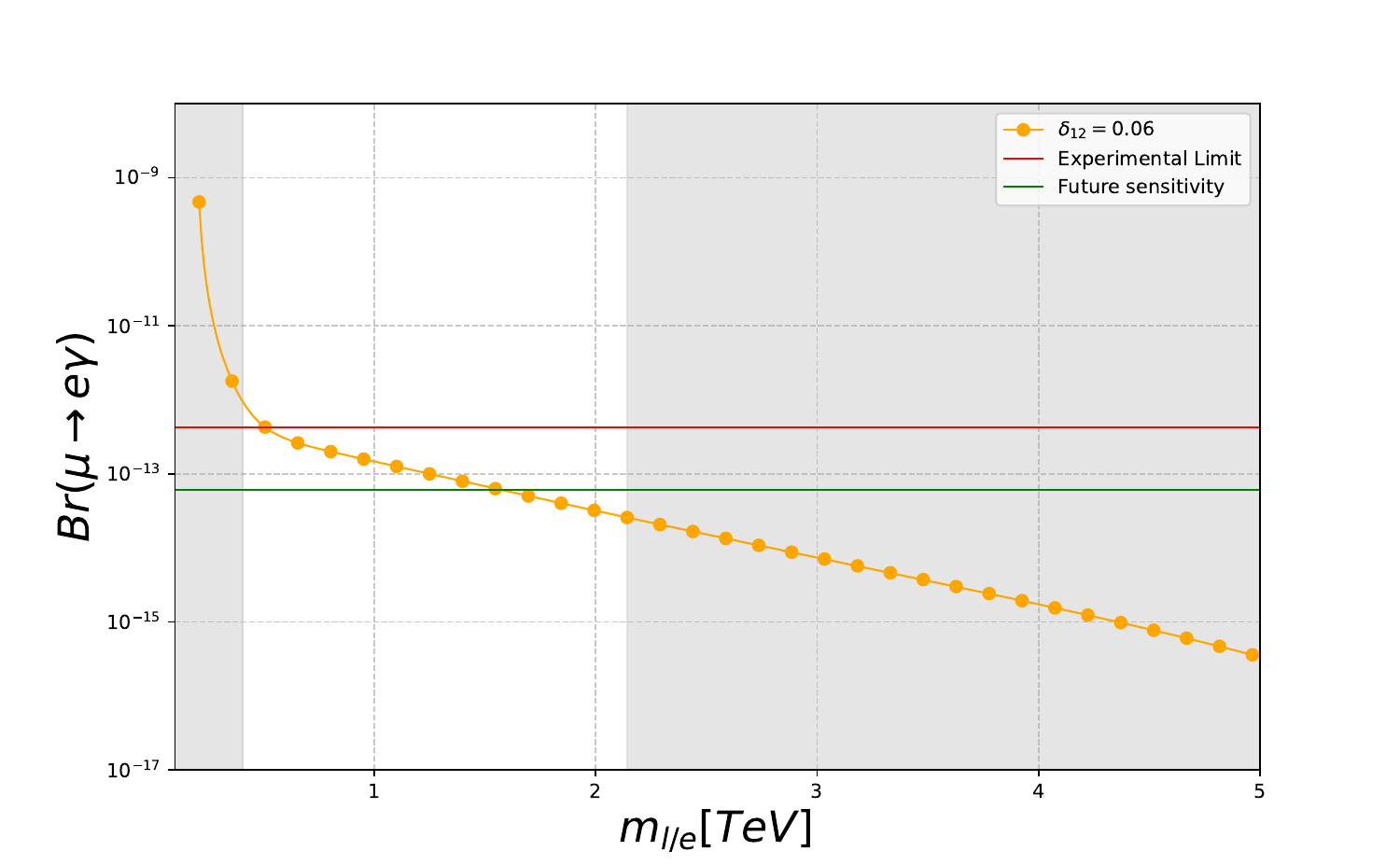}}
  \subfigure[]{\label{fig:sub2}\includegraphics[width=0.32\textwidth]{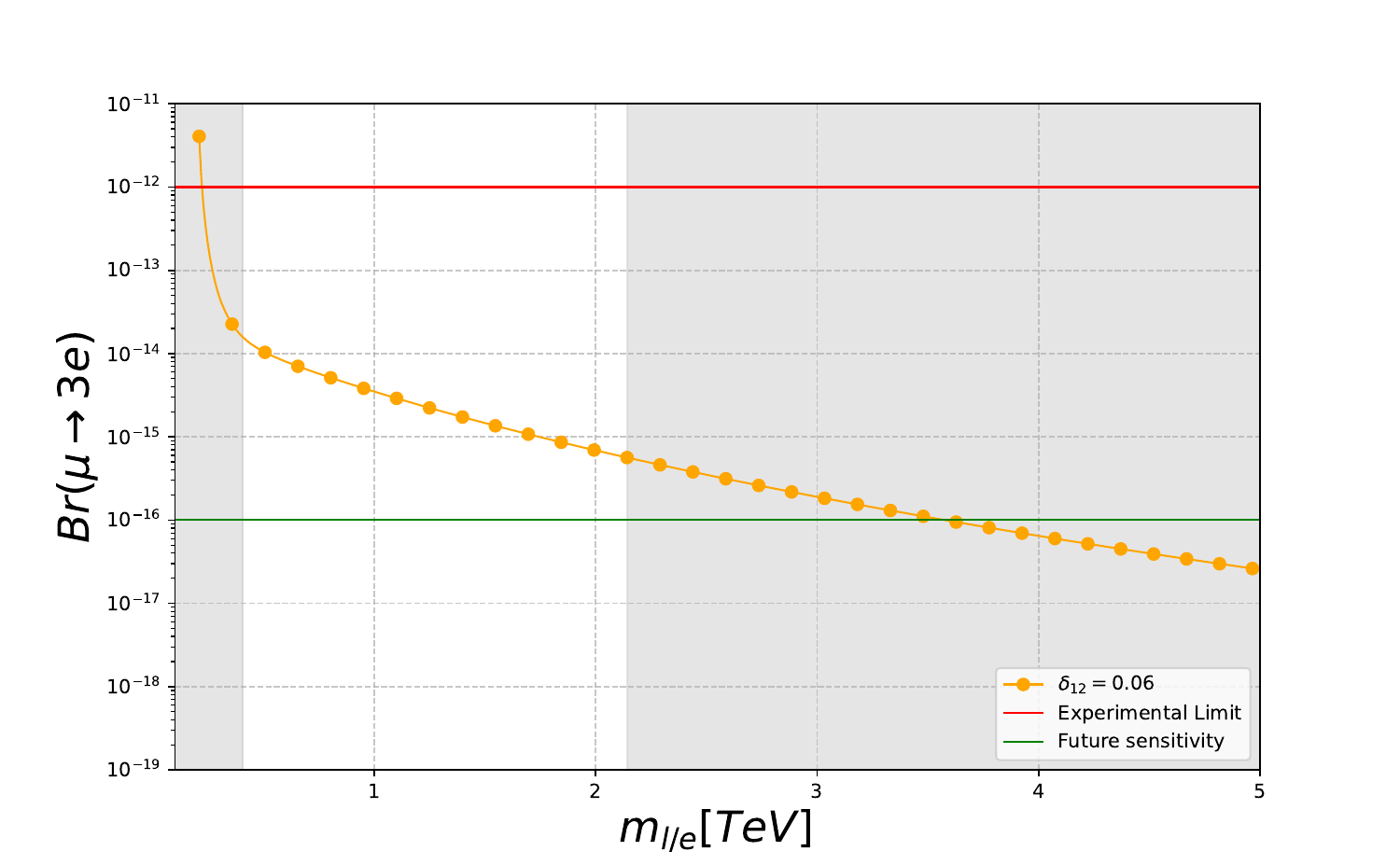}}
  \subfigure[]{\label{fig:sub3}\includegraphics[width=0.32\textwidth]{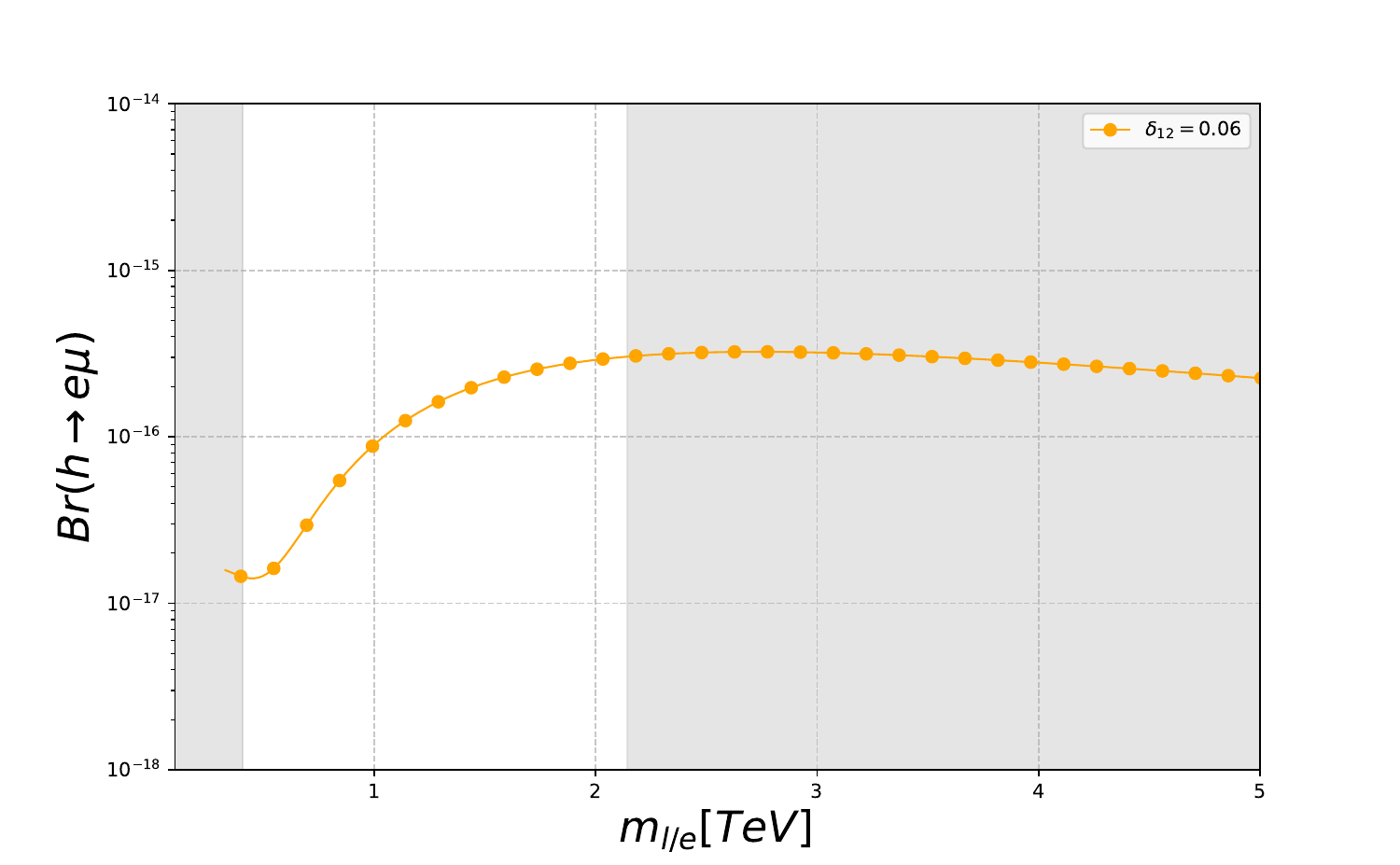}}

  \subfigure[]{\label{fig:sub1}\includegraphics[width=0.32\textwidth]{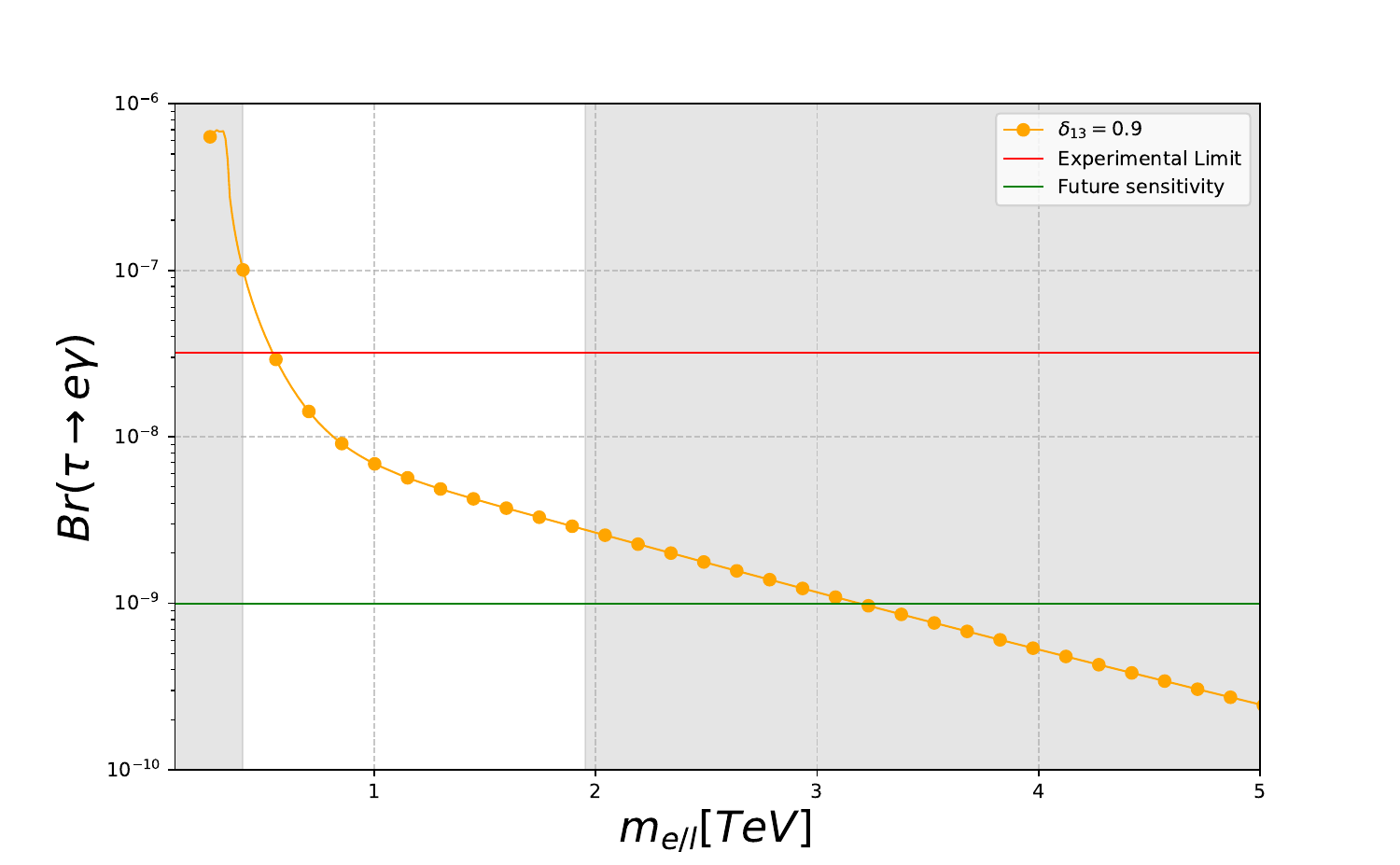}}
  \subfigure[]{\label{fig:sub2}\includegraphics[width=0.32\textwidth]{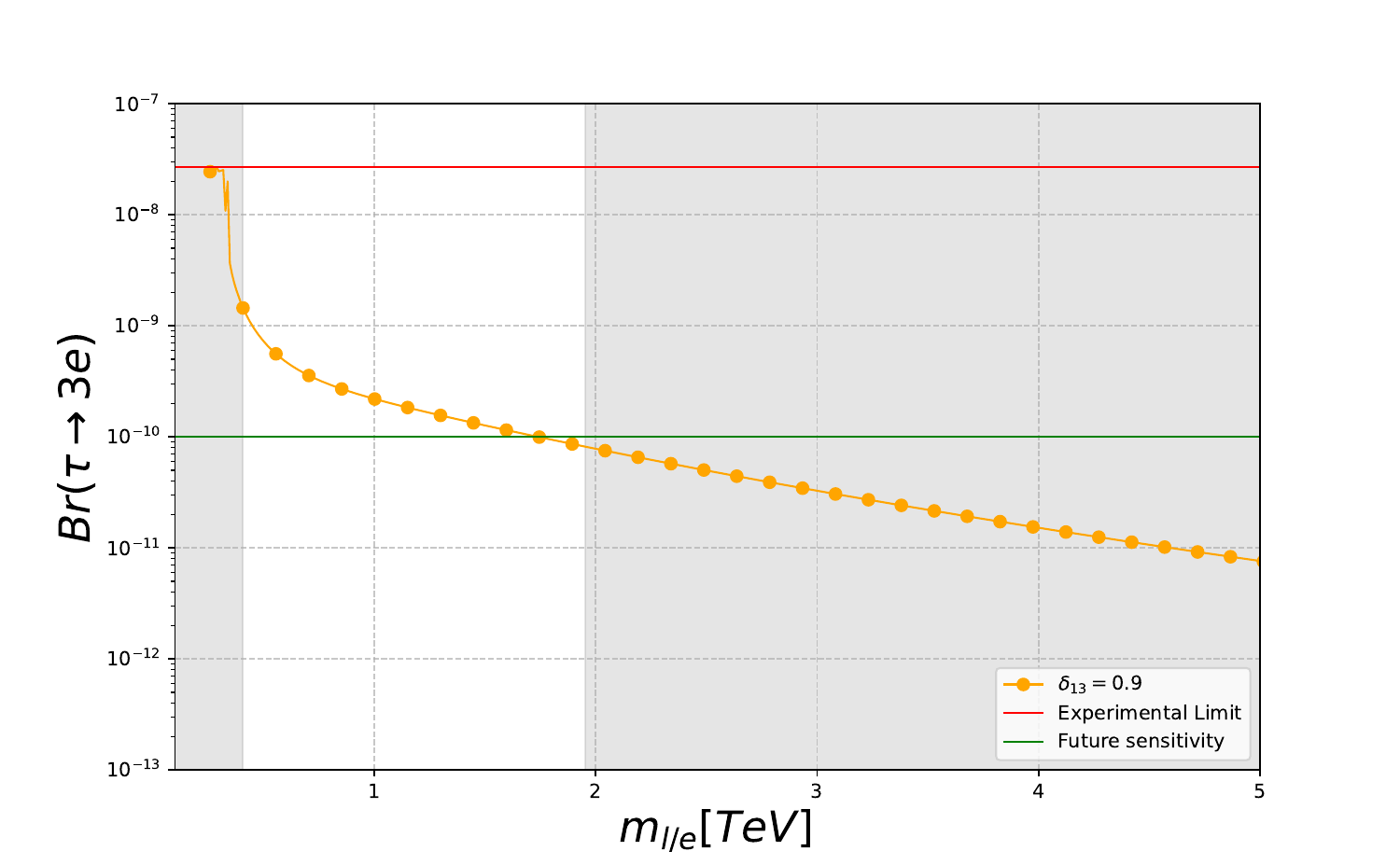}}
  \subfigure[]{\label{fig:sub3}\includegraphics[width=0.32\textwidth]{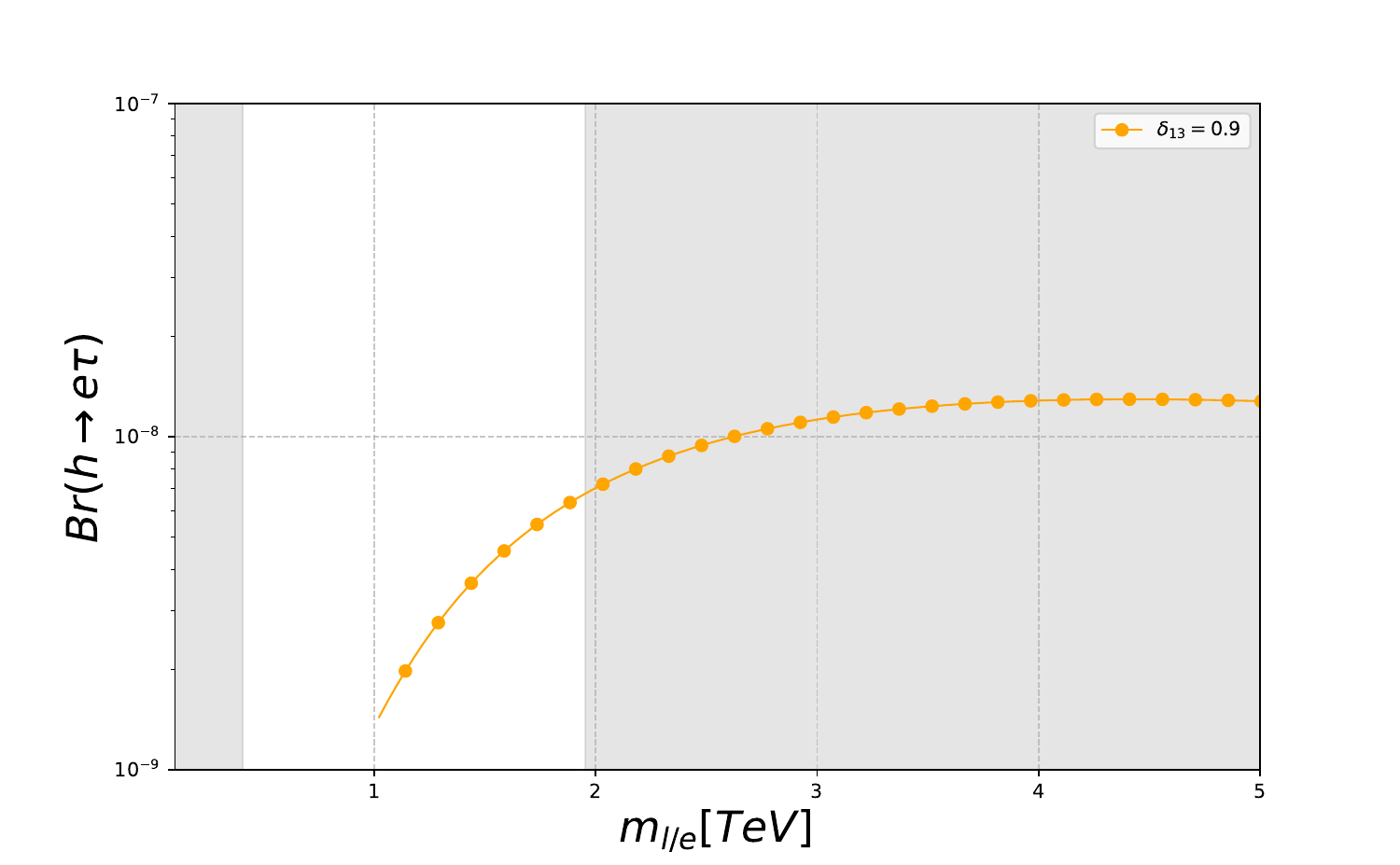}}

  \subfigure[]{\label{fig:sub1}\includegraphics[width=0.32\textwidth]{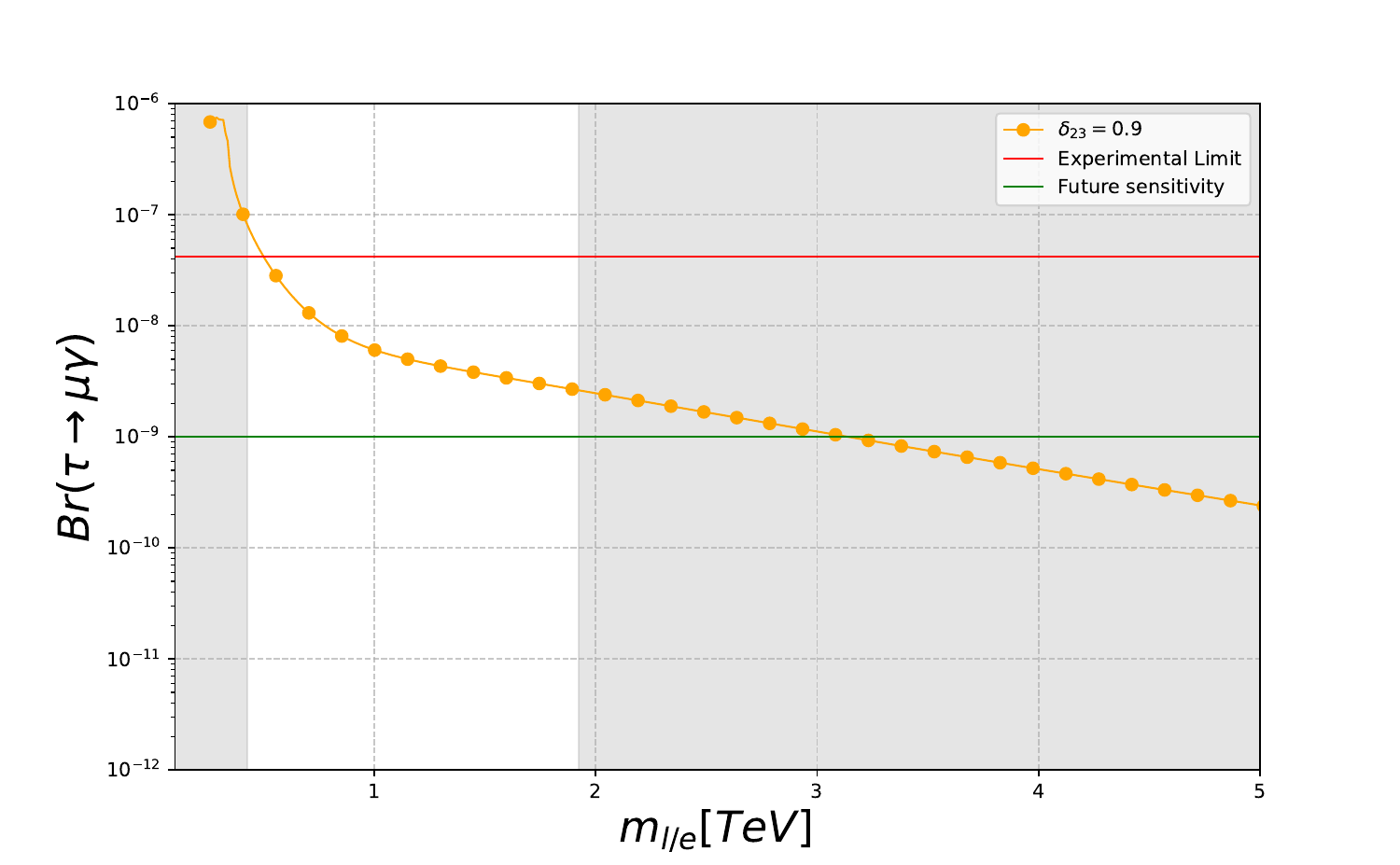}}
  \subfigure[]{\label{fig:sub2}\includegraphics[width=0.32\textwidth]{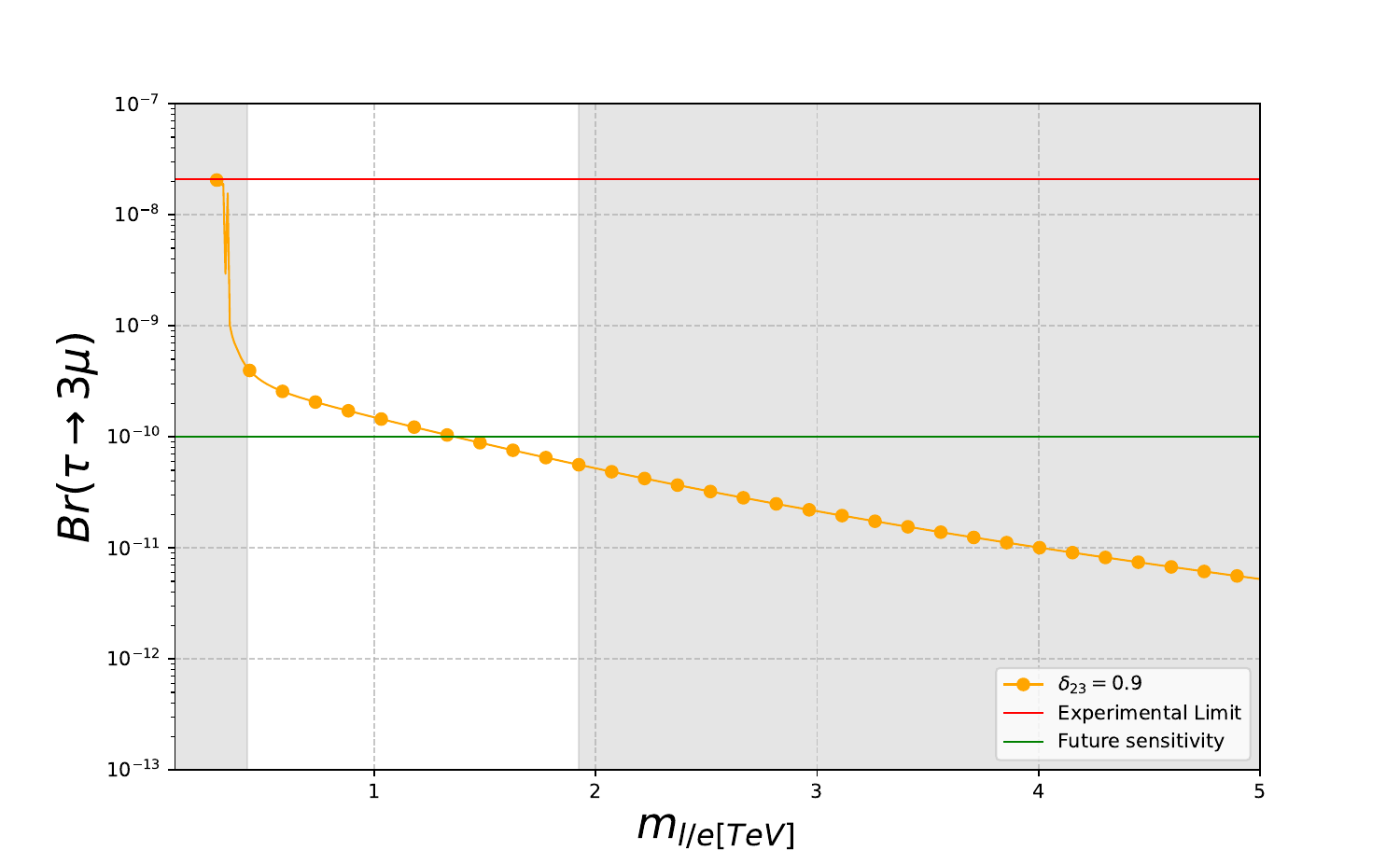}}
  \subfigure[]{\label{fig:sub3}\includegraphics[width=0.32\textwidth]{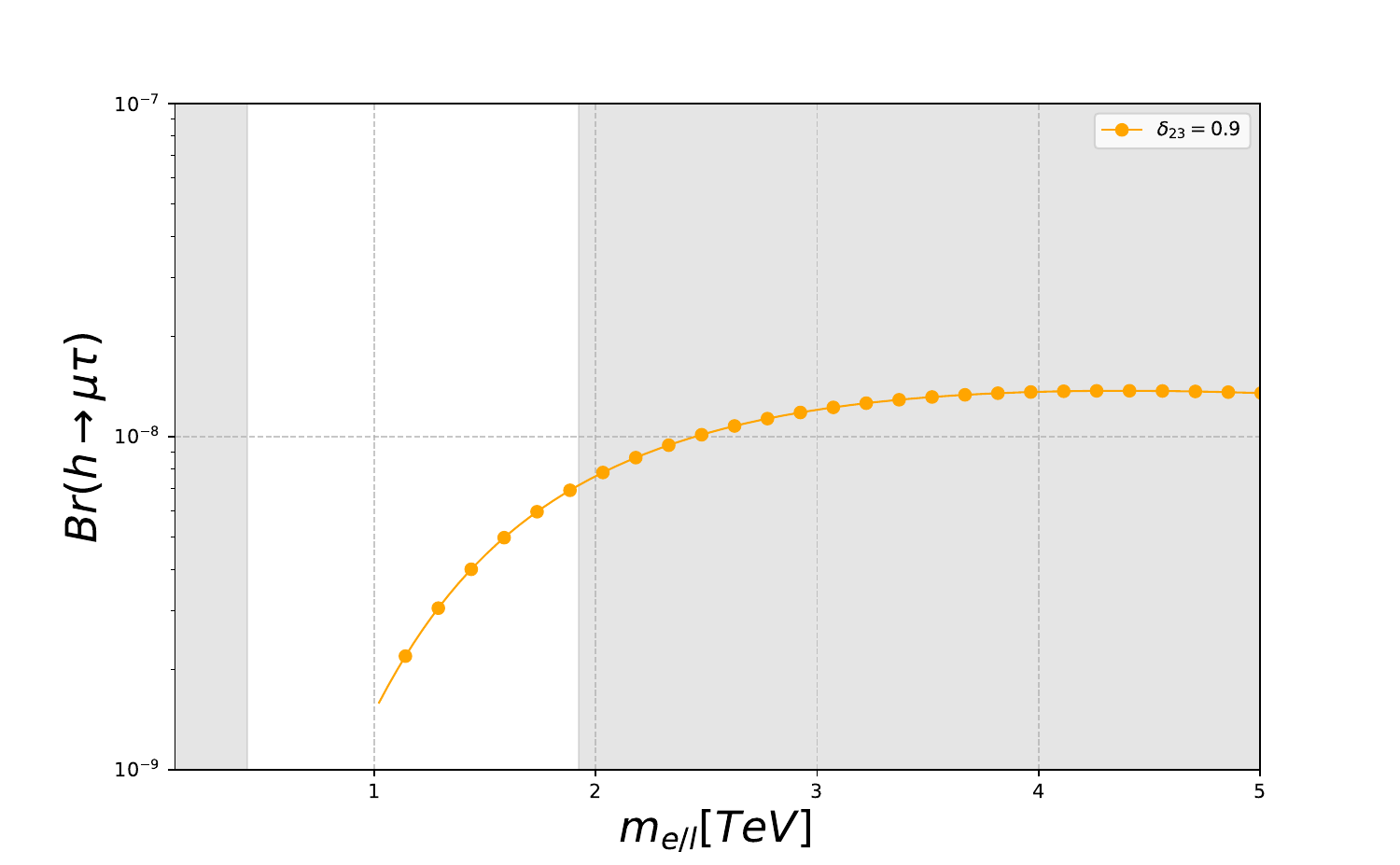}}

  \caption{The influence of Slepton mass for LFV with shadowed area represents exclusion by Higgs mass at 5$\sigma$ where the red and green lines denote the present limits and future sensitivities respectively.}
  \label{fig:SL}
\end{figure}

Furthermore, our analysis reveals that both the flavor mixing parameters $\delta_{ij}$ and the slepton mass contribute significantly to LFV observables. We then picture  the contribution  of LFV observables versus $m_{\tilde{l}/\tilde{e}}$. In Fig.~\ref{fig:SL} we found that the LFV observables  Br($\ell_{i}\to\ell_{j} \gamma$) and Br($\ell_{i}\to 3\ell_{j}$) decrease as the mass of Sleptons increases. That beacuse the mass parameters $m_{\tilde{l}/\tilde{e}}$ will appear  in the denominator terms of  expression for the decay width. However, we observed that Higgs LFV exhibits the opposite trend, it increases with $m_{\tilde{l}/\tilde{e}}$. As shown in Fig.~\ref{fig:self}.(a,b), the mass of Sleptons can affect the Higgs one-loop self-energy, As the Slepton mass increases, the Higgs mass corrections from loop diagrams decrease. Using Eq.~(\ref{eq:deacyh}), since the Higgs mass appears in the denominator, this leads to an increase in the Br($h\to\ell_i^{+}\ell_j^{-}$).

\begin{figure}
  \centering
  \subfigure[]{\label{fig:sub1}\includegraphics[width=0.32\textwidth]{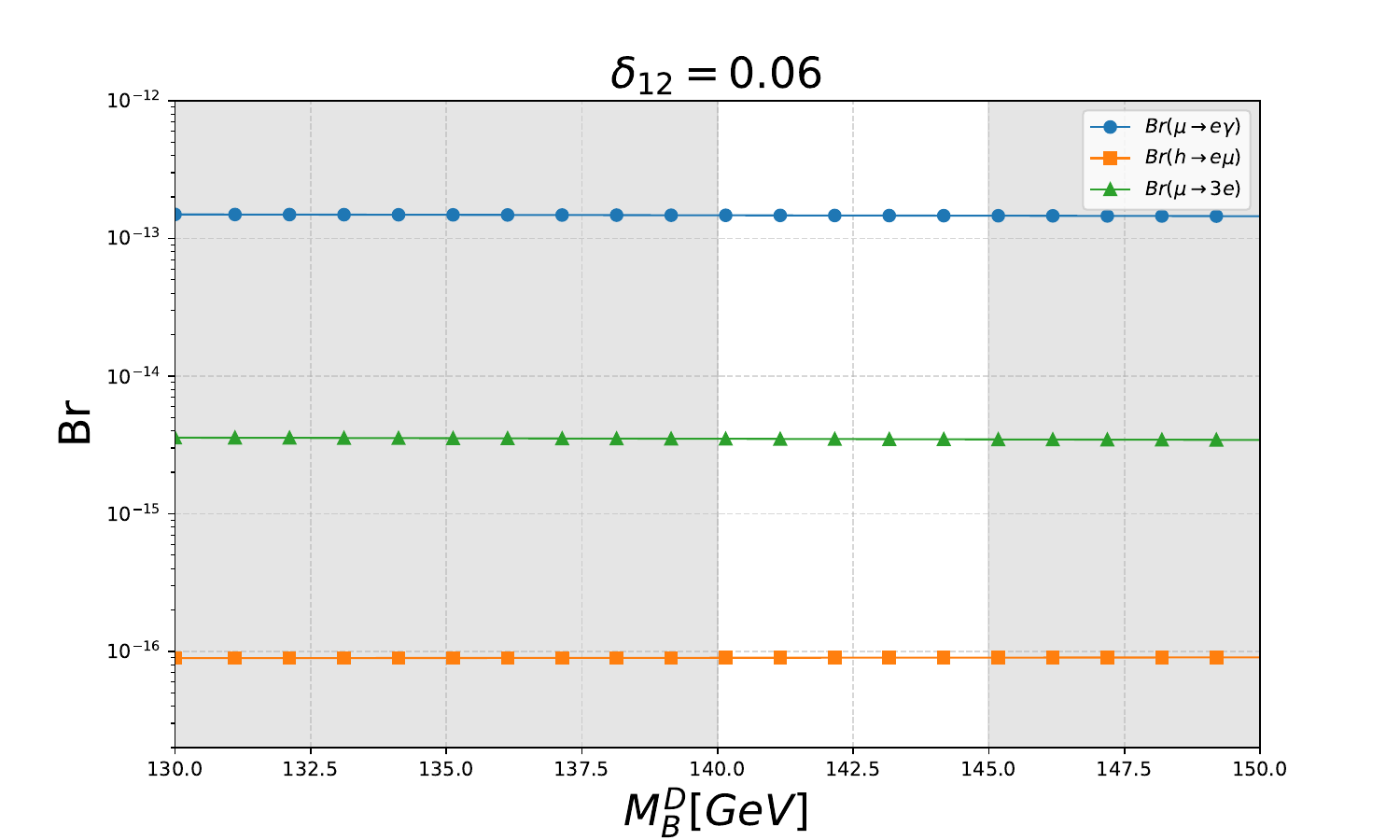}}
  \subfigure[]{\label{fig:sub1}\includegraphics[width=0.32\textwidth]{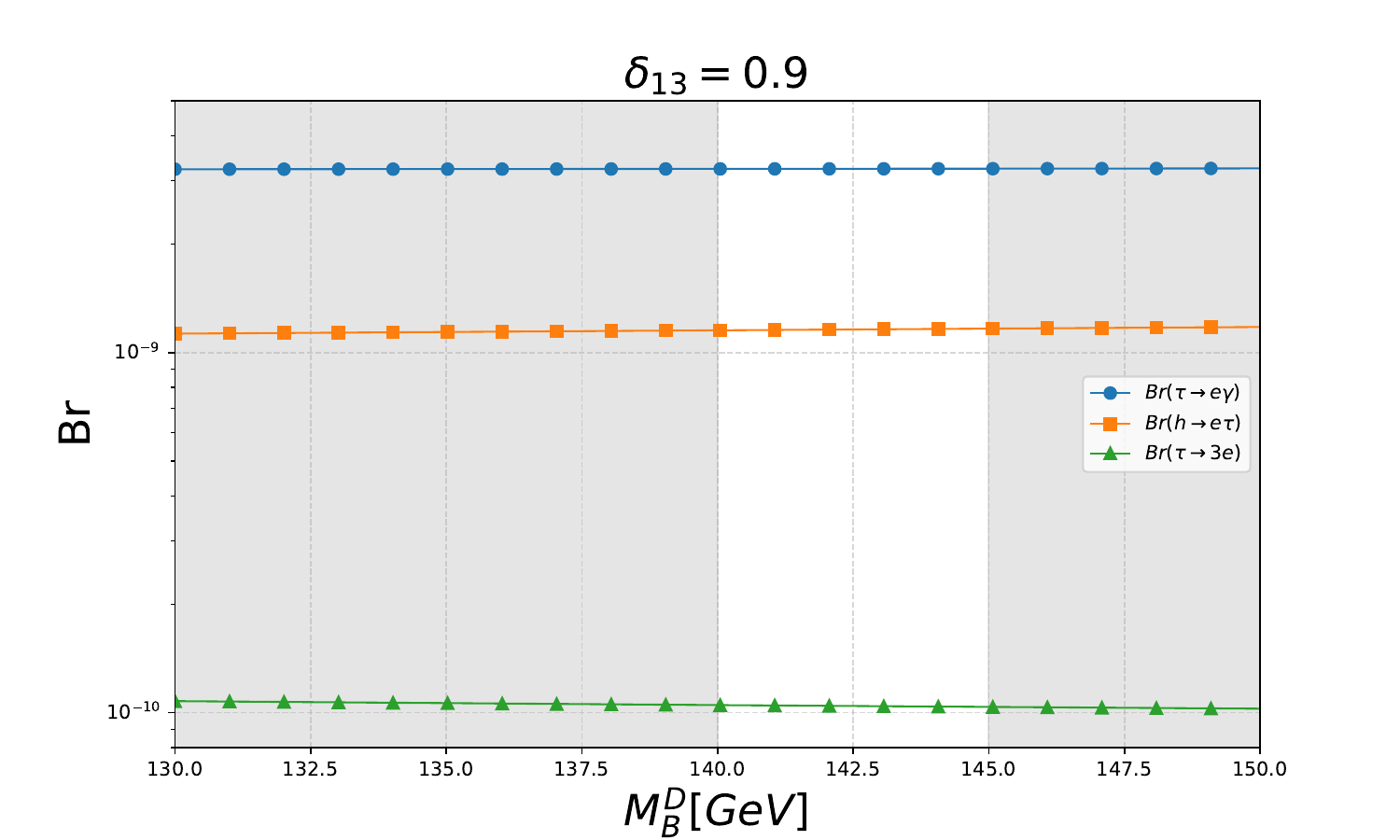}}
  \subfigure[]{\label{fig:sub1}\includegraphics[width=0.32\textwidth]{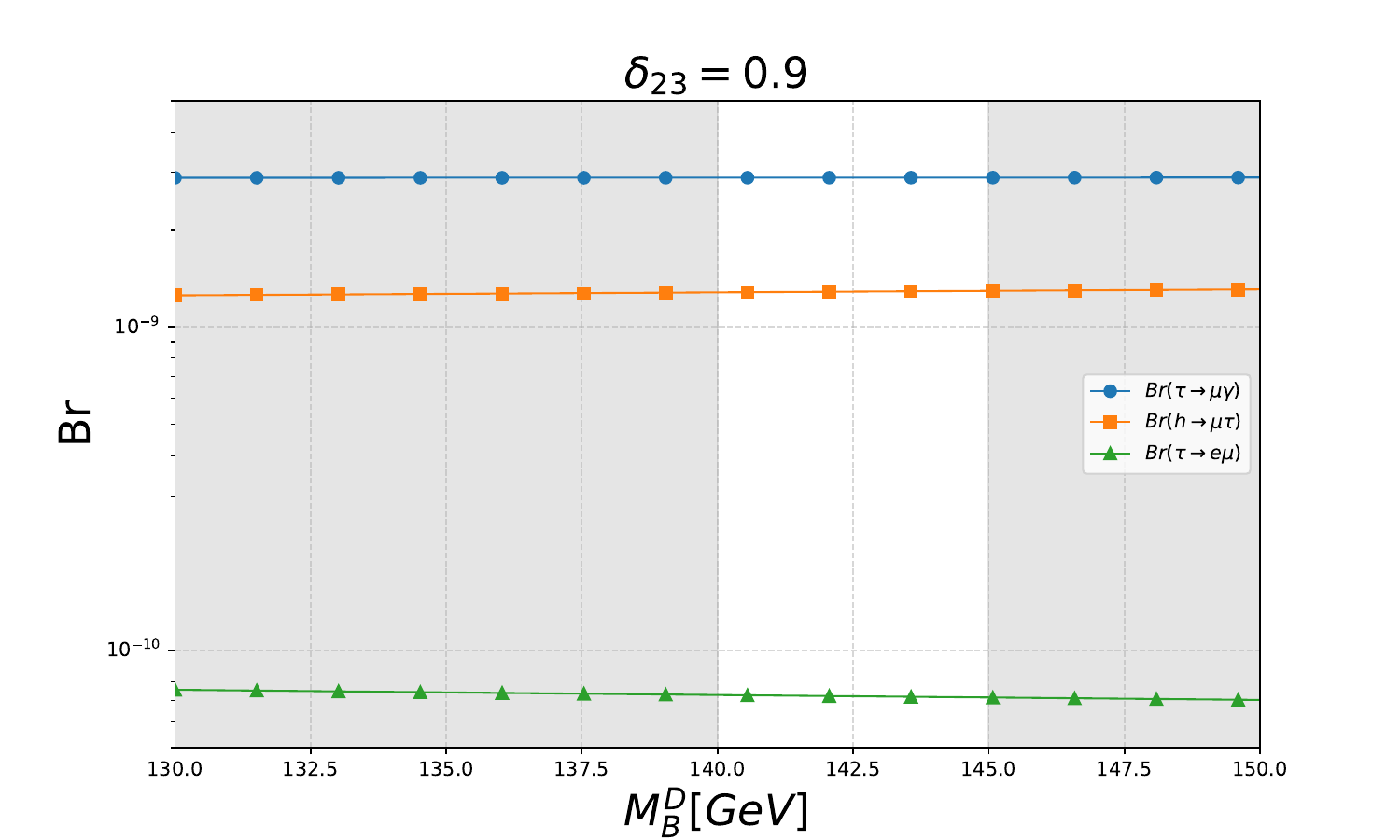}}

  \subfigure[]{\label{fig:sub1}\includegraphics[width=0.32\textwidth]{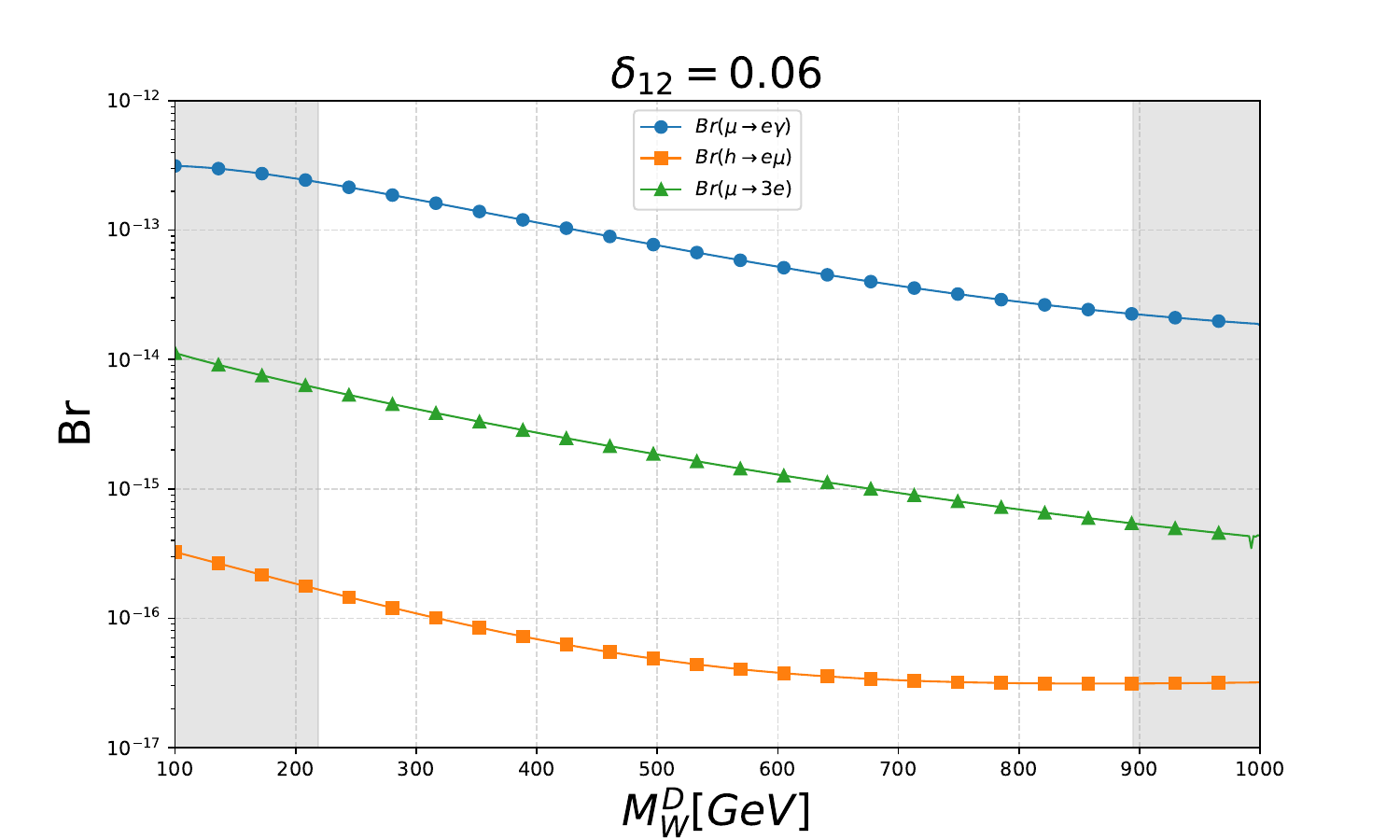}}
  \subfigure[]{\label{fig:sub1}\includegraphics[width=0.32\textwidth]{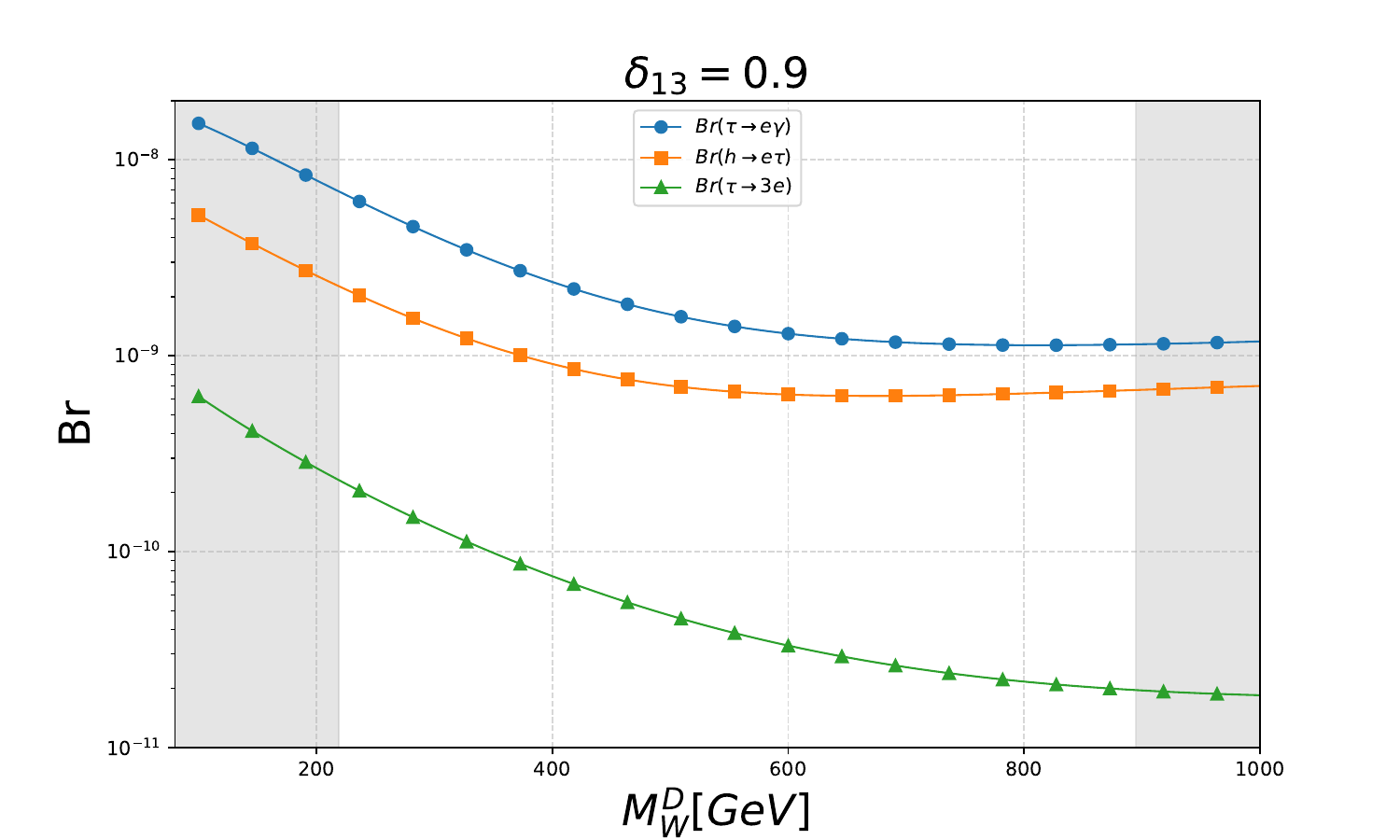}}
  \subfigure[]{\label{fig:sub1}\includegraphics[width=0.32\textwidth]{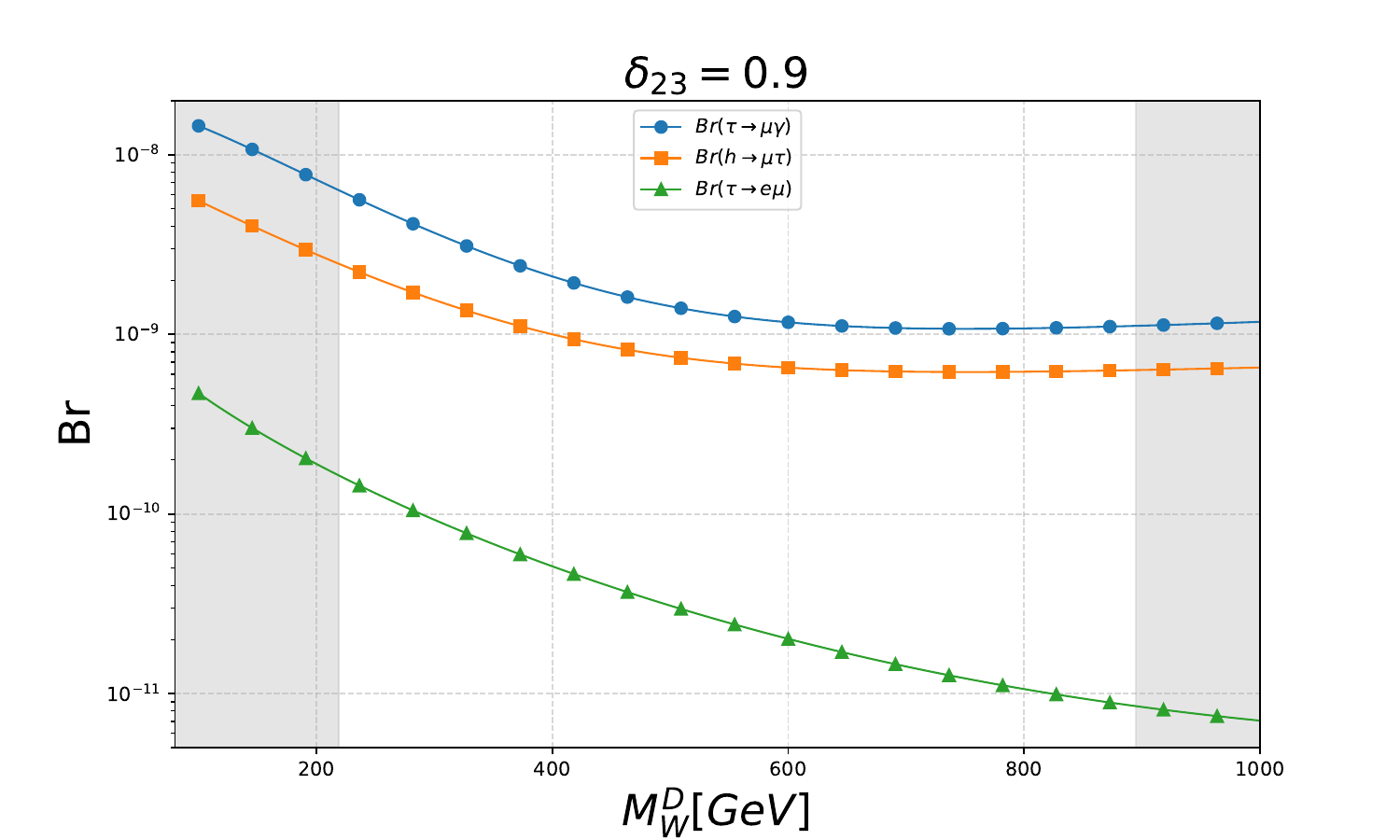}}

  \subfigure[]{\label{fig:sub1}\includegraphics[width=0.32\textwidth]{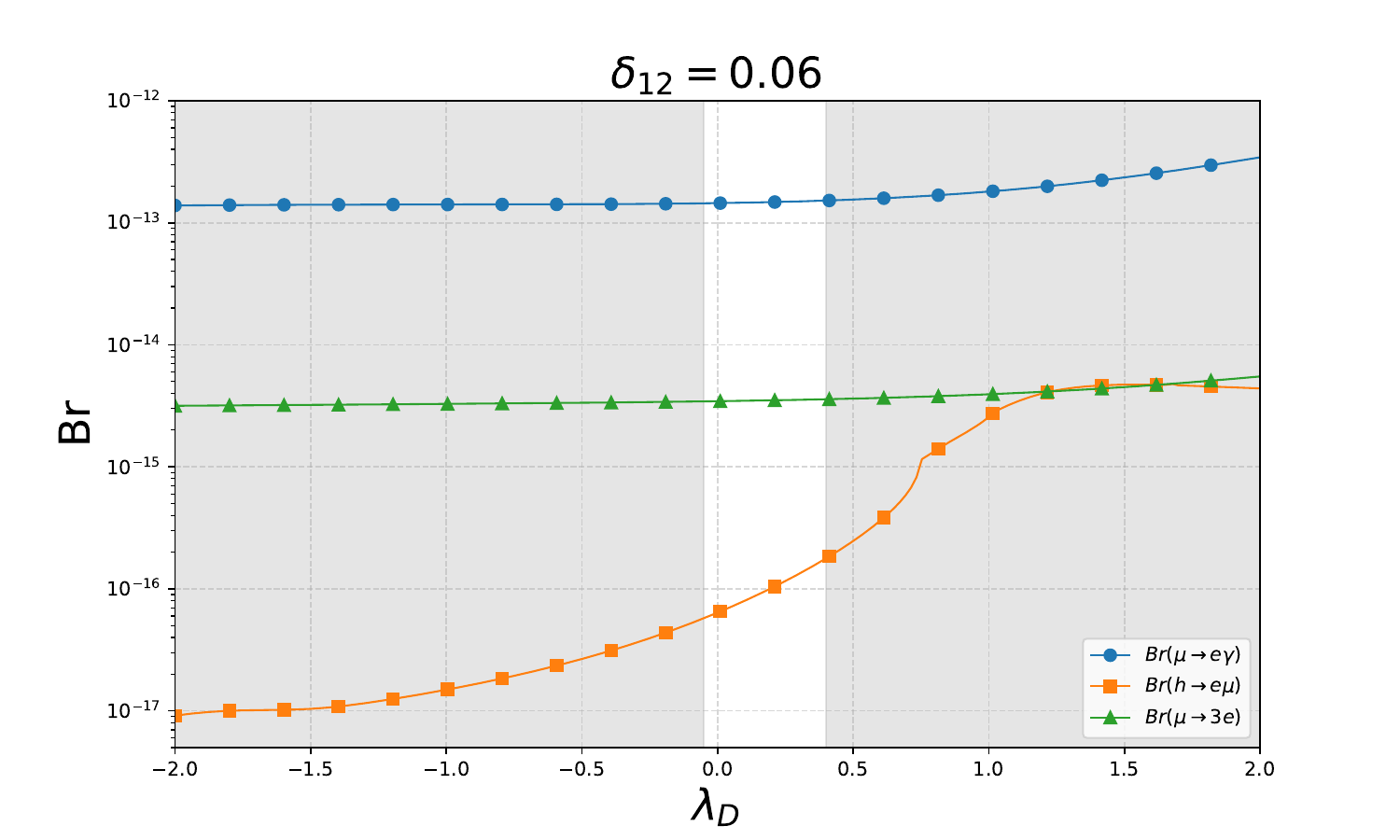}}
  \subfigure[]{\label{fig:sub1}\includegraphics[width=0.32\textwidth]{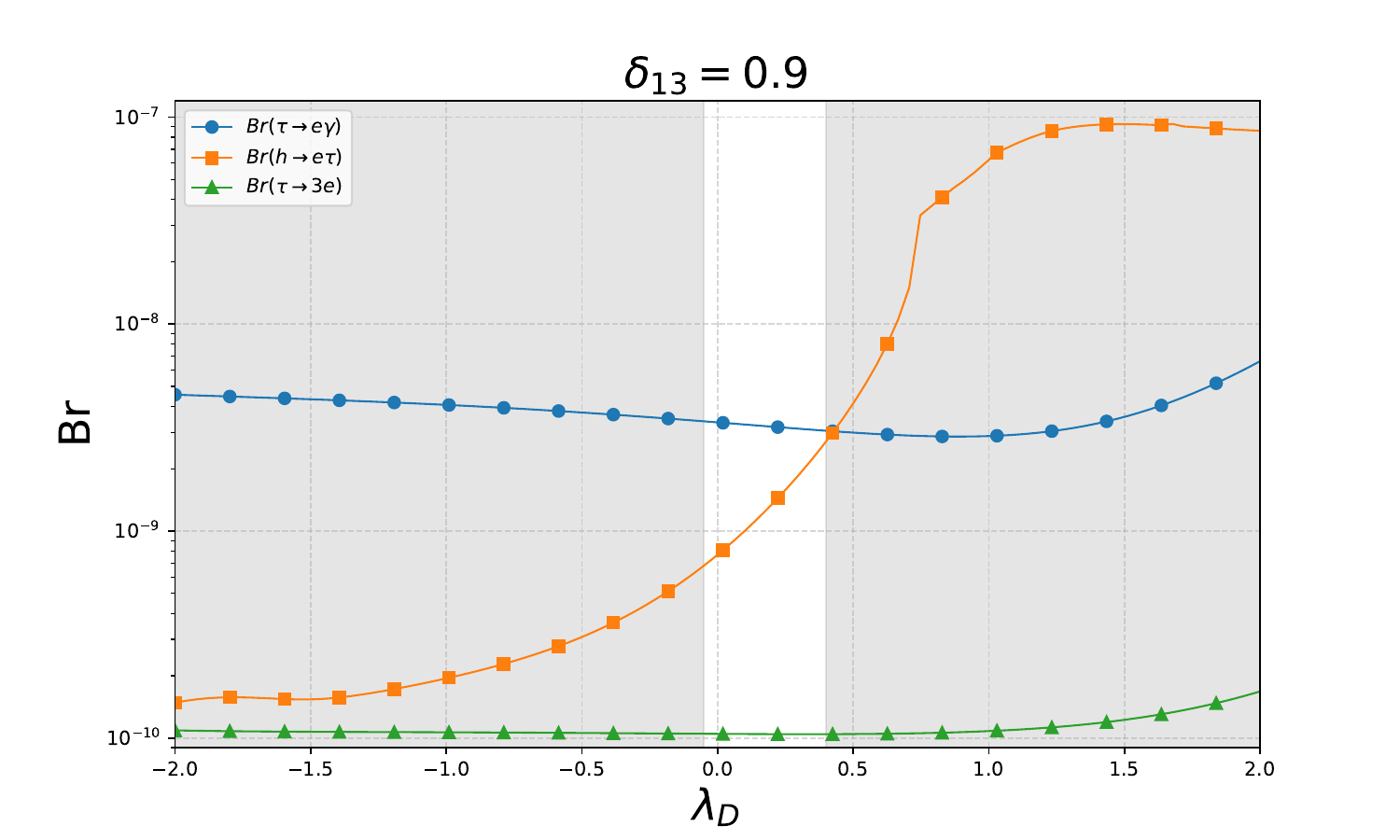}}
  \subfigure[]{\label{fig:sub1}\includegraphics[width=0.32\textwidth]{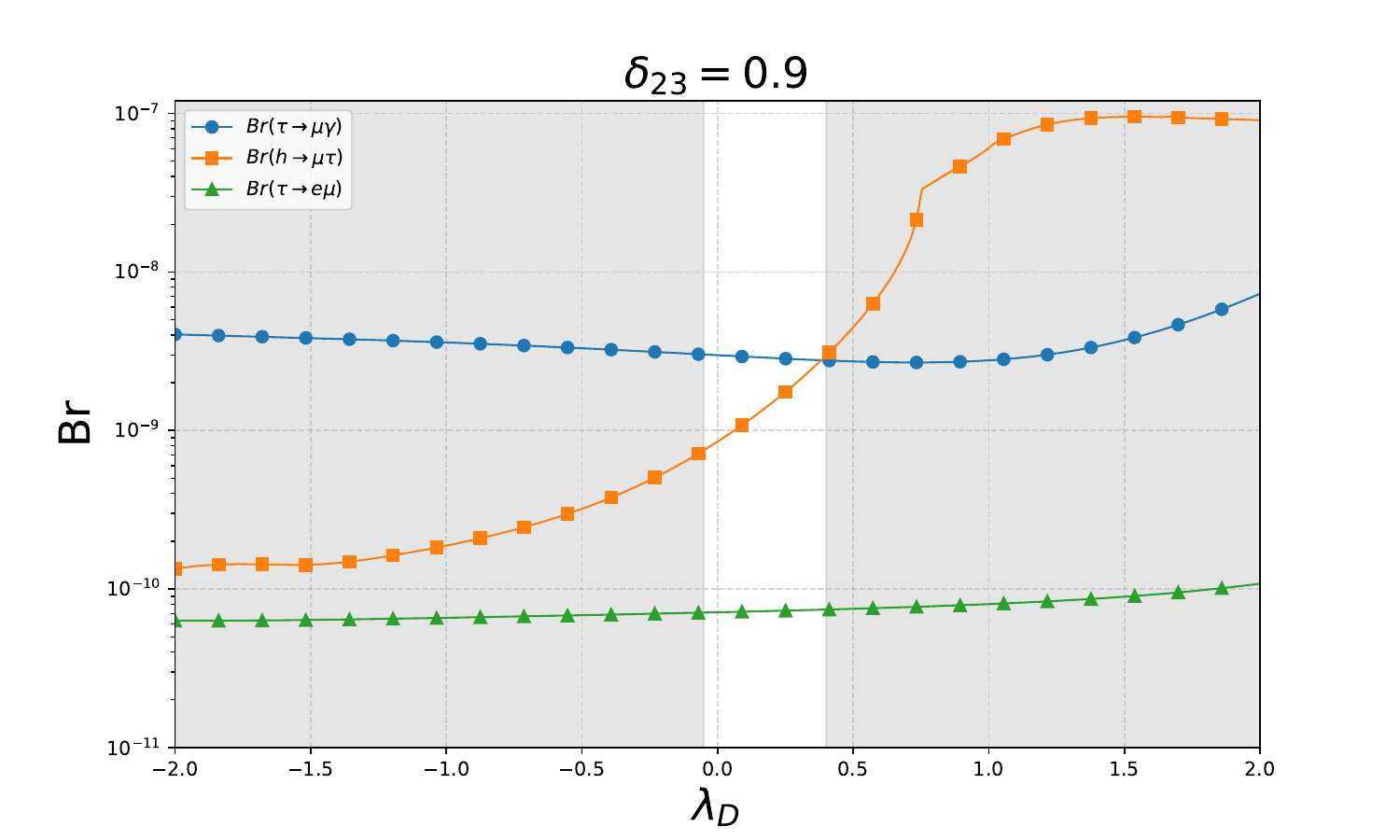}}

  \subfigure[]{\label{fig:sub1}\includegraphics[width=0.32\textwidth]{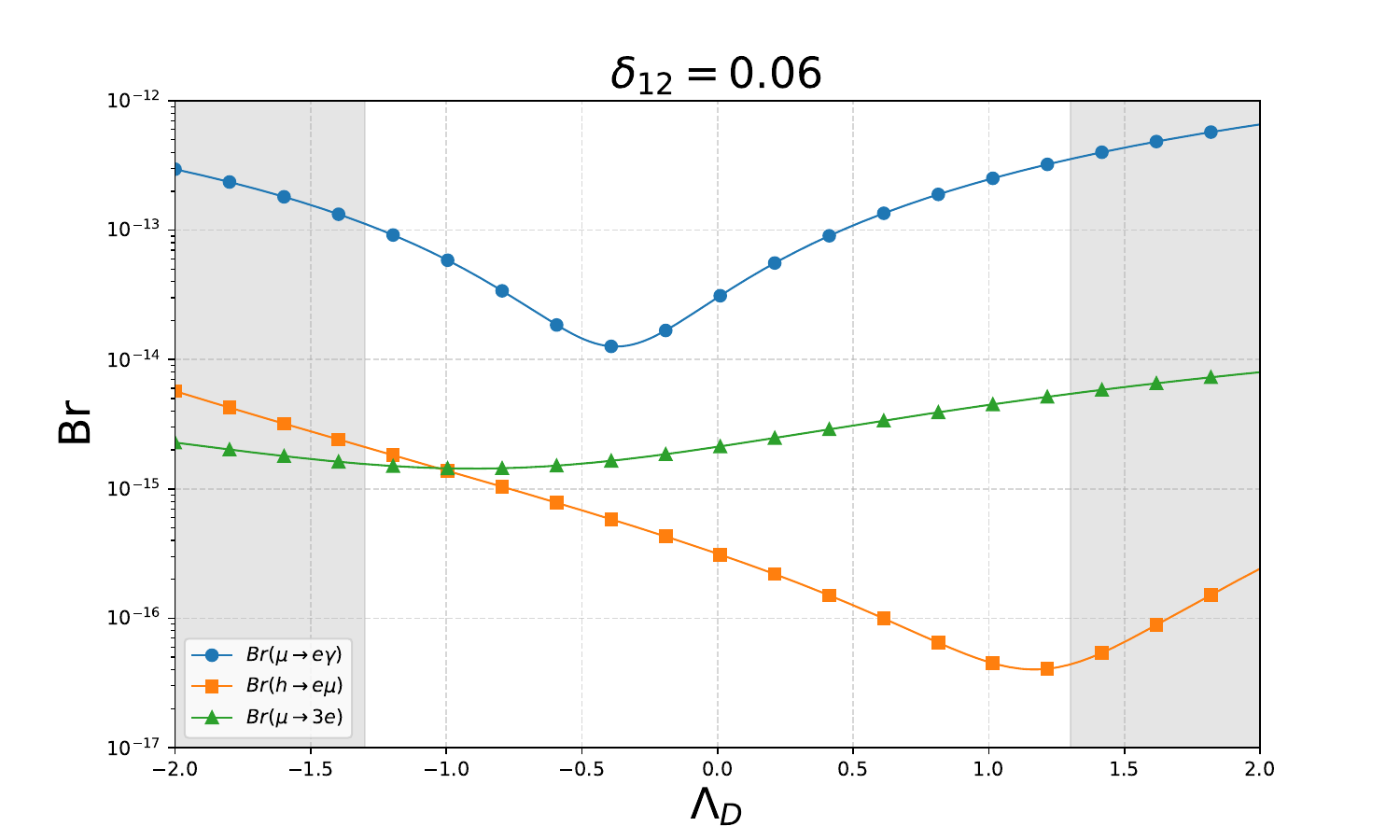}}
  \subfigure[]{\label{fig:sub1}\includegraphics[width=0.32\textwidth]{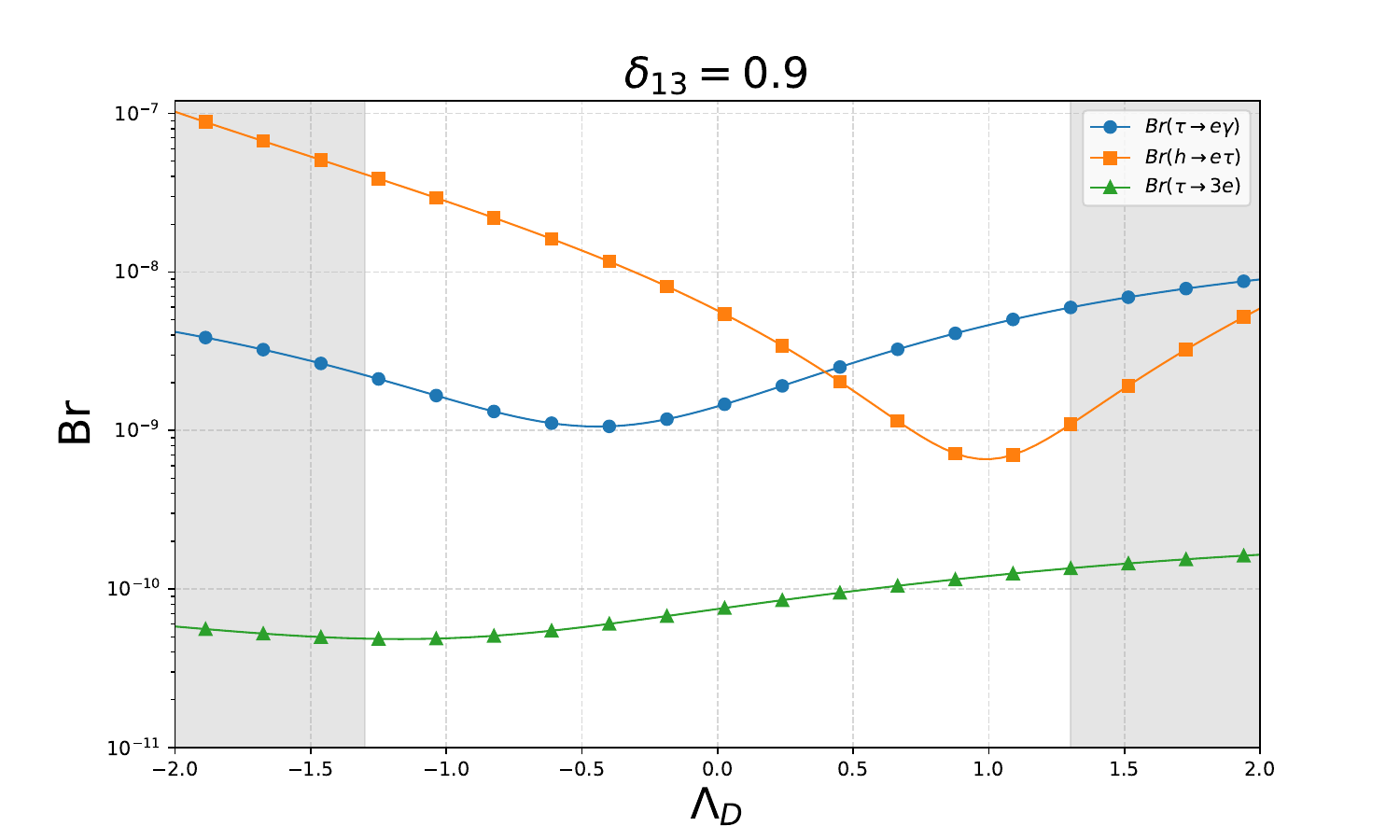}}
  \subfigure[]{\label{fig:sub1}\includegraphics[width=0.32\textwidth]{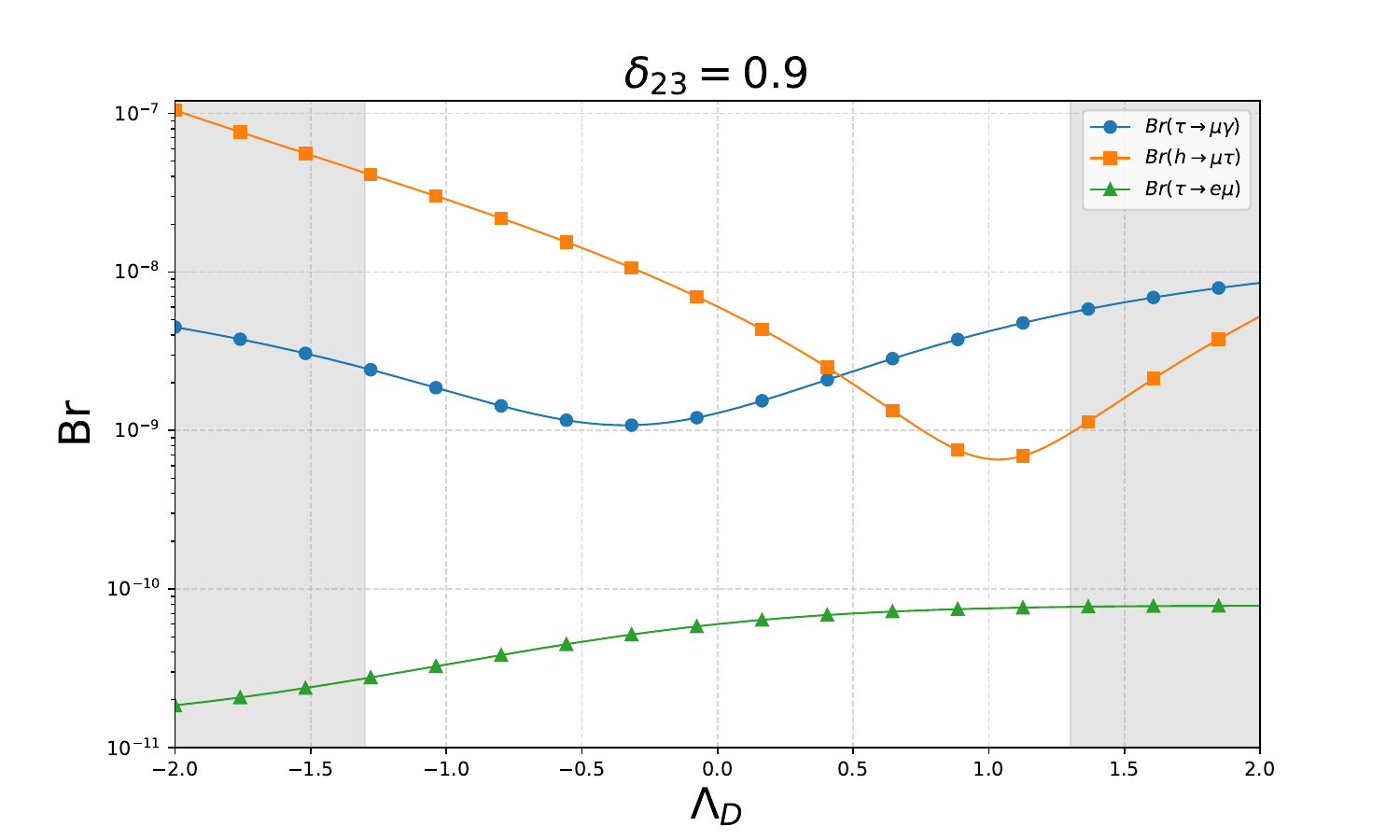}}

  \caption{The influence of Yukawa-like parameter $\lambda_D$,$\Lambda_D$ and Dirac gaugino parameters $M^D_B$ and $M_W^D$ with shadowed area represents exclusion by Higgs mass at 5$\sigma$. }
  \label{fig:new parameters}
\end{figure}

We then illustrated the impact of the MRSSM parameters on these LFV processes which shown in Fig.~\ref{fig:new parameters}. We find that within the allowed Higgs mass range, $M^B_D$ has no significant impact on these nine LFV processes. That is beacuse The term $g_1v_SM^B_D$ appears in the scalar particle mass matrix. Due to the constraints of $v_S$ at 1 TeV and the tadpole equation, the range of $M^B_D$  is significantly restricted. But the parameter $M^W_D$  in contrast, the parameter range of $M^W_D$ is larger than that of $M^B_D$. This is because  $M^W_D$is coupled to the T field, and due to the smallness of the T field's VEV, the  contribution of $M^W_D$ to the Higgs mass is not significant. They also appear in the tree-level Neutralino  Chargino Sleptons and Squarks mass matrixs. Since $M^B_D$ and $M^W_D$  affects the masses of particles, it can influence the LFV processes by altering  these particles in the decay loops. We see that in Fig.~\ref{fig:new parameters} (d,e,f), the LFV  decrease as $M_W^D$ increase.

Yukawa-like parameter $\lambda$ and $\Lambda$ appear in the Higgs Neutralino and Chargino mass matrix. So that these parameters will alse affect the self-energy correction of Neutralino, Chargino, Sneutrino and Sleptons, thereby influencing LFV. The impact of these parameters would be complex. Fig.~\ref{fig:new parameters}(g,h,i,j,k,l)  shows consistency with our analysis. In  Higgs mass matrix, the parameter $\lambda_D$ appears in terms like $\lambda_Dv_S$, which can significantly impact the  tree-level Higgs mass, consequently, influence the Higgs LFV.

The impact of the newly introduced parameters, $Y_x$ and $m_{\tilde{\nu}}$, on the LFV process are not very significant, as shown in Fig.~\ref{fig:new add parameters}. This is because the Yukawa coupling $Y_{\nu}$ is small. The right-handed Neutrinos, whose masses are influenced by $Y_x$, participate in the LFV processes through the coupling $Y_{\nu}$ to the left-handed Neutrinos. Therefore, their contribution is suppressed by the smallness of $Y_{\nu}$.

The impact of the right-handed Sneutrino mass parameter  $m_{\tilde{\nu}}$  on LFV processes is still negligible. That is beacuse  the contribution from right-handed Sneutrino vertices to LFV is suppressed by the coupling \( Y_\nu \).

\begin{figure}
  \centering
  \subfigure[]{\label{fig:sub1}\includegraphics[width=0.32\textwidth]{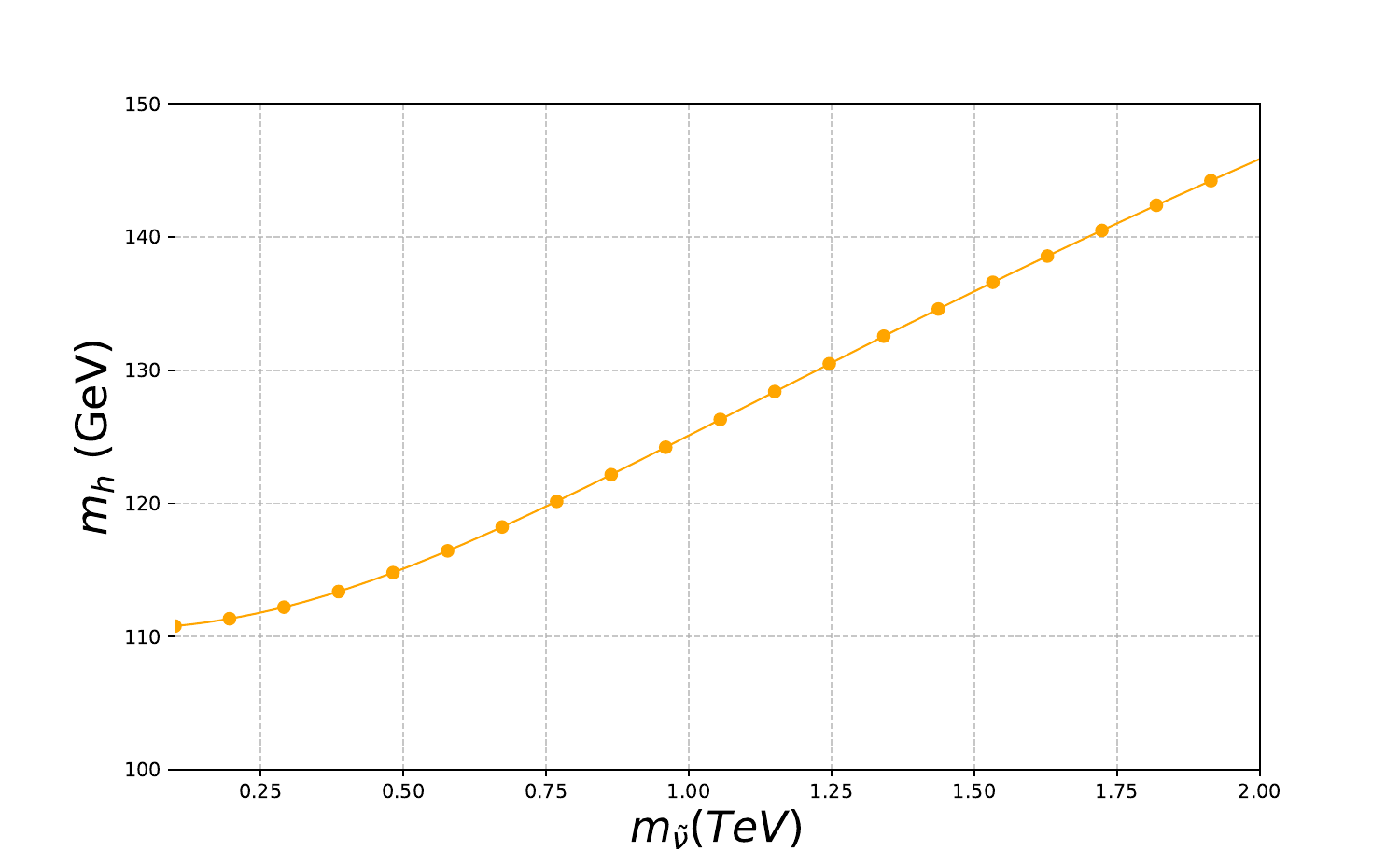}}
  \subfigure[]{\label{fig:sub1}\includegraphics[width=0.32\textwidth]{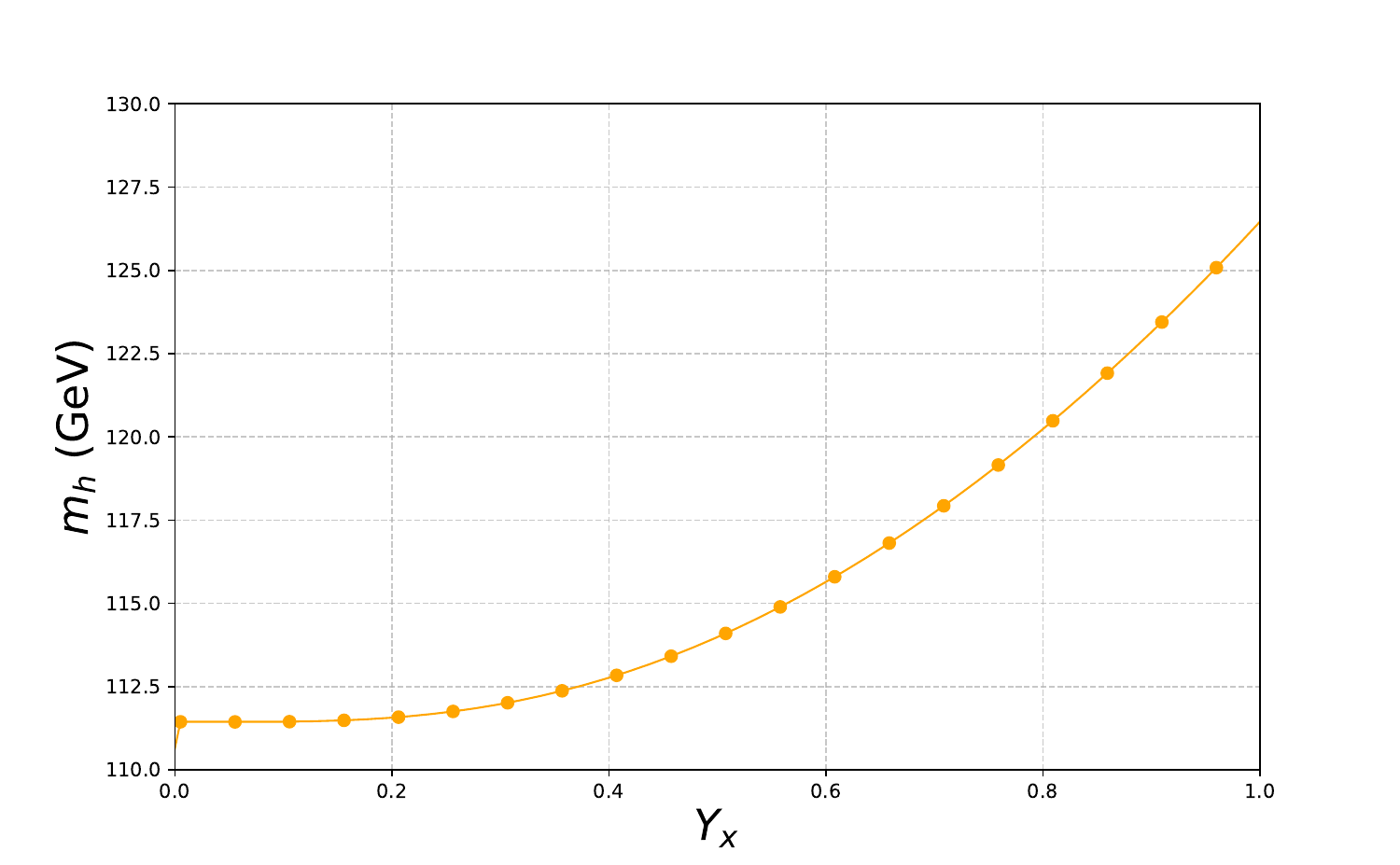}}

  \caption{The influence of newly parameters $Y_x$ and $m_{\tilde{\nu}}$ to Higgs mass}
  \label{fig:new  add parameters higgs}

\end{figure}

Since Sneutrinos appears in the Higgs self-energy correction, as shown in Fig.~\ref{fig:new  add parameters higgs}, these two new parameters do have a slightly  significant impact on the Higgs mass.

\section{Conclusion}

In this work, we analyzed the MRSSMSeesaw model, which extends the MRSSM by incorporating right-handed Neutrino Field to generate Neutrino masses via the Type-I seesaw mechanism. We focus  on LFV processes, including rare decays such as \(\ell_i \to \ell_j \gamma\), \(\ell_i \to 3\ell_j\), and Higgs decays \(h \to \ell_i \ell_j\). We derived the mass matrices for Higgs bosons, Neutralinos, Charginos, Sleptons, Neutrinos and Sneutrinos.

Our numerical analysis showed that LFV deacy rates are highly sensitive to the Slepton mixing parameters \(\delta_{ij}\). The experimental limit on  \(\text{Br}(\mu \to e\gamma)\) constrains \(\delta_{12} < 0.075\), while Higgs LFV decay rate  Br($h \to e\mu$) as  up to  \(\mathcal{O}(10^{-15})\) and Br(\(h \to e\tau\)), Br( \(h \to \mu\tau\)) reach  up to  \(\mathcal{O}(10^{-9})\). We found that \(\text{Br}(\ell_i \to \ell_j \gamma)\) and \(\text{Br}(\ell_{i} \to 3\ell_{j})\) decrease with increasing Slepton masses, while Higgs LFV deacy rates increase due to suppressed loop corrections to the Higgs mass. Because   $M^W_B$, $M^W_D$ and Yukawa-like parameters \(\lambda\) and \(\Lambda\) appear  at tree level mass  matrix of  Higgs Neutralinos, Charginos, Sneutrinos and Sleptons , they will have significant impact on the considered  LFV processes. Due to the smallness of $Y_{\nu}$, the new parameters $Y_x$ and $m_{\tilde{\nu}}$ negligibly affect LFV processes, but slightly influence the Higgs mass via Sneutrino loops.

In conclusion, our study provides insights into the MRSSMSeesaw model's phenomenology, through the analysis of LFV mixing parameters, we find that the non-diagonal elements which correspond to the generations of the initial lepton and final lepton are  the main sensitive parameters and LFV sources. Most of the parameters are able to break the experimental upper limit, providing new ideas for the search of new physics.

\begin{acknowledgments}
		
The work has been supported by the National Natural Science Foundation of China (NNSFC) with Grants No. 12075074, No. 12235008,  Natural Science
Foundation of Guangxi Autonomous Region with Grant No. 2022GXNSFDA035068, Hebei Natural Science Foundation with Grant No. A2022201017, No. A2023201041, the youth top-notch talent support program of the Hebei Province.		
      
\end{acknowledgments}


\end{document}